%% file: thesis.tex
\documentclass[12pt,epsfig,cite,a4paper,ref,pdftex,openany]{book}

\pdfoutput=1
\usepackage{amssymb,epsfig,here}
\usepackage{fancybox,colordvi,graphicx}
\usepackage{fancyhdr}
\usepackage{a4wide}
\usepackage{thesis}
\usepackage{multirow}
\usepackage{url}

\renewcommand{\sectionmark}[1]%
    {\markright{\thesection\ #1}}

\newcounter{saveenumi}

\begin{document}

\begin{titlepage}
\smallskip
\smallskip
\smallskip
\centerline{ \Large THE ANDRZEJ SOLTAN INSTITUTE FOR NUCLEAR STUDIES }
\smallskip
\smallskip
\smallskip
\vspace{7cm}
\centerline{ \Large \bf Development of electromagnetic calorimeter detectors}
\smallskip
\centerline{ \Large \bf and simulations for spectroscopic measurements of charmonium}
\smallskip
\centerline{ \Large \bf with PANDA}
\bigskip
\medskip
\centerline{ \large \bf  Dmytro Melnychuk} \vspace{0.15
\textheight} { \leftskip=0.5\textwidth \hsize=\textwidth \noindent
Ph.D. thesis written under the scientific \newline
supervision of \newline
Dr. hab. Boguslaw Zwieglinski.
\par
} \vfill \centerline{\Large Warsaw 2009}
\end{titlepage}

\tableofcontents

\setcounter{secnumdepth}{4}

\input{introduction}
\input{physics_motivation}
\input{panda_experiment}
\input{crystal_apd}
\input{monte_carlo}
\input{conclusions}
\input{acknowledgments}
\bibliographystyle{plain}
\bibliography{references}
\end{document}

%% file: introduction.tex
\chapter*{Introduction}\label{chap:Introduction}
\addcontentsline{toc}{chapter}{Introduction}
Charmonium, the bound system of charmed quark and antiquark, is studied experimentally since 1974, when the $\jpsi$ meson was discovered by two groups at SLAC and BNL. Since that time eight charmonium states below the mass of open charm ($\DDbar$) threshold were observed. The theoretical description of charmonium started with the non-relativistic potential model, which was motivated by the large mass of the $c$ quark. Improved by incorporation of spin-dependent terms and relativistic corrections it provides a good agreement with the mass spectra of charmonium below the $\DDbar$ threshold. The recent progress in approaches based on Lattice QCD and Effective Field Theory (NRQCD) also allowed a good theoretical description of the charmonium system. However, there are still many open question in this area, which can only be answered with high precision and high luminosity measurements.

The PANDA experiment, which is a part of the future Facility for Antiproton and Ion Research (FAIR) at Darmstadt, will provide possibility for the precise spectroscopy of charmonium produced in antiproton-proton annihilations. The cooled antiproton beam circulated in the High Energy Storage Ring (HESR) with a momentum between 1.5 \gevc and 15 \gevc will provide a peak luminosity up to $2 \cdot 10^{32} \, cm^{-2}s^{-1}$. In contrast to the $e^{+}e^{-}$ annihilations which excite only the states with quantum numbers of the photon, $J^{PC}=1^{--}$, in the $p\overline{p}$ collisions the charmonium states with all the quantum numbers can be formed directly. Therefore their masses and widths can be measured with high precision determined by the parameters of the beam and not limited by the detector resolution.

The subject of this thesis is the study of electromagnetic transitions in charmonium with PANDA. The possible registration of the $h_c$ state in charmonium, observed recently by the E760 and E835 Fermilab experiments, is used for demonstration of the physics performance of the PANDA detector. The measurement of the angular distribution of $\gamma$-rays from radiative transitions can be used for the verification of its $J^{PC}$ quantum numbers. The electromagnetic calorimeter of the PANDA detector is the crucial component for this studies from the point of view of signal registration and background suppression.

In \Refchap{chap:PhysMotivation} an overview of the charmonium physics is given, which includes the status of experimental studies as well as development of the theoretical description of charmonium. \Refchap{chap:panda} gives an overview of the current design status of the PANDA detector. The electromagnetic calorimeter (EMC) is described in more detail because of its interest for the presented work.

In \Refchap{sec:measurements} the results of measurements of energy resolution of $PbWO_4$ scintillators with the Avalanche Photodiode (APD) readout are presented. Low energy protons and $\gamma$-rays produced in the radiative capture reaction $^{11}B(p,\gamma)^{12}C$ were used for the response studies.

\Refchap{chap:phys:MonteCarlo} presents the results of Monte Carlo simulations which permit to evaluate the performance of the PANDA detector in clarifying several questions related to physics of the $h_c$ charmonium state. Chapter 5 contains conclusions.

%% file: physics_motivation.tex
\chapter{Physics motivation}\label{chap:PhysMotivation}
\section{Physics of charmonium}
\subsection{Quark model and QCD}
The Standard Model of particle physics is $U(1)\otimes SU(2) \otimes SU(3)$ field theory
and according to it all the matter is composed from the 6 quarks and 6 leptons divided into 3 generations and gauge bosons corresponding to 3 groups of symmetry of the Standard Model. One of the main components of the Standard Model is Quantum Chromodynamics or QCD. QCD describes interaction between quarks mediated by massless gauge bosons - gluons. QCD is responsible for 99$\%$ of the mass of proton and as a result of the baryonic matter of Universe.

\begin{figure}
\begin{center}
\includegraphics[width=0.6\swidth]{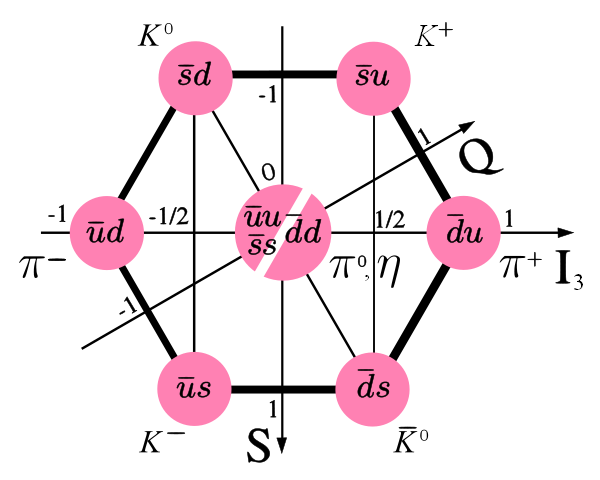}
\caption{Octet of the lightest spin 0 mesons, located on the plane strangeness (S) - isospin $I_{3}$ projection. $Q$ is charge axis.}
\label{fig:phys:meson_nonet}
\end{center}
\end{figure}

The notion of quark was introduced before establishing of QCD in quark model by Murray Gell-Mann in 1964 \cite{bib:GellMann:1964nj}, who proposed together with George Zweig the scheme of classification of mesons and baryons known at that time. The lightest mesons and baryons were classified according to the multiplets of $SU(3)$ flavour group in the following way. In case of the three flavours of quarks they lie in the fundamental representation, 3 (called the triplet) of flavour SU(3). The antiquarks lie in the complex conjugate representation $3^{*}$. Mesons consisted from a quark and an anti-quark and nine states made out of quark-antiquark pairs can be decomposed into a trivial representation (singlet) and an adjoint representation (octet):
\begin{equation}
3 \otimes 3^{*}=8 \oplus 1,
\end{equation}
which are filled with the 9 lightest spin 0 mesons (\Reffig{fig:phys:meson_nonet}).

\begin{figure}
\begin{center}
\includegraphics[width=0.6\swidth]{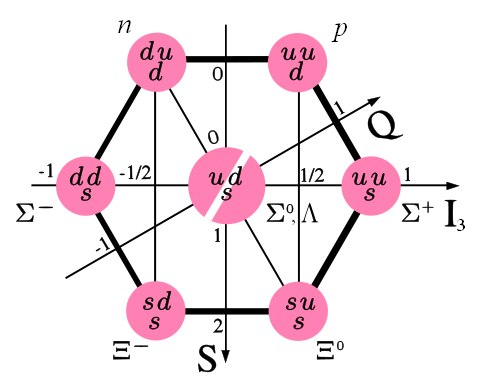}
\caption{Octet of the lightest spin 1/2 baryons, located on the plane strangeness (S) - isospin $I_{3}$ projection. $Q$ is charge axis.}
\label{fig:phys:baryon_octet}
\end{center}
\end{figure}

In case of baryons, the basis quark states can be considered as the six states of three flavours and two spins per flavour (the so called approximate spin-flavour SU(6) symmetry). The spin-flavour part of a barion wavefunction is decomposed in the following way:
\begin{equation}
6 \otimes 6 \otimes 6 = 56_S \oplus 70_M \oplus 70_M \oplus 20_A,
\end{equation}
where the 56 symmetric states are decomposed under flavour SU(3) into:
\begin{equation}
56 = 10^{\frac{3}{2}}\oplus 8^{\frac{1}{2}};
\end{equation}
i.e. a decouplet of spin-3/2 and an octet of spin-1/2 baryons. The octet of lightest spin-1/2 baryons in presented in \Reffig{fig:phys:baryon_octet}.

The quantum number of color was proposed for an explanation of the $\Delta^{++}$ baryon properties \cite{bib:Greenberg:1964pe}. It has spin $S=3/2$ and is composed of the three $u$ quarks with parallel spins and 0 relative angular momentum. In order to have an antisymmetric total wave-function to conform with Pauli principle there should be a hidden quantum number, which was called color. It can have three values: red, green and blue and the combination of all three colors gives white, i.e. a color neutral object. This corresponds to the case of baryons, where all three quarks have different color quantum numbers. In case of mesons each of them is built from the quark-antiquark pair in which a quark carries color and an anti-quark carries the corresponding anti-color. As a result all the observable particles are color-neutral.

The existence of three colors came after the measurements of the ratio of electron-positron annihilation into hadrons and muon pairs:
\begin{equation}
R=\frac{\sigma(e^{+}e^{-}\rightarrow hadrons)}{\sigma(e^{+}e^{-}\rightarrow\mu^{+}\mu^{-})}.
\end{equation}
In the lowest order it is equal to $N\sum_{i}q_{i}^{2}$, where $N$ is the number of colors and the sum is taken over quarks with $2 m_{q}<E$, $E$ - energy of the process. In the energy region below the $b$-quark production, the sum is equal $\frac{30}{9}$ for $N=3$, which is in good agreement with experimental data.

The additional step between discovery of quarks, color quantum number and the QCD is establishing of dynamical role of color octet gluons, which was proposed by Fritzsch and Gell-Mann \cite{bib:Fritzsch:1972jv}. Although the word Lagrangian was not mentioned in their work, they made complete description of two terms that make up the QCD Lagrangian.

QCD is the relativistic quantum field theory of strong interactions and describes interaction of quarks and gluons according to the QCD Lagrangian:
\begin{equation}
\label{eq:theory:QCD_lagrangian}
\mathcal{L}_{QCD}=-\frac{1}{4}G_{a}^{\mu \nu}G^{a}_{\mu \nu}+ \sum_{f}\overline{q_{f}}[i\gamma^{\mu}D_{\mu}-m_{f}]q_{f},
\end{equation}
where
\begin{equation}
G_{a}^{\mu \nu}=\partial^{\mu} A^{\nu}_{a}-\partial^{\nu} A^{\mu}_{a}+g_s f_{a}^{bc}A^{\mu}_{b}A^{\nu}_{c}
\end{equation}
is the gluon field strength tensor, and
\begin{equation}
D^{\mu}=\partial^{\mu}-i\frac{g_s}{2}A^{\mu}_{a}\lambda^{a}
\end{equation}
is the gauge covariant derivative, $g_s$ is the gauge constant, $f$ denotes the quark flavour, $\gamma^{\mu}$ are the Dirac $\gamma$-matrices, $A^{\mu}_{a}$ correspond to the gluon field with $a$ running from 1 to 8, $\lambda^{a}$ correspond to $3 \times 3$ matrices - generators of $SU(3)$ group, $f_{a}^{bc}$ are structure constants of $SU(3)$ group and $\mu$, $\nu$ are Lorentz indexes running from 1 to 4.

QCD has some similarities to QED with the essential differences coming from it being non-abelian theory (SU(3) group, which is a base group of the QCD symmetry is non-abelian in contrast to U(1) symmetry of QED). This property made implications to the basic interaction vertex. In QED the vertex connects only two fermions (electrons) with a gauge boson (photon), whereas in QCD in addition to a two quark-gluon vertex there are tree- and four-gluon interaction vertices, which describe interaction between gluons.

The gauge constant is related to an effective coupling constant of the strong interaction by:
\begin{equation}
\alpha_s=\frac{g_s^2}{4 \pi}.
\end{equation}
The coupling constant $\alpha_s$ depends on the energy scale $\mu$ and in the lowest order can be expressed as:
\begin{equation}
\label{eq:theory:alpha_s}
\alpha_s(\mu)=\frac{2 \pi}{[11-(2/3)n_f]\ln(\mu/\Lambda)},
\end{equation}
where $n_f$ - number of quark flavours with the mass less than $\mu$ and $\Lambda$ - QCD scale parameter. At the mass of $Z^{0}$ boson $\alpha_s(Z^{0})=0.1172 \pm 0.0020$.

\begin{figure}
\begin{center}
\includegraphics[width=0.7\swidth]{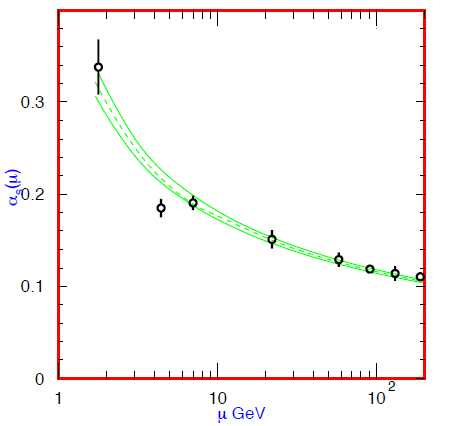}
\caption{Summary of the measured values of $\alpha_s(\mu)$ \cite{bib:pdg}.}
\label{fig:phys:alpha_s}
\end{center}
\end{figure}

As can be seen from the above expression (\Refeq{eq:theory:alpha_s}) and from the measured values (\Reffig{fig:phys:alpha_s}), $\alpha_s$ decreases for higher energies or small distances and approaches zero at $\mu \rightarrow \infty$, a property called the asymptotic freedom.

Although the Feynman rules can be derived from the QCD Lagrangian (\Refeq{eq:theory:QCD_lagrangian}), such a technique can be used for the calculation of strong processes only at large momenta, where coupling constant becomes very small. At small momenta, and correspondingly large coupling constant the perturbative methods do not work, in that case different effective theories or numerical solutions such as Lattice QCD are used to calculate strong processes. Their application for description of the charmonium is discussed in \Refsec{sec:phys:charm_theory}.

\subsection{History of charmonium physics}\label{sec:theory:history_charmonium}
The review of charmonium physics outlined below is based on the review articles by Bianco et al.\cite{bib:Bianco:2003vb}, which covers theory of charmonium, Bettoni and Calabrese \cite{bib:Bettoni:2005bb} which describes the current status of experiments, QWG report \cite{bib:Brambilla:2004wf} which covers both theoretical and experimental aspects of charmonium physics, reviews given in several PhD theses (e.g. \cite{bib:joffe-2005}, \cite{bib:Gollwitzer:1993rp}) on charmonium physics and on original papers which are referred to in the text.

The first proposal for existence of the fourth quark flavour appeared in 1964 in the work of Bjorken and Glashow \cite{bib:Bjorken:1964gz}. They proposed the quark flavour which they termed "charm" to combine quarks in doublets by analogy with lepton doublets ($e^{-}$,$\nu_{e}$), ($\mu^{-}$,$\nu_{\mu}$) and as a consequence extension of the SU(3) flavour group to SU(4). The argument for such an extension comes from the requirement of renormalizability of the theory. The triangle diagram, which is represented by fermion loop with three external spin-one lines, creates the so-called Adler-Bell-Jackiw anomaly. The non-conservation of an axial current produced by this diagram leads to infinities to higher orders. The only way to get rid of them is to cancel them among the different fermion loops. In the Standard Model this requires that electric charges of all the fermions - quarks and leptons should add up to zero. For electron, muon, up, down and strange quarks their charges add up to -2. This means that the fourth quark with the three colors is needed with the charge $+\frac{2}{3}$ to make the sum equal 0.

The experimental evidence for the existence of the fourth quark came from the Cabibbo theory \cite{bib:Cabibbo:1963yz}. He proposed a model which explained suppression of the strangeness-changing semi-leptonic weak decay ($\Delta S=1$) with respect to strangeness conserving decay ($\Delta S=0$). According to his idea in weak interactions the flavour eigenstate $d$ does not participate but rather a mixture of $d$ and $s$ quarks, $d_{c}=d \cos(\theta_{c}) + s \sin(\theta_{c})$. The angle $\theta_{c}$ is called Cabibbo angle with an experimental value of 0.25. However, this model predicted the existence of a neutral current with $\Delta S=1$, which leads, in particular, to the contribution to $K_{L}\rightarrow \mu^{+} \mu^{-}$ several orders of magnitude larger than experimentally observed.

The solution to this problem was proposed by Glashow, Iliopoulos and Maiani \cite{bib:Glashow:1970gm} in 1970. They proposed the existence of the fourth quark with a charge +2/3 and a flavour called \textit{charm} which participates in weak interaction in a doublet with the state $s_{c}=s\cos(\theta_{c})-d \sin(\theta_{c})$ orthogonal to the $d_c$ state; i.e. the two quark doublets which participate in weak interaction are:

\begin{equation}
{u \choose d_{c}=d \cos \theta_{c}+s \sin \theta_{c}} {c \choose s_{c}=s \cos \theta_{c} -d \sin \theta_{c}}.
\end{equation}

Due to $d_{c}$ and $s_{c}$ orthogonality the strangeness-changing ($\Delta S=1$) term in the neutral current,
\begin{equation}
J_{\mu}^{(+)}=\ldots+(s_{c} \overline{d_{c}} + \overline{s_{c}}d_{c})\sin \theta_{c} \cos \theta_{c} \,,
\end{equation}
vanishes.

But in spite of these arguments, the existence of the fourth quark and validity of the quark model itself became widely accepted only with the discovery of the $\jpsi$ meson in 1974. The $\jpsi$ was discovered almost simultaneously by two groups from the USA, neither of which was searching for charm.

\begin{figure}
\begin{center}
\includegraphics[width=1.0\swidth]{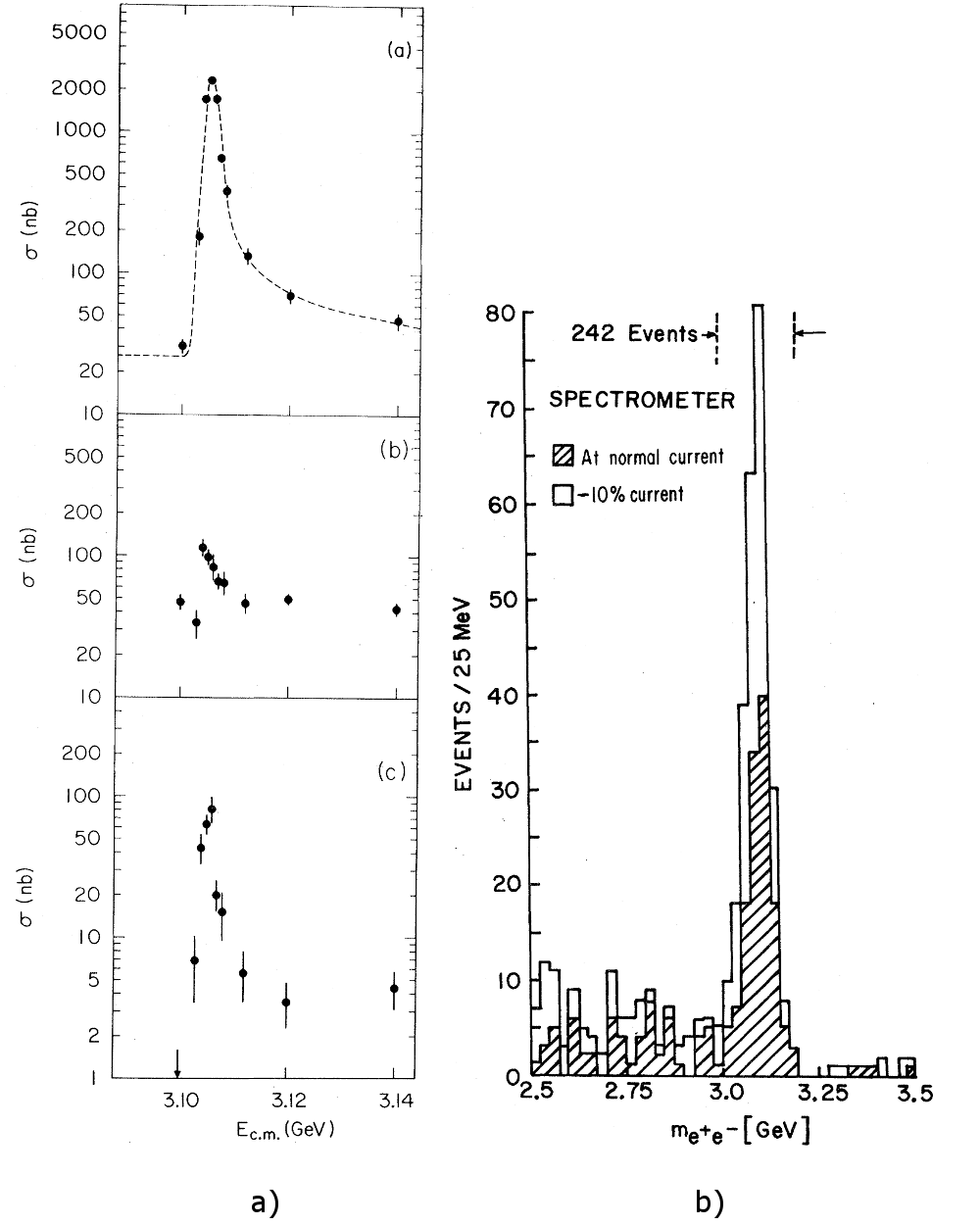}
\caption{$\jpsi$ discovery plots: by the SLAC group in $e^{+}e^{-}$ collisions \cite{bib:Augustin:1974xw}(a) and by the BNL group in $p + Be \rightarrow e^{+} e^{-} + X$ reaction \cite{bib:Aubert:1974js}(b).}
\label{fig:phys:jpsi_discovery}
\end{center}
\end{figure}

The group of Richter at SLAC's SPEAR $e^{+}e^{-}$ collider, reported the resonance called $\psi$ which was observed as a peak in cross-section of the reaction $e^{+}e^{-}\rightarrow e^{+}e^{-}, \mu^{+} \mu^{-}, hadrons$ \cite{bib:Augustin:1974xw}. The reported mass of the resonance was $3.105 \pm 0.003$ \gev and its width $\Gamma < 1.3$ \mev.
The corresponding discovery plot is presented in \Reffig{fig:phys:jpsi_discovery}(a). The three plots top to bottom show enhancements in the cross-section for $e^{+}e^{-}, \mu^{+} \mu^{-}$ and hadron final states, respectively.

In the Brookhaven National Laboratory (BNL) the group led by Ting reported a particle called $J$ in the $e^{+}e^{-}$ invariant mass in the reaction
\begin{equation}
p + Be \rightarrow e^{+} e^{-} + X
\end{equation}
in the fixed target experiment at 28 GeV \cite{bib:Aubert:1974js}. The reported mass of the enhancement is $M(e^{+}e^{-})=3.1$ \gev. The plot is presented in \Reffig{fig:phys:jpsi_discovery}(b).

Immediately after announcing the discovery of new states by these two groups the ADONE $e^{+}e^{-}$ collider at Frascati, which was designed for a maximum center of mass energy of 3.0 \gev, boosted current in their magnets over the design limits and confirmed existence of the 3.1 \gev resonance \cite{bib:Bacci:1974za}. The papers announcing new resonance discovery appeared in the December 1974 issue of the Physical Review Letters and the particle has been henceforth called $\jpsi$. Another confirmation of $\jpsi$ came from DESY in the wide angle elastic $e^{+}e^{-}$ scattering experiment \cite{bib:Braunschweig:1974mw}.

In all the quoted experiments the observed width of the peak was dominated by experimental resolution. The proper width of the $\jpsi$ state could be determined from the measurements of the total integrated cross-section and the leptonic branching ratio, both of which were measured experimentally. From the Breit-Wigner formula the $\jpsi$ width has been estimated at the level of one hundred keV, which is much smaller than the \mev width dominated by beam resolution. Most resonances in strong interaction known at that time were much wider, having widths of the order of \mev and larger. The narrow width of the discovered $\jpsi$ particle led to its interpretation as a bound system of the fourth "charm" quark-antiquark pair $c\overline{c}$. Such an interpretation was proposed by Appelquist, and Politzer \cite{bib:Appelquist:1975pa} and De Rujula and Glashow \cite{bib:DeRujula:PhysRevLett.34.46}.

\begin{figure}
\begin{center}
\includegraphics[width=0.8\swidth]{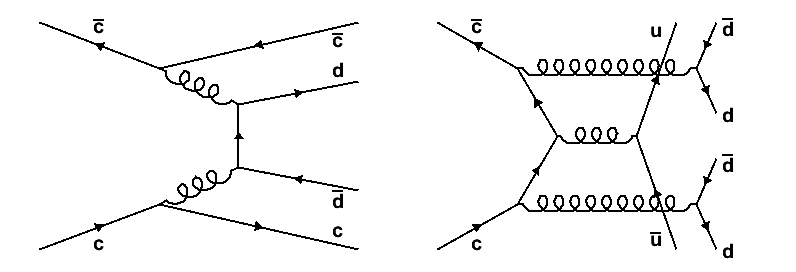}
\caption{Feynman diagrams demonstrating OZI rule. The left diagram demonstrates the OZI-allowed decay of a charmonium state into a $D^{+}D^{-}$ pair, right diagram corresponds to an OZI-suppressed decay of a charmonium state into light hadrons.}
\label{fig:phys:ozi}
\end{center}
\end{figure}

\begin{figure}
\begin{center}
\includegraphics[width=0.9\swidth]{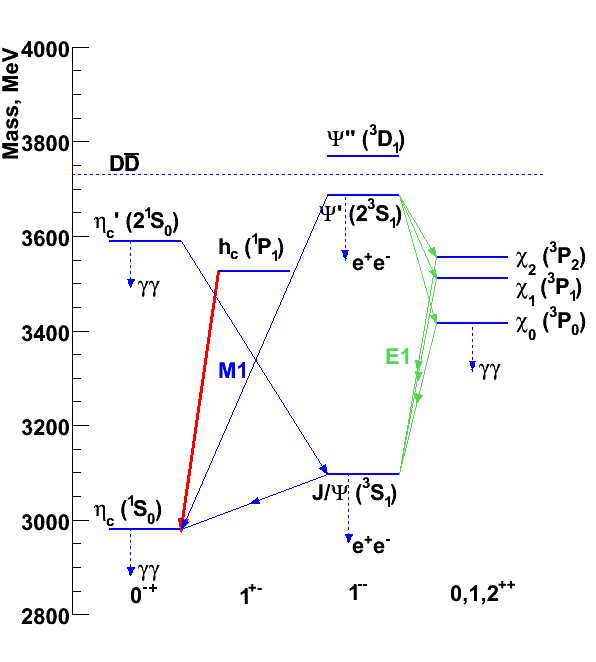}
\caption{The spectrum of charmonium states below the \DDbar threshold.}
\label{fig:phys:CharmoniumSpectrum}
\end{center}
\end{figure}

The small width of $\jpsi$ can be interpreted in terms of the OZI (Okubo-Zweig-Iizuka) rule \cite{bib:Okubo:1963fa}, \cite{bib:Zweig:1964jf}, \cite{bib:Iizuka:1966fk}. The OZI rule states that strong processes, which can  be described by Feynman diagrams containing only disconnected quark lines between the initial and final states are strongly suppressed in comparison with diagrams with connected quark lines. Two examples related to the processes with a charmed quark-antiquark pair are presented in \Reffig{fig:phys:ozi}. The left diagram corresponds to the OZI-allowed decay of the charmonium state. It is assumed that the mass exceeds the $D \overline{D}$ threshold, which corresponds for example to the $\psi''$ decay (see \Reffig{fig:phys:CharmoniumSpectrum}). The right diagram corresponds to the OZI-suppressed $\jpsi$ decay into light hadrons.

The OZI rule can be explained qualitatively in the following way. Processes with disconnected quark lines in the initial and final states are connected via gluons. All the intermediate gluons must combine in a way to preserve all the strong interaction quantum numbers. Since a gluon carries a color quantum number and mesons are colorless there should be at least 2 intermediate gluons, connecting the initial and final states. On the other hand vector mesons, such as $\jpsi$, have charge parity quantum number $C=-1$, which requires an odd number of gluons. As a result, at least 3 intermediate gluons participate in $\jpsi$ strong decay. Since the decay mesons are massive, the gluons must be very energetic ("hard") and the coupling constant is small in comparison with the OZI-allowed decay where gluons can be "soft" and the coupling constant is large.

The first radial excitation of charmonium, $\psi'$, was found at SLAC \cite{bib:Abrams:1974yy} soon after the $\jpsi$ observation. This state also appeared to be very narrow with the observed width consistent with beam resolution ($\approx 1$ \mev). Because the mass of $\psi'$ is elevated (3686 \mev), but still below the $D\overline{D}$ threshold, the narrowness of the resonance was also interpreted as a result of the OZI rule.

\begin{figure}
\begin{center}
\includegraphics[width=0.8\swidth]{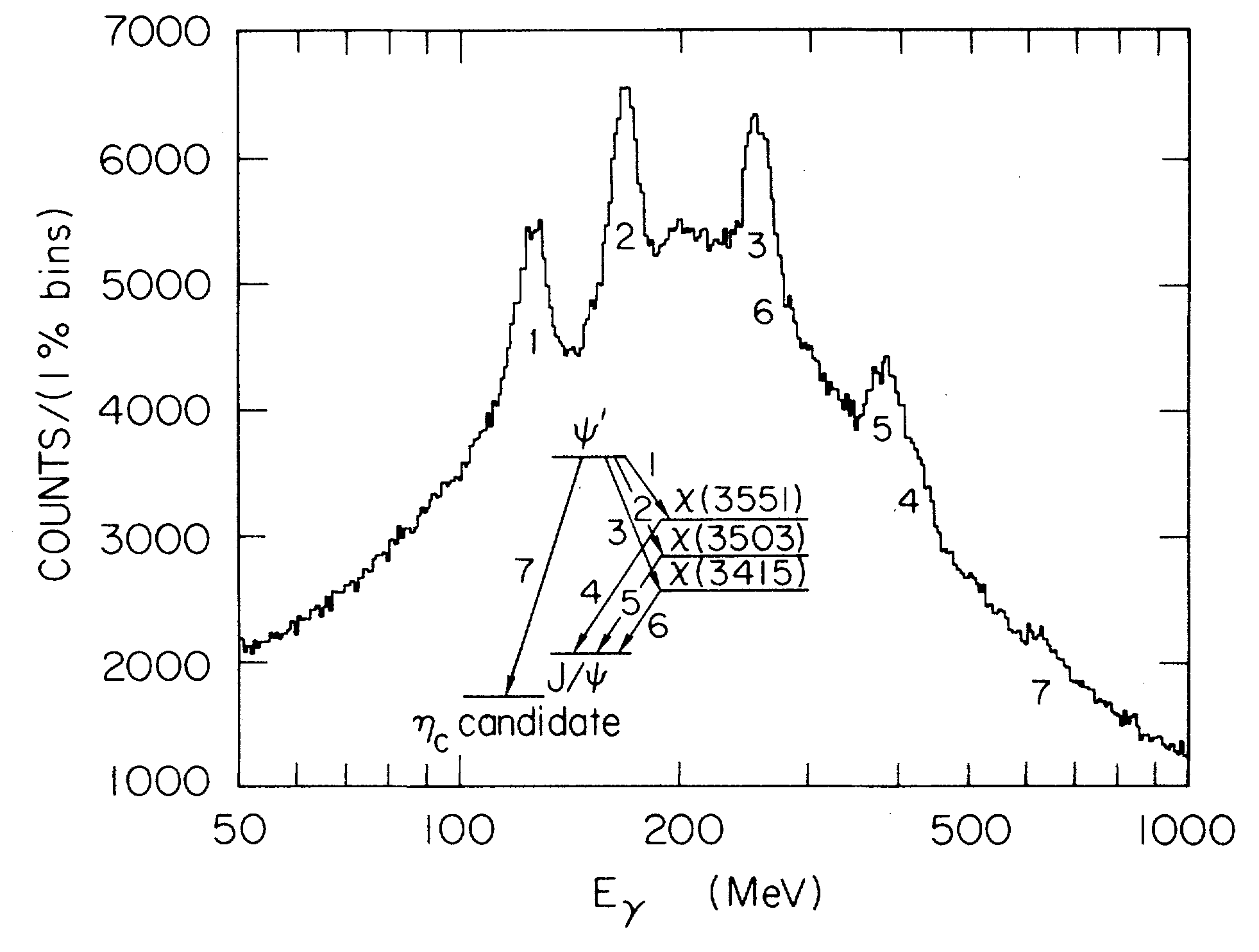}
\caption{The inclusive photon spectrum from $\psi'$ and $\chi_{cJ}$ radiative decays \cite{bib:Partridge:PhysRevLett.45.1150} (Crystal Ball).}
\label{fig:phys:CrystalBall}
\end{center}
\end{figure}

Other charmonium states ($\chi_{c0}$, $\chi_{c1}$, $\chi_{c2}$, $\eta_{c}$) were measured at SLAC with the series of improved detectors Mark I, Mark II, Mark III and  Crystal Ball. The $\chi_{cJ}$ and $\eta_{c}$ states observed in the radiative decay of $\psi'$ in the Crystal Ball experiment \cite{bib:Partridge:PhysRevLett.45.1150} are presented in \Reffig{fig:phys:CrystalBall}.

The radial excitation of the charmonium ground state, $\eta_{c}'$, was firmly established in 2002 by the Belle experiment \cite{bib:Choi:2002na} as a peak in the invariant mass distribution of $K_{S} K^{-} \pi^{+}$ with the mass 3654 \mev in a sample of exclusive $B \rightarrow K K_{S} K^{-} \pi^{+}$ decays collected with the Belle detector at the KEKB asymmetric $e^{+}e^{-}$ collider. The $\eta_{c}'$ was confirmed soon afterwards by the CLEO \cite{bib:Asner:PhysRevLett.92.142001} and Babar \cite{bib:Aubert:PhysRevLett.92.142002} experiments.

The $h_c$ is the last among the observed discrete charmonium states. This is a major topic of the present thesis, therefore a separate chapter (\Refsec{sec:phys:hc_observation}) is devoted to the history of its discovery.

\Reffig{fig:phys:CharmoniumSpectrum} shows the spectrum of the charmonium states below the $D\overline{D}$ threshold discovered so far. The states of charmonium are classified according to the standard spectroscopic notation $n^{2s+1}L_{J}$, where $n$ numerates the radial excitations and $L$ denotes the orbital angular momenta S, P, D, ... for L = 0, 1, 2, ... respectively. Each constituent quark has spin $\frac{1}{2}$ therefore the total spin quantum number can be either S=0 or 1. J is the quantum number of the total momentum, a vector sum of the spin and orbital momenta. An alternative notation used to denote the charmonium states is $J^{PC}$, where $P$ is the space parity and $C$ - is the charge conjugation parity. For a quark-antiquark system these can be expressed as:

\begin{equation}
\label{eqn:phys:p_parity}
P=(-1)^{L+1},
\end{equation}
and
\begin{equation}
\label{eqn:phys:c_parity}
C=(-1)^{L+S},
\end{equation}
respectively.

According to Eqs. \ref{eqn:phys:p_parity} and \ref{eqn:phys:c_parity} the states with natural spin-parity $P=(-1)^J$ should have $S=1$ and $CP=+1$, respectively. The states with natural spin-parity and $CP=-1$ ($0^{+-}$, $1^{-+}$, $2^{+-}$,  etc) are forbidden in the $q\overline{q}'$ model as well as the state $J^{PC}=0^{--}$. However, exotic quantum numbers can be realized for \textit{exotic} mesons (glueballs, hybrids, diquark-antidiquarks, etc).

All the known charmonium states below the $D\overline{D}$ threshold with their masses and widths are summarized in \Reftbl{tab:phys:charmonium_summary}.

\begin{table}
\begin{center}
\begin{tabular}{|l|l|c|c|c|}
\hline
Notation& $n^{2s+1}L_{J}$& $J^{PC}$ & Mass, MeV & Width, MeV \\
\hline
$\eta_{c}$ & $1^{1}S_{0}$& $0^{-+}$& $2980.3 \pm 1.2$ & 26.7 \\
$J/\psi$ & $1^{3}S_{1}$& $1^{--}$& $3096.916 \pm 0.011$ & 0.0932 \\
$\chi_{c0}$ & $1^{3}P_{0}$& $0^{++}$& $3414.75 \pm 0.31$ & 10.2 \\
$\chi_{c1}$ & $1^{3}P_{1}$& $1^{++}$& $3510.66 \pm 0.07$ & 0.89 \\
$\chi_{c2}$ & $1^{3}P_{2}$& $2^{++}$& $3556.20 \pm 0.09$ & 2.03 \\
$h_{c}$ & $1^{1}P_{1}$& $1^{+-}$& $3525.93 \pm 0.27$ & $< 1$ \\
$\eta_{c}'$ & $2^{1}S_{0}$& $0^{-+}$& $3637 \pm 4$ & 14 \\
$\psi'$ & $2^{3}S_{1}$& $1^{--}$& $3686.09 \pm 0.04$ & 0.317 \\
\hline
\end{tabular}
\caption{Charmonium states below the $D\overline{D}$ threshold.}
\label{tab:phys:charmonium_summary}
\end{center}
\end{table}

\begin{figure}
\begin{center}
\includegraphics[width=0.6\swidth]{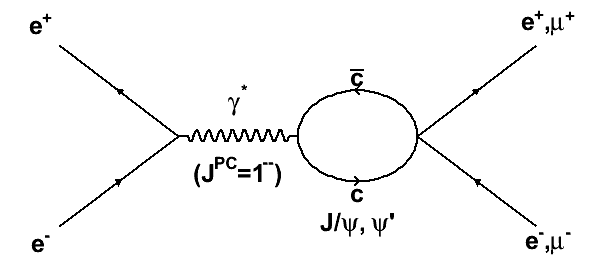}
\caption{Formation of the vector ($1^{--}$) charmonium in $e^{+}e^{-}$ annihilations, decaying into lepton pairs.}
\label{fig:phys:epem_charmonium}
\end{center}
\end{figure}

The first two discovered charmonium states had $J^{PC}=1^{--}$, i.e. the quantum numbers of the photon. Only these states can be directly formed in $e^{+}e^{-}$ annihilations. The Feynman diagram for the corresponding process is shown in \Reffig{fig:phys:epem_charmonium}. All the other states accessed in $e^{+}e^{-}$ annihilations are reached in the decay of such vector states, in particular via radiative transitions. For example, the energy balance in the cascade transitions $\psi'\rightarrow \chi_{c 0,1,2} + \gamma_1 \rightarrow \jpsi + \gamma_1 + \gamma_2$, via the putative states $\chi_{cJ}$ permitted to assign them to charmonium in the Crystal Ball experiment \cite{bib:Partridge:PhysRevLett.45.1150} (see \Reffig{fig:phys:CrystalBall}). Other charmonium states, besides $J^{PC}=1^{--}$, can in principle be obtained in $e^{+}e^{-}$ annihilations with two or three intermediate photons. However, a process with two intermediates photons contains additional factor $\alpha^{2}$ in probability and correspondingly the event rate a factor $10^{-4}$ smaller than in processes with one intermediate photon. The technique using $e^{+}e^{-}$ collisions was the only means to study charmonium until the middle of the 80-ties.

\begin{figure}
\begin{center}
\includegraphics[width=1.0\swidth]{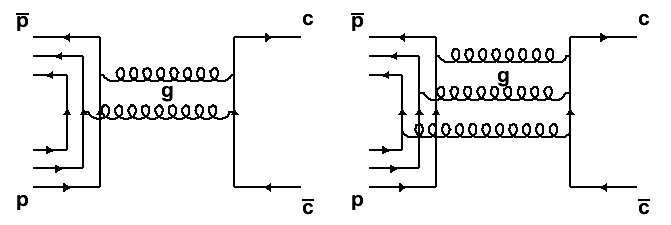}
\caption{Formation of charmonium in 2- and 3-gluon $p\overline{p}$ annihilation.}
\label{fig:phys:ppbar_charmonium}
\end{center}
\end{figure}

An alternative mechanism of charmonium production in proton-antiproton annihilations for charmonium spectroscopy was demonstrated for the first time in the R704 experiment \cite{bib:Baglin:1986yd} at CERN. This technique made possible the formation of states with all $J^{PC}$ values via intermediary of an appropriate number of gluons or $q\overline{q}$ pairs. An example of the process with two or three gluon annihilation is shown in \Reffig{fig:phys:ppbar_charmonium} and corresponds to formation of the states with $C=+1$ an $C=-1$, respectively, due to negative internal charge parity of a gluon. This technique became successfully applicable thanks to the development of stochastic cooling of the antiproton beam. In R704 for the first time the $0^{-+}$, $1^{++}$ and $2^{++}$ charmonium states were formed directly. The technique developed at CERN was later successfully used at Fermilab in the two subsequent experiments E760 and E835 \cite{bib:E835:Garzoglio:2004kw}. A significant improvement has been achieved in resolution of $non-1^{--}$ charmonium states.
The main difficulty in application of the $p\overline{p}$ technique in charmonium spectroscopy is dealing with huge hadronic background. The total $p\overline{p}$ cross-section in the energy range of charmonium is nearly 60 mb, whereas the highest peak cross-section for an individual charmonium state production ($\jpsi$) is about 270 nb \cite{bib:Armstrong:1992wu}.

It is worthwhile to mention here the states above the $D\overline{D}$ threshold at 3.73 \gev. This region is rather poorly known. The first experiments with $e^{+}e^{-}$ collisions reported a number of wide vector states, which were assigned to $\psi(2S)$, $\psi(3S)$, $\psi(4S)$ \cite{bib:Gaiser:1985ix} but not all of them were confirmed in a much more precise BES experiment. In addition to vector states this region should contain first radial excitations of the singlet and triplet vector states as well as the narrow D-states. Only one of them, $\Psi(3770)$, which is considered to be largely $^3D_1$, has been observed so far \cite{bib:Rapidis:1977cv}.

Moreover, in the last years in analyses of high statistics data samples by CLEO, Babar and Belle experiments a number of new states labeled X, Y and Z were discovered with masses in the range of charmonium \cite{bib:Godfrey:2008nc}. Some of them can be interpreted as charmonium states, for example $X(3872)$ has a possible interpretation as $1^3D_2$ or $1^3D_3$ with alternative interpretations as charmonium hybrid or a $D^0\overline{D^{0*}}$ molecule. The other two states Z(3931) and X(3940) have the most probable interpretation as $\chi_{c2}(2P)$ and $\eta_c(3S)$, respectively, while the nature of Y(3940), Y(4260), Y(4320), Y(4660) and Z(4430) is an open question. To conclude, the situation above $D\overline{D}$ still needs further work to be fully understood.

\section{Theoretical approaches to charmonium}\label{sec:phys:charm_theory}
In description of charmonium spectra two different approaches are used: phenomenological and QCD-oriented. The phenomenological approach tries to model the main features of QCD which are relevant to describe the spectrum of charmonium; it is represented mainly by potential models. The QCD-oriented approaches use QCD-based calculations, whether directly (lattice QCD) or within different approximations (effective theories).
\subsection{Phenomenological approach. The potential model}\label{sec:phys:potential_model}
QCD motivated potential models have been used for describing charmonium spectrum since the discovery of charmonium states. However, their relation to QCD is rather loose and the model quantities like quark mass, potential parameters etc. cannot be reliably related to the Standard Model parameters. Nevertheless, the application of potential models to the physics of charm was rather beneficial and helped a lot in interpretation of the experimental results.
Non-relativistic potential models are based on the assumption that the quark-antiquark interaction can be approximated with the aid of a static potential. Such an assumption is based on the expectation that for sufficiently heavy quarks the characteristic time scale associated with the relative motion of the constituent quarks is much larger than that associated with the gluonic or the sea quark degrees of freedom. In this case the adiabatic approximation applies and the effect of gluon and sea quarks can be represented by an average instantaneous interaction potential between the heavy quark sources. Despite some doubts on the existence of a static potential between two quarks, calculations based on this assumption were successful in predicting the masses of bound states of charmonium. The non-relativistic treatment of charmonium is possible due to relatively high charmed quark mass, which leads to $v^{2}/c^{2}\approx 0.25$ in the bound states.

The non-relativistic potential model is based on the Schr\"{o}dinger equation:
\begin{equation}
[T + V ] \Psi = E \Psi,
\end{equation}
where $T$ is the kinetic energy term and $V(r)$ is the potential energy term.
The potential should reproduce two main properties of QCD: the asymptotic freedom at small distances and the confinement at large distances.
The quark-gluon interaction in QCD is similar to the electron-photon interaction in QED with the Born term for $q\overline{q}$ interaction of the Coulomb form at small distances. This was first recognized in 1975 by Appelquist and Politzer \cite{bib:Appelquist:1975pa}. The difference from QED is that the gluon self-coupling results in a slow decrease of the strong coupling constant at small distances where quark-antiquark potential due to one gluon exchange has the form:
\begin{equation}
V(r)=-\frac{4}{3}\frac{\alpha_s(r)}{r},
\end{equation}
with the factor 4/3 arising from SU(3) color factors.

Appelquist and Politzer proposed the term "orthocharmonium" for the $\jpsi$ state in analogy with $^3S_1$ orthopositronium and predicted the existence of $^1S_0$ paracharmonium. Later they were able to predict \cite{bib:Appelquist:PhysRevLett.34.365} the complete spectrum of bound states of charmonium based on the potential intermediate between Coulomb and harmonic oscillator in its radial dependence.

However the most popular among the potential models became the Cornell potential proposed by Eichten et al. in 1975 \cite{bib:Eichten1975:PhysRevLett.34.369}, which combines one-gluon exchange Coulomb term and the additional term proportional to $r$, known as the confinement term. At distances of roughly 1 $fm$, which correspond to energy scale of $\Lambda_{QCD}$ the $\alpha_s$ increases significantly and one-gluon-exchange is no longer a good approximation. Qualitatively the chromoelectric lines of force can be imagined to bunch together into a flux tube which leads to a distance independent force or the potential $V(r)=\sigma r$. Such a term satisfies confinement, i.e. the experimental fact that colored free quarks are never experimentally observed.
The Cornell potential has the form:
\begin{equation}
\label{eq:theory:cornell_potential}
V(r)=-\frac{k}{r}+\frac{r}{a^2},
\end{equation}
where the constant $1/a^2$ is of order $\sim 1$ GeV/fm.

Soon afterwards many alternative formulations of the quark-antiquark potential appeared.
Quigg and Rosner proposed the logarithmic potential \cite{bib:Quigg:1979vr} in 1979:
\begin{equation}
V(r)=C\ln(r/r_0),
\end{equation}
which was motivated by the experimental observation that radial excitations of the S ground state are nearly equal for charmonium (589 MeV) and bottonium (563 MeV). The independence on the quark mass can be satisfied with the logarithmic potential.

Richardson proposed in 1978 \cite{bib:Richardson:1978bt} to construct the potential in momentum space in order to take into account the momentum dependence of $\alpha_s$. The proposed form of the potential was:
\begin{equation}
V(q^2)=-\frac{4}{3}\frac{12 \pi}{33-2n_f}[q^2 \times \ln(1+q^2/\Lambda^2)]^{-1},
\end{equation}
where $n_f$ is the number of fermion flavors with mass below $q$. An advantage of this potential is that only one parameter $\Lambda_{QCD}$ is engaged. It reproduces $1/r$ dependence at small distances and yields linear behavior in $r$ at large distances.

Additional potentials were proposed by Buchmuller and Tye \cite{bib:Buchmuller1981:PhysRevD.24.132}, Martin \cite{bib:Martin:1980jx} and Celmaster and Georgi \cite{bib:Celmaster1978:PhysRevD.17.879}. It should be stressed, however, that practically all the potentials are identical in the region between 0.2 $fm$ and 1 $fm$, which is the physical radius of the charmonium bound states.
An additional remark here in place is that calculations of a static spin-independent quark-antiquark potential from Lattice QCD are fitted perfectly with the Cornell potential (\Refeq{eq:theory:cornell_potential}) with parameters $k=0.322$ and $1/a^2 = 0.54 \, \gev/fm$ \cite{bib:Bali:2000gf}.

However the models with spin-independent potentials were not accurate enough to predict mass differences between states in the same orbital angular momentum or spin multiplets. Additional spin and orbital momentum dependent terms were introduced into the Hamiltonian to describe charmonium spectra in more details.

The development of spin-dependent potential started with the work of Eichten et al. \cite{bib:Eichten1975:PhysRevLett.34.369}, Pumplin et al. \cite {bib:Pumplin:1975cr} and Schnitzer \cite{bib:Schnitzer:1975ux}. A systematic investigation of spin-dependent forces in heavy quark systems was performed in 1979 by Eichten and Feinberg \cite{bib:Eichten:1979pu}.

The spin-dependent part of the potential is separated into three terms:
\begin{itemize}
\item \textbf{Spin-orbit} term which splits the states with the same orbital momentum depending on the $<L \cdot S>$ value (fine structure):
\begin{equation}
V_{LS}= <L\cdot S> (3 \frac{dV_{V}}{dr}-\frac{dV_{S}}{dr})/(2m^2_c r),
\end{equation}
where $V_{S}$ and $V_{V}$ are the scalar and vector components of the non-relativistic potential, respectively.
The expectation value of the $L\cdot S$ operator is equal to:
\begin{equation}
<L\cdot S>=\frac{1}{2}[J(J+1)-L(L+1)-S(S+1)].
\end{equation}

\item \textbf{Spin-spin} term describes the interaction between spins of the quarks. It is responsible for splitting between the spin singlet and spin triplet states (hyperfine structure):
\begin{equation}
V_{SS}= \frac{2<S_{1} \cdot S_{2}>}{3 m_{c}^{2}} \nabla^{2}V_{V}(r),
\end{equation}
where the expectation value:
\begin{equation}
<S_{1} \cdot S_{2}>=\frac{1}{2}[S(S+1)-\frac{3}{2}].
\end{equation}

\item \textbf{Tensor} term contains the tensor effect of the vector potential:

\begin{equation}
V_{T}= \frac{<T_{12}>}{2 m_{c}^{2}} (\frac{1}{r} \frac{dV_{V}}{dr}-\frac{d^{2}V_{V}}{dr^{2}}),
\end{equation}
where
\begin{equation}
T_{12}=\frac{1}{3}[3(S\cdot \hat{r})(S\cdot \hat{r})-S^2],
\end{equation}
with the expectation value
\begin{equation}
<T_{12}>=\frac{-<L\cdot S>^2-1/2 <L \cdot S> +1/3<L^2><S^2>}{(2L+3)(2L-1)}.
\end{equation}

\end{itemize}

The tensor and spin-orbit interactions are responsible for mass splitting of the $\chi_{cJ} (1^3P_J)$, $J=0,1,2$ states. The spin-spin interaction, which splits the vector and pseudoscalar states, is responsible for mass differences between $\jpsi$ and $\eta_c$ and between $\psi'$ and $\eta_c'$ (see \Reffig{fig:phys:CharmoniumSpectrum} and \Reftbl{tab:phys:charmonium_summary}).

Additional important point is that spin-dependent terms in the Hamiltonian are more singular than $r^{-2}$ and therefore are illegal operators in the Schr\"{o}dinger equation. It was shown in \cite{bib:Godfrey:1985xj} that the relativistic potential $V(p,r)$ differs from its non-relativistic limit in that the coordinate $\mathbf{r}$ (which in the non-relativistic limit is the relative distance $\mathbf{r_{12}}=\mathbf{r_1} -\mathbf{r_2}$) becomes smeared over the distance of the order of inverse quark mass. The smearing of the potential has the consequence of taming all the singularities in the spin-dependent potential, so that the contributing terms become legal operators. The usually implemented form of the smearing function is:

\begin{equation}
\rho_{ij}(\mathbf{r'}-\mathbf{r})=\frac{\sigma_{ij}^3}{\pi^{3/2}}e^{-\sigma_{ij}^2(\mathbf{r'}-\mathbf{r})^2},
\end{equation}

with
\begin{equation}
\sigma^{2}_{ij}=\sigma_0^2 \left[ \frac{1}{2}+\frac{1}{2}\left[\frac{4 m_i m_j}{(m_i+m_j)^2}\right]^4\right] + s^2 \left[\frac{2 m_i m_j }{m_i+m_j}\right]^2,
\end{equation}
where $\sigma_0$ = 1.8 \gev and $s$ = 1.55.
\subsection{QCD sum rules}
One of the first calculational techniques successfully applied for mass predictions of the charmonium states is QCD sum rule. In general sum rule technique is applicable when sums or integrals over observables can be related to normalisation conditions reflecting unitarity etc. or to a quantity that can be calculated in the underlying theory. QCD sum rule introduced in 1979 by Shifman, Vainstein and Zakharov \cite{bib:Shifman:1978bx} allows to express low energy hadronic quantities through basic QCD parameters. QCD vacuum is typified by large fluctuating fields whose strength is characterised by the quark and gluon condensates $<0|q\overline{q}|0>$, $<0|G^{(a)}_{\mu \nu}G^{(a) \mu \nu}|0>$. These quantities parameterize non-perturbative dynamics and vanish in the perturbation theory. When a pair of quarks is injected into the vacuum, the subsequent formation of a hadron can be calculated via dispersion relations. The heavy quark-antiquark bound states, charmonium and bottonium, appear to be sensitive only to the gluon condensate. The success of QCD sum rule technique proved in the correct prediction of the mass of charmonium ground state $\etac$ in 1978 \cite{bib:Shifman:1978zq} and the $1P$ charmonium states in 1981 \cite{bib:Reinders:1981si}. However no other charmonium states were calculated with this technique due to large power corrections in the expansion of vacuum polarization operator.
\subsection{pQCD}
The perturbative calculations in QCD (pQCD) are valid for small running coupling constant, i.e. as it follows from asymptotic freedom in QCD, at large momenta. This technique has been applied for charmonium annihilation calculations under the assumption that annihilation is the process occurring at short distances. Also this technique was applied for the calculation of strong radiative corrections to the electromagnetic, radiative (decay to $\gamma \gamma$) and hadronic decays of the S and P states in quarkonia by Barbieri et al. \cite{bib:Barbieri:1975am}. However the pQCD predictions met with several problems. The first one is that in contrast to positronium in charmonium $m(q\overline{q})\neq 2 m(q)$, as a result there is an ambiguity whether to use $m(q)$ or $m(q\overline{q})/2$ in calculations, which affects the calculated hadronic decay width. The second problem is in that the coupling constant is large enough, $\alpha_s \sim 0.3$, for the charmed quarks so that the lowest order gluon radiative corrections can reach up to 100 \percent. However radiative corrections calculated with pQCD are in some cases in good agreement with experiment, for example the radiative width of $\eta_c$ is calculated with precision of 10\percent \cite{bib:Barbieri:1975am}.
\subsection{Lattice QCD}
The lattice QCD technique was proposed by Wilson in 1974 \cite{bib:Wilson:1974sk}. The main intention behind lattice QCD was to define an entirely non-perturbative regularisation scheme for QCD, based on the principle of local gauge invariance. The four-dimensional space-time continuum is replaced by a discrete lattice with the spacing $a$ between lattice sites. Distances $\sim a$ or smaller cannot be treated in this way, in the other way stated, the finite spacing introduces an ultra-violet cut-off $\sim \pi/a$ on momenta in the lattice QCD. In the more recent calculations asymmetric lattices have been used, in which lattice spacing for time, $a_t$, is chosen smaller than for space, $a_s$, with the ratio $a_s/a_t \sim 3$. The typical lattice spacing is $a_s \sim$ 0.07 $fm$ and the overall size of the lattice is 1-4 $fm$.

Observables are calculated taking their expectation values in the path integral approach: calculated quantities are averaged over all the possible configurations of gauge fields on the lattice and weighted with the respective exponent of the action. From the point of view of numerical calculations it is convenient to work in Euclidean space-time in which the path integral measure can be defined. Most results that are obtained in Euclidean space can be related to the space-like region of the Minkowski world and can, in principle, be analytically continued into the time-like region of interest. However, with the results obtained on a discrete set of points such a continuation is not straightforward. Fortunately, the mass spectrum remains unaffected by the rotation to imaginary time, as long as reflection positivity holds, which is the case for most lattice actions used for the charmonium mass calculations.

Past lattice results were often obtained in the "quenched" approximation, i.e. no quark-antiquark pairs were allowed to be excited from the QCD vacuum. Recently, the progress in computers and computational techniques allowed to do unquenched calculations with $u\overline{u}$, $d\overline{d}$ and $s\overline{s}$ pairs excited from the vacuum.

\subsection{Effective theories. NRQCD}
From the point of view of QCD the description of hadrons containing two heavy quarks is a rather challenging problem and a proper relativistic treatment of the bound state based on the Bethe-Salpeter equation has proved difficult. A non-relativistic treatment of the heavy quarkonium dynamics, which is suggested by the large mass of a heavy quark has clear advantages. The velocity of quarks in the bound state provides a small parameter in which the dynamical scales can be ordered and the QCD amplitudes dynamically expanded. A non-relativistic bound state is characterised by at least three scales: the scale of the mass $m$ (hard), the scale of the momentum transfer $p \sim mv$ (soft) and the scale of kinetic energy of the quark and anti-quark in the center of mass frame $E \sim p^2/m \sim m v^2$ (ultrasoft). In a non-relativistic system ($v \ll 1$) the above scales are hierarchically ordered: $m \gg mv \gg mv^2$. The wide span of energy scales involved makes the lattice QCD calculations extremely challenging because of the conflicting requirements that the space-time grid should be large in comparison with the largest length of the problem ($1/mv^2$) and the lattice spacing be smaller compared with the smallest length $1/m$. The hierarchy of scales has the advantage that it can be used to substitute QCD with a simpler but equivalent Effective Field Theory (EFT). An EFT is the quantum field theory with the following properties: a) it contains the relevant degrees of freedom to describe phenomena that occur in a certain limited range of energies and momenta and b) it contains an intrinsic scale that sets the limit of applicability of the EFT.

The prototype of an EFT for heavy quarks is the Heavy Quark Effective Theory (HQET), which is an EFT of QCD suitable to describe systems with only one heavy quark. These systems are characterised by two energy scales: $m$ and $\Lambda_{QCD}$. HQET is obtained by integrating out the scale $m$ and use of an expansion in powers of $\Lambda_{QCD}/m$. Following the above discussion, one may conclude that the bound states made of two heavy quarks are characterised by more than two scales. Integrating out only the scale $m$ leads to an EFT called Nonrelativistic QCD (NRQCD) that still contains the lower scales as dynamical degrees of freedom. Starting from NRQCD two approaches can be followed for spectrum computation: direct lattice calculations or further integration of the soft scale to arrive at an EFT in which only the ultrasoft degrees of freedom remain dynamical, an approach called pNRQCD. The lattice calculations with NRQCD and their application in charmonium physics are reviewed in \cite{bib:Bali:2000gf}.

\section{$h_c$ meson}\label{sec:phys:hc}
The present chapter is devoted to the $h_c$ singlet state of charmonium ($1^{1}P_{1}$). It is the most elusive charmonium state with mass below the \DDbar threshold, since its experimental observation met with many difficulties. Its quantum numbers make this state not directly accessible in $e^+e^-$ annihilations. Also, it cannot be reached in the radiative E1 transition from $\psi'$, since the C-parity conservation forbids the transition from $1^{--}$ to $1^{+-}$. Its observation at the B-factories in $B$ decay via an intermediate $\eta_c(2S)$ state, which can decay via an E1 transition to $h_c$, is suppressed because of a large hadronic width of the $\eta_c(2S)$.

In this chapter the main arguments behind the interest in precise measurements of the mass and width of $h_c$ are presented, the theoretical predictions of mass and decay width are summarized, and previous attempts of the experimental observation of $h_c$ are reviewed. The results of a Monte-Carlo study of the PANDA detector performance in the foreseen $h_c$ measurements are presented in \Refchap{chap:phys:MonteCarlo}.

\subsection{Spin dependence of the confinement potential}
The main argument besides the interest in precision mass measurement of $h_c$ is determination of the mass difference (hyperfine splitting) between the singlet ($^1P_1$) and the centroid of the triplet P-states ($^3P_J$). In the framework of potential models the hyperfine splitting arises from the spin-spin term of the charmonium potential:
\begin{equation}
\Delta_{HF}=-V_{SS}=M_{h_c}-M_{\chi_{cog}},
\end{equation}

where $M_{\chi_{cog}}$ is the mass centroid of the triplet $^3P_J$ states:
\begin{equation}
\label{eq:theory:mcog}
M_{\chi_{cog}}=\frac{\sum (2J+1)M_{\chi_J}}{\sum (2J+1)}=\frac{M_{\chi_{0}}+3M_{\chi_{1}}+5M_{\chi_{2}}}{9}.
\end{equation}

Since the spin-spin potential is a contact one ($V_{SS}\sim \nabla^2 V_{V}(r)$, see \Refsec{sec:phys:potential_model}) its expectation value is finite only for the wavefunction finite at the origin.
Therefore, it gives rise to the hyperfine splitting between the triplet and singlet states only in $L=0$, which is reflected, for example, in the mass difference between $\jpsi$ and $\eta_c$. The hyperfine splitting for the P-wave states should be zero and the higher order corrections should provide no more than a few MeV deviation from this result \cite{bib:Appelquist:1978aq}, \cite{bib:Godfrey:2002rp}. This is a consequence of the assumption that the long-range confinement potential is a pure Lorentz scalar. The magnitude of hyperfine splitting larger than a few MeV would be an indication of a vector component of the confinement potential, but the experimental observations of $h_c$ performed so far indicate that the latter should not be significant.

\subsection{Mass predictions}
Mass predictions for the singlet $^1P_1$ state in charmonium might serve as a validation test of the theoretical approaches mentioned above in \Refsec{sec:phys:charm_theory}. Most of the $h_c$ mass predictions are based on the potential model calculations including variation of the strong coupling constant $\alpha_s$ and the charmed quark mass $m_c$. The most commonly used form is that of the Cornell potential \cite{bib:Eichten1975:PhysRevLett.34.369} and the potential proposed by Richardson \cite{bib:Richardson:1978bt}. The linear confinement term is usually assumed to be scalar but some authors consider an admixture of the vector confinement. In some cases relativistic corrections at the level of $v^2/c^2$ are taken into account. \Reftbl{tab:phys:hc_mass} summarizes all the predictions of $h_c$ mass expressed as the hyperfine splitting $\Delta_{HF}=M_{\chi_{cog}}-M_{h_c}$. It is based on the summaries presented in \cite{bib:Godfrey:2002rp} and \cite{bib:joffe-2005}. The value $(M_{\chi_{cog}})_{exp}=3525.3\pm 0.1$ \mev is known \cite{bib:pdg} from experiment.

\begin{table}
\begin{center}
\begin{tabular}{|l|c|c|p{8cm}|}
\hline
Author& Year & $\Delta_{HF}, MeV$ & Approach \\
\hline
Eichten \cite{bib:Eichten:1979pu} & 1979 & 0 & Potential model with Cornell potential and scalar confinement term. \\
Ono \cite{bib:Ono:1982ft}& 1982 & +9.8 & Potential model with smeared hyperfine interaction. \\
McClary \cite{bib:McClary:1983xw}& 1983 & -5.2 & Potential model with smeared hyperfine interaction and relativistic corrections. \\
Moxhay \cite{bib:Moxhay:1983vu}& 1983 & 0 & Potential model with long-range longitudinal color electric field. \\
Grotch \cite{bib:Grotch:1984gf}& 1984 & +(6-17)& Potential model with Lorentz scalar and vector confinement terms. Predictions depend on the fraction of scalar versus vector contribution.\\
Godfrey \cite{bib:Godfrey:1985xj} & 1985 & +8 & Potential model with smeared short-range hyperfine interaction.\\
Pantaleone \cite{bib:Pantaleone:1985uf}& 1986& -3.6 & pQCD calculation ($\alpha_s=0.24$).\\
Gupta \cite{bib:Gupta:1986xt}& 1986 & -2& Potential model with 1-loop QCD corrections.\\
Igi \cite{bib:Igi:1987rt}& 1987& +24.1 & Potential model with short-distance 2-loop QCD calculation.\\
Pantaleone & 1988& -1.4 & pQCD calculation ($\alpha_s=0.33$).\\
Lichtenberg \cite{bib:Lichtenberg:1991fw} & 1992& +4 & Potential model with pQCD corrections at short distances. \\
Halzen \cite{bib:Halzen:1992wm} & 1992 & -0.7 $\pm$ 0.2 & pQCD calculation ($\alpha_s=0.28$). \\
Bali \cite{bib:Bali:2000gf} & 1997 & -1.5 $\pm$ 2.5 & Lattice QCD \\
Manke \cite{bib:Manke:2000dg}& 2000 & +(1.7-4.0) & Lattice QCD \\
Okamoto \cite{bib:Okamoto:2001jb} & 2002& -1.5 $\pm$ 2.6 & Lattice QCD\\
\hline
\end{tabular}
\caption{Summary of the $h_c$ mass predictions expressed in terms of the predicted hyperfine splitting $\Delta_{HF}=M_{\chi_{cog}}-M_{h_c}$.}
\label{tab:phys:hc_mass}
\end{center}
\end{table}

\subsection{Decay modes of $h_c$}
Theoretical calculations predict $h_c$ as a narrow state with $\leq$ 1 \mev total width. The possible decay modes of $h_c$ are electromagnetic or hadronic transitions to lower charmonium states or a decay into light hadrons.
All predictions for the partial and total widths for the $h_c$ decay are summarized in \Reftbl{tab:phys:hc_width}. Below some details concerning width calculations are discussed.

The radiative transitions in charmonium occur, as in atoms, from the higher to lower mass states with the emission of a photon. The electromagnetic transitions are divided into electric dipole transitions (E1) and magnetic dipole transitions (M1). The E1 transitions obey the $\Delta L=\pm 1, \, \Delta S =0$ selection rule, for the M1 transitions $\Delta L=0, \, \Delta S = \pm 1$. For $h_c$ the E1 transition to $\eta_c$ is possible with the width given in non-relativistic limit by:
\begin{equation}
\Gamma(h_c \rightarrow \eta_c + \gamma)= \frac{4}{9}e^2 \alpha k^3 |E_{if}|^2,
\end{equation}
where $k=(M_i^2-M_f^2)/2 M_i$ is the photon momentum, $|E_{if}|$ is the transition dipole matrix element:
\begin{equation}
|E_{if}|=\int_0^{\infty} \mathrm{d}r r^3 R_i(r) R_f(r)
\end{equation}
where $R_i(r) \, [R_f(r)]$ are the radial parts of the charmonium initial [final] wave functions, which are model dependent. Predictions for E1 decay width have been made by several authors, most of the predictions relate the partial width for $h_c$ decay to the already measured partial widths for other charmonium states. For example, Bodwin et al. \cite{bib:Bodwin:1992ye} predict the $h_c$ decay width by scaling with photon energy the radiative decay of $\chi_{1,2,3}$ to $\jpsi$.
Renard \cite{bib:Renard:1976up} also predicts a few \kev for the M1 transitions to the triplet $P$ states in addition to the E1 width.
Barnes et al. \cite{bib:Barnes:2005pb} evaluated the radiative width using two models, the relativised Godfrey-Isgur model and a nonrelativistic potential model for the wave functions.

The hadronic transitions of $h_c$ to $\jpsi$ are allowed energetically with the emission of one or two $\pi^0$ýû. The $\jpsi \pi \pi$ width is predicted to be larger than $\jpsi \pi^{0}$. The predictions for $\jpsi \pi \pi$ and $\jpsi \pi^{0}$ by Kuang et al. \cite{bib:Kuang:1988bz} were made in the framework of QCD multipole expansion. In the work of Chen and Yi \cite{bib:Chen:1992fr} the same approach is used, taking into account also the 2-loop corrections.

The partial width of $h_c$ decay into light hadrons is actually the width of decay into three gluons ($ggg$) or two gluons plus a gamma ($gg\gamma$), where gluons hadronize. In most works scaling of the partial width from the known decay is used, as for example Bodwin et al. \cite{bib:Bodwin:1992ye} use the $\chi_1$ partial width into hadrons for their prediction. It is important to mention here that $h_c$ decay into light hadrons is OZI suppressed, which is typical for a charmonium state (see \Refsec{sec:theory:history_charmonium}).

The total width prediction is the sum of partial widths. Some authors quote the sum of their own predictions adding the results of other authors for the missing channels.

\begin{table}
\begin{center}
\begin{tabular}{|l|c|c|c|c|c|c|}
\hline
Author& Year & $\Gamma(\eta_c \gamma)$ & $\Gamma (\jpsi \pi^{0})$ & $\Gamma(\jpsi \pi^{0} \pi^{0})$ & $\Gamma(hadrons)$ & $\Gamma(total)$ \\
& & (keV) & (keV) & (keV) & (keV) & (keV) \\
\hline
Renard \cite{bib:Renard:1976up}  & 1976 & 240 & & &370& 500-1000\\
Galkin \cite{bib:Galkin:1990vp}  & 1990 & 559 & & & &\\
Novikov \cite{bib:Novikov:1977dq}& 1977 & 975 & & &60-350&\\
McClary \cite{bib:McClary:1983xw}& 1983 & 483 & & & &\\
Chao \cite{bib:Chao:1992hd}      & 1992 & 385 & & & &\\
Bodwin \cite{bib:Bodwin:1992ye}  & 1992 & 450 & & & 530& 980\\
Casalbuoni \cite{bib:Casalbuoni:1992yd}& 1992 & 450& & & &\\
Ko \cite{bib:Ko:1992fk}          & 1992 & 400 & 1.6& & & \\
Gupta \cite{bib:Gupta:1993pd}    & 1993 & 341.8& & & &\\
Barnes \cite{bib:Barnes:2005pb}  & 2005 & 352-498 & & & &\\
Kuang \cite{bib:Kuang:1988bz}    & 1988 & & 2& 4.12& 53.7& 394\\
Chemtob \cite{bib:Chemtob:1989nk}& 1989 & & 0.0061& 52.6& &\\
Chen \cite{bib:Chen:1992fr}      & 1992 & &0.29-0.58& 4.1-7.1& 19-51&360\\
\hline
\end{tabular}
\caption{Partial and total width predictions for the $h_c$ decay.}
\label{tab:phys:hc_width}
\end{center}
\end{table}

\subsection{Experimental observations} \label{sec:phys:hc_observation}
The first attempts for observation of the $h_c$ date back to 1985 and the R704 experiment at the CERN Intersecting Storage Rings (ISR), where the proton-antiproton annihilation has been used to produce the charmonium states. The expected high hadronic background was an argument to chose the $\jpsi \rightarrow e^+e^-$ decay as a tag for charmonium formation in the R704 experiment. To detect high mass $e^+e^-$ pairs R704 was constructed as a nonmagnetic spectrometer with two arms positioned symmetrically relative to the beam axis to detect lepton pairs. A 2.3 $\sigma$ signal was claimed with the observation of 5 $p\overline{p}\rightarrow \jpsi+X$ events \cite{bib:Baglin:1986yd} near the $\chi_c$ center of gravity. This signal was interpreted as the $h_c\rightarrow \jpsi \pi^0$ decay with the $h_c$ mass $3525.4 \pm 0.8 \pm 0.4 MeV/c^2$.

\begin{figure}
\begin{center}
\includegraphics[width=0.8\swidth]{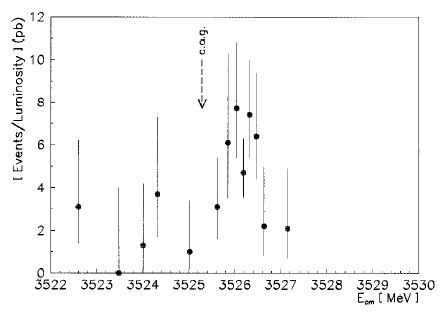}
\caption{Evidence for the $h_c \rightarrow \jpsi \pi^0$ observation in the E760 experiment \cite{bib:Armstrong:1992ae}. c.o.g. indicates the position of the center of gravity (\Refeq{eq:theory:mcog}) of the $\chi_{cJ}$ states.}
\label{fig:phys:hc_e760}
\end{center}
\end{figure}

After shutdown of the ISR, the E760 experiment was proposed at the antiproton source complex of Fermilab to continue the study of $h_c$ in proton-antiproton annihilations. An antiproton beam (up to $4.0 \cdot 10^{11}$ accumulated $\overline{p}$) intersecting with an internal hydrogen jet target produced instantaneous luminosity in the range $(3\div9)\cdot10^{30} \,cm^{-2}s^{-1}$. A system of stochastic cooling compensated the effects of scattering and energy loss in the target to keep the beam energy spread at the level $\Delta p /p < 2.5 \cdot 10^{-4}$. The search for $h_c$ has been performed in the vicinity of the $\chi_c$ center-of-gravity in small energy steps ($\leq 500$ \kev). The E760 experiment was a non-magnetic spectrometer optimised for the detection of the electromagnetic final states with an electromagnetic calorimeter, composed of lead-glass scintillators, covering polar angles from $2^{\circ}$ to $70^{\circ}$ and having full azimuthal coverage.
The following decay modes of $h_c$ have been searched for:
\begin{equation}
\label{eq:theory:hc_gammagamma}
p + \overline{p} \rightarrow h_c \rightarrow \eta_c \, \gamma \rightarrow (\gamma \gamma) \gamma,
\end{equation}
\begin{equation}
p + \overline{p} \rightarrow h_c \rightarrow J/\psi \, \pi^{0} \rightarrow (e^+e^-) \pi^0,
\end{equation}
\begin{equation}
p + \overline{p} \rightarrow h_c \rightarrow J/\psi \, 2 \pi \rightarrow (e^+e^-) 2 \pi.
\end{equation}
According to theoretical predictions (see \Reftbl{tab:phys:hc_width}) the first decay mode should be the dominant one, but due to low branching ratio for the transition $\eta_c \rightarrow \gamma \gamma$, the event rates for all three decay modes were assumed comparable. The evidence for the resonance in the $p + \overline{p} \rightarrow J/\psi \, \pi^{0}$  channel has been demonstrated with the 59 observed events \cite{bib:Armstrong:1992ae}. The observation plot is presented in \Reffig{fig:phys:hc_e760}. The mass of the resonance was estimated at $M_R=3526.2 \pm 0.15 \pm 0.2 \, MeV/c^2$. An upper limit on the resonance width $\Gamma< 1.1 \, MeV$ was set at the 90 \percent confidence level. No events were found to fit the reaction $p + \overline{p} \rightarrow J/ \psi \, \pi^+ \pi^-$ or $p + \overline{p} \rightarrow J/ \psi \, \pi^0 \pi^0$.

The E835 experiment, which was an upgrade of E760, continued the study of $h_c$ in two runs with the collected luminosities of 141.4 $pb^{-1}$ in Run-I (1997) and 113.2 $pb^{-1}$ in Run-II (2000). The most important modification of the detector with respect to E760 was an upgrade of electronics of the calorimeter, which included pulse shaping and installation of a time-to-digital converter for each calorimeter channel to reduce pileups and replacement of the forward calorimeter with lead-glass modules. The two reactions $p + \overline{p} \rightarrow h_c \rightarrow \eta_c \, \gamma $ and $p + \overline{p} \rightarrow h_c \rightarrow J/\psi \, \pi^{0}$ were investigated.

\begin{figure}
\begin{center}
\includegraphics[width=0.7\swidth]{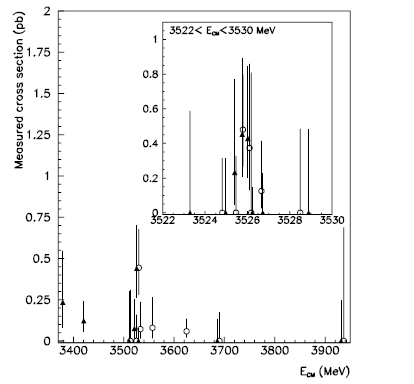}
\caption{Observation of the $h_c \rightarrow \eta_c \gamma$ in the E835 experiment \cite{bib:sim:hc_E835}. The inset shows the $h_c$ search region in an expanded scale.}
\label{fig:phys:hc_e835}
\end{center}
\end{figure}

With the 6 times larger collected statistics in comparison with E760 no evidence of the $h_c \rightarrow \jpsi \pi^0$ signal was observed. In conclusion the resonance reported by E760 was not confirmed. However the evidence of the signal in the $h_c \rightarrow \eta_c \gamma$ mode was observed with $\eta_c \rightarrow \gamma \gamma$ (\ref{eq:theory:hc_gammagamma}) with the statistical significance (P-value) 0.001 \cite{bib:sim:hc_E835}. The observation plot is presented in \Reffig{fig:phys:hc_e835}. The reported mass is $M_{R}=3525.8 \pm 0.2 \pm 0.2 \, \mathrm{MeV}$ and the resonance width $\Gamma < 1 \, \mathrm{MeV}$. The signal was observed in both runs with comparable strength.

The most recent observation of $h_c$ was made by the CLEO experiment \cite{bib:sim:hc_CLEO} at the Cornell Electron Storage Ring in the reaction:
\begin{equation}
e^+e^- \rightarrow \psi(2S) \rightarrow \pi^0 h_c, \, h_c \rightarrow \eta_c \gamma , \, \pi^0 \rightarrow \gamma \gamma.
\end{equation}
The study has been performed for both CLEO III and CLEO-c detector configurations, which are characterised by 93\percent solid angle coverage for both charged and neutral product particles.

Two methods of the analysis have been used: exclusive analysis with the reconstruction of $\eta_c$ decay which has higher signal purity and inclusive analysis with higher signal yield. Both methods identify $h_c$ as an enhancement in the spectrum of neutral pions from the decay $\psi(2S)\rightarrow \pi^0 h_c$. For the exclusive analysis $\eta_c$ was reconstructed in the following channels: $K^0_S K^{\pm} \pi^{\mp}$, $K^0_L K^{\pm} \pi^{\mp}$, $K^+K^-\pi^+\pi^-$, $\pi^+\pi^-\pi^+\pi^-$, $K^+K^-\pi^0$ and $\pi^+\pi^-\eta$. The reported \cite{bib:sim:hc_CLEO} mass is $M(h_c)=3524.4 \pm 0.6 \pm 0.4 \, MeV$ and the significance of the observation is greater than $5\sigma$.

\begin{figure}
\begin{center}
\includegraphics[width=0.8\swidth]{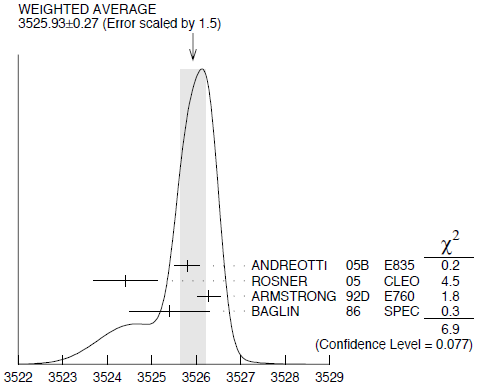}
\caption{PDG \cite{bib:pdg} plot for the $h_c$ mass averaging.}
\label{fig:phys:hc_pdg}
\end{center}
\end{figure}

The current PDG mass of $h_c$, $M(h_c)=3525.93\pm 0.27$ \mev \cite{bib:pdg}, is based on the four observations mentioned above. The plot demonstrating the mass averaging from PDG \cite{bib:pdg} \Reffig{fig:phys:hc_pdg} shows that there are some discrepancies between the observations, which motivate improved experimental search with higher statistics and better resolution.

\section{Angular distributions of photons in radiative transitions in charmonium}\label{sec:theory:angular_dist}
Formation of charmonium as a resonance in the $p\overline{p}$ annihilation provides an experimental tool to determine the electromagnetic radiative multipole structure for the transition between the resonance and the selected final state. The study of a radiative angular distribution provides information on the relative annihilation amplitudes through the possible helicity states. In those cases, in which only one multipole and one helicity state is allowed by the conservation laws, the angular distribution can be used to verify $J^{PC}$ characteristics of the newly discovered states, such as $h_c$. The two independent formalisms can be applied for the description of radiative angular distributions, helicity amplitudes and transition multipoles.

\subsection{Helicity formalism}
Here a short summary of the helicity formalism is provided, based on \cite{bib:McTaggart:1998sf}.
The formalism of helicity amplitudes was proposed by Jacob and Wick in 1959 \cite{bib:Jacob:1959at} to describe the collisions of particles with spin. In this formalism states are labeled with the projection of the total angular momentum along the direction of particle momentum, which avoids relativistic complications resulting from division of the total angular momentum into spin and orbital parts. Additional useful property of the helicity formalism is that for the two-particle states, in the center-of-mass frame, the orbital angular momentum is oriented perpendicular to the direction of relative motion of the two particles. With the helicity defined as:
\begin{equation}
\Omega=\frac{\overrightarrow{J} \cdot \overrightarrow{p}}{|\overrightarrow{p}|},
\end{equation}
the component of the spin along the direction of relative motion will be the component of the total angular momentum along this direction, since $\overrightarrow{J}=\overrightarrow{L}+\overrightarrow{S}$ and $\overrightarrow{L}\cdot \overrightarrow{p}=0$. Moreover, the helicity quantum number is invariant under the ordinary rotation, however if the momentum vector is rotated by 180\degrees the helicity quantum number changes sign.

\subsection{Prediction for the angular distribution of $\gamma$ in  $p\overline{p} \rightarrow \eta_c \gamma$}
\begin{figure}
\begin{center}
\includegraphics[width=0.6\swidth]{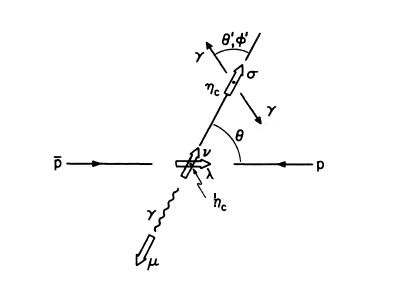}
\caption{Angles and helicities involved in the process $p\overline{p} \rightarrow h_c \rightarrow \gamma_1 \eta_c \rightarrow \gamma_1 \gamma \gamma$; $\gamma_1$ is the photon emitted in the $h_c$ decay.}
\label{fig:phys:hc_decay}
\end{center}
\end{figure}

The angular distribution of $\gamma_1$ in the $p\overline{p} \rightarrow h_c \rightarrow \gamma_1 \eta_c \rightarrow \gamma_1 \gamma \gamma$, based on the helicity formalism, is provided in \cite{bib:Olsson:1986aw}. The joint angular distribution for this process can be expressed as:
\begin{equation}
W(\theta ; \theta ', \phi ')= \sum_{\lambda} B_{|\lambda|}^2 \sum_{\mu , \mu ' = \pm 1} d^{1}_{\lambda \mu} (\theta) d^{1}_{\lambda \mu '} (\theta) A_{0}^{2},
\end{equation}
with the angles and helicities as indicated in \Reffig{fig:phys:hc_decay}. Here the angles $\theta '$, $\phi '$ specify the $\eta_c \rightarrow \gamma \gamma$ decay in the $\eta_c$ rest frame. $W(\theta ; \theta ', \phi ')$ does not depend on the angles $\theta '$ and $\phi '$ since the pseudoscalar state $\eta_c$ decays into $\gamma \gamma$ isotropically.

A normalised joint angular distribution is defined as:
\begin{equation}
\hat{W}(\theta ; \theta ', \phi ') =C \frac{W(\theta ; \theta ', \phi ')}{B_{0}^{2}+2B_{1}^{2}}.
\end{equation}

A constant $R$ is introduced defining the fractional contribution of the helicity-one amplitude into the initial production process:
\begin{equation}
R=\frac{2B_{1}^{2}}{B_{0}^{2}+2B_{1}^{2}}.
\end{equation}

For each transition the joint distribution $\hat{W}(\theta ; \theta ', \phi ')$ can be expressed in terms of the observables $\{K_{i}\}$ and the elementary trigonometric functions. $\{K_{i}\}$ can be expressed in terms of the helicity amplitudes $A_{i}$, which in turn can be written in terms of the multipole transition amplitudes $a_{i}$. The general relation between the two sets of amplitudes is:
\begin{equation}
A_{\nu} = \sum_{k} a_{k} [\frac{2k+1}{2j+1}]^{1/2} \langle k ,1; 1, \nu -1| j' , \nu \rangle,
\end{equation}
where $<;|>$ is the Clebsch-Gordan coefficient.

The $h_c \rightarrow \eta_c \gamma_1$ decay proceeds via the $E1$ amplitude only, so that $a_{1}=A_{0}=1$.
The angular distribution is:
\begin{equation}
\frac{32 \pi^2}{3} \hat{W}(\theta ; \theta ', \phi ') = (K_1 + K_2 \cos^{2} \theta),
\end{equation}
with
\begin{equation}
K_{1}= 1 - \frac{1}{2}R,
\end{equation}
and
\begin{equation}
K_{2}= \frac{3}{2}R - 1,
\end{equation}
so that
\begin{equation}
\hat{W}(\theta)=\hat{W}(\pi/2) (1+ \alpha \cos ^2 \theta),
\end{equation}
with
\begin{equation}
\alpha = \frac{3R-2}{2-R}.
\end{equation}

Because of C-parity conservation the $B_1$ helicity state does not enter into $h_c$ production and $R=0$. Therefore, the angular distribution of $\gamma$-rays in the $h_c$ decay has particularly simple form:
\begin{equation}
\hat{W}(\theta)=\hat{W}(\pi/2)\sin ^2 \theta.
\end{equation}

%% file: panda_experiment.tex
\chapter{The PANDA experiment}\label{chap:panda}
The PANDA (Pbar ANnihilations at DArmstadt) is a hadron physics experiment which is planned as part of the future Facility for Antiproton and Ion Research (FAIR) at Darmstadt. The cooled antiproton beam with a momentum between 1.5 \gevc and 15 \gevc will be provided for the PANDA experiment in High Energy Storage Ring (HESR). The antiprotons will interact with an internal hydrogen target to reach a peak luminosity up to $2 \cdot 10^{32} \, cm^{-2}s^{-1}$.
\section{The PANDA physics program}
Experimental program of the PANDA is concentrated on the studies of strong interaction and hadron structure performed with the aid of an antiproton beam. The major topics of the PANDA physics program are summarized below.
\subsection{Charmonium spectroscopy}
Precise measurements of the properties (energies, total and partial widths) of the charmonium states below the \DDbar threshold have fundamental importance for understanding of QCD. The theoretical predictions for the charmonium spectrum have been performed in several theoretical approaches outlined in \Refsec{sec:phys:charm_theory} starting with the simple non-relativistic potential model and ending with the latest calculations using the effective field theoretical approach (NRQCD). Most of them can be verified with high precision with the experimental results provided by PANDA. Moreover, the recent discoveries of new states (X, Y, Z) at the B factories renewed interest in physics of charmonium  and stimulated intense experimental and theoretical activities. All charmonium states can be formed directly in $\pbarp$ annihilation, in particular those with quantum numbers other than $J^{PC}=1^{--}$ and their masses determined with precision of several 100 \kev, which is 10 to 100 times better than achieved in the $e^{+}e^{-}$ experiments.

\subsection{Search for gluonic excitations (hybrids and glueballs)}
Glueballs and hybrids are the bound states predicted by QCD in which the gluonic degree of freedom explicitly manifests itself. Glueballs are the states of pure glue. Prediction of their mass spectrum was one of the early achievements of the lattice QCD.
There is about 15 glueballs in the mass range accessible with the PANDA experiment \cite{bib:Morningstar:1999rf}. Glueballs with exotic quantum numbers are called oddballs; they cannot mix with the "normal" mesons. As a consequence, they are predicted to be rather narrow and easy to identify experimentally. The hybrids consist of a \qqbar pair and an excited glue. The additional gluonic degree of freedom manifests itself as a contribution to quantum numbers. In the simplest scenario this corresponds to adding the quantum numbers of a gluon to a simple \qqbar pair.
Charmonium hybrids are expected to exist in the 3-5 \gev mass region. The lowest lying charmonium hybrid is expected with $J^{PC}=1^{-+}$ and the mass of 4.34 \gev \cite{bib:Bernard:1997ib}.

\subsection{Study of hadrons in nuclear matter}
The change of hadron masses inside nuclear medium has been proposed as the manifestation of partial restoration of the chiral symmetry breaking
predicted by QCD due to finite density, and thus to be an indicator of changes in quark condensates. In particular, the attractive mass shifts reflecting reduced quark pair condensate at finite density have been predicted for vector mesons. Numerous experiments have been devoted to deduce
in-medium mass shifts or nuclear potentials for hadrons in the light quark sector ($u$,$d$,$s$), both for the vector and for the pseudoscalar mesons. The energy range of HESR and detection capability of PANDA will permit to extend this type of studies into the sector of charmed hadrons for both open and hidden charm.

Another open question where PANDA can contribute is the $\jpsi N$ dissociation cross section on a series of nuclear targets. Knowledge of this parameter is important for interpretation of the \jpsi suppression effect as the signature of quark-gluon plasma formation in ultra-relativistic heavy ion collisions.

\subsection{Open charm spectroscopy}
The D meson, consisting of the heavy and light constituents, is a very interesting object since it combines the aspect of the heavy quark as a static color source on the one hand, and the aspect of chiral symmetry breaking and restoration due to the presence of the light quark on the other hand. The phenomenological quark model was able to describe the excitation spectra of heavy-light systems for the lowest states. However the series of new observations in the open charm spectrum starting with the unexpected discovery of a narrow $D_s(2317)$ state by BaBar \cite{bib:Aubert:2003fg} attracted much interest since the new states do not fit well into the quark model predictions. Suggestions appeared that the scalar and the axial-vector $D_s$ mesons have an exotic non-$c\overline{s}$ structure, with possible tetraquark \cite{bib:Terasaki:2003qa} or $DK$ molecule interpretation \cite{bib:Barnes:2003dj}. Precise measurements of the widths and decay modes will allow to distinguish between different interpretations.

\subsection{Hypernuclear physics}
Hypernuclei are the systems in which one or more protons or neutrons are replaced by the hyperons. Although single and double $\Lambda$-hypernuclei were discovered decades ago, only 6 double $\Lambda$-hypernuclei have been discovered up to now. The \pbar beam at FAIR will allow efficient production of hypenuclei with more than one strange barion. These studies are expected to provide valuable information on hyperon-hyperon interaction. At PANDA not only $\Xi^{-}$ but also $\Omega^-$ atoms can be studied for the first time supplying unique information on hyperon-nucleus interaction.

\subsection{Electromagnetic Processes}
The experimental setup of PANDA will offer the opportunity to study also a certain class of hard exclusive processes such as Wide Angle Compton Scattering (WACS) and determination of electromagnetic formfactors in the time-like region from the process $p\overline{p} \rightarrow e^{+} e^{-}$ and/or $\mu^+ \mu^-$. These processes will be accompanied with a significant hadronic background, therefore the particle identification capability of PANDA will be essential to execute them successfully to complement the excellent performance of the Electromagnetic Calorimeter used to detect lepton pairs.

\section{FAIR and HESR storage ring}\label{sec:panda:hesr}
\begin{figure}
\begin{center}
\includegraphics[width=0.8\swidth]{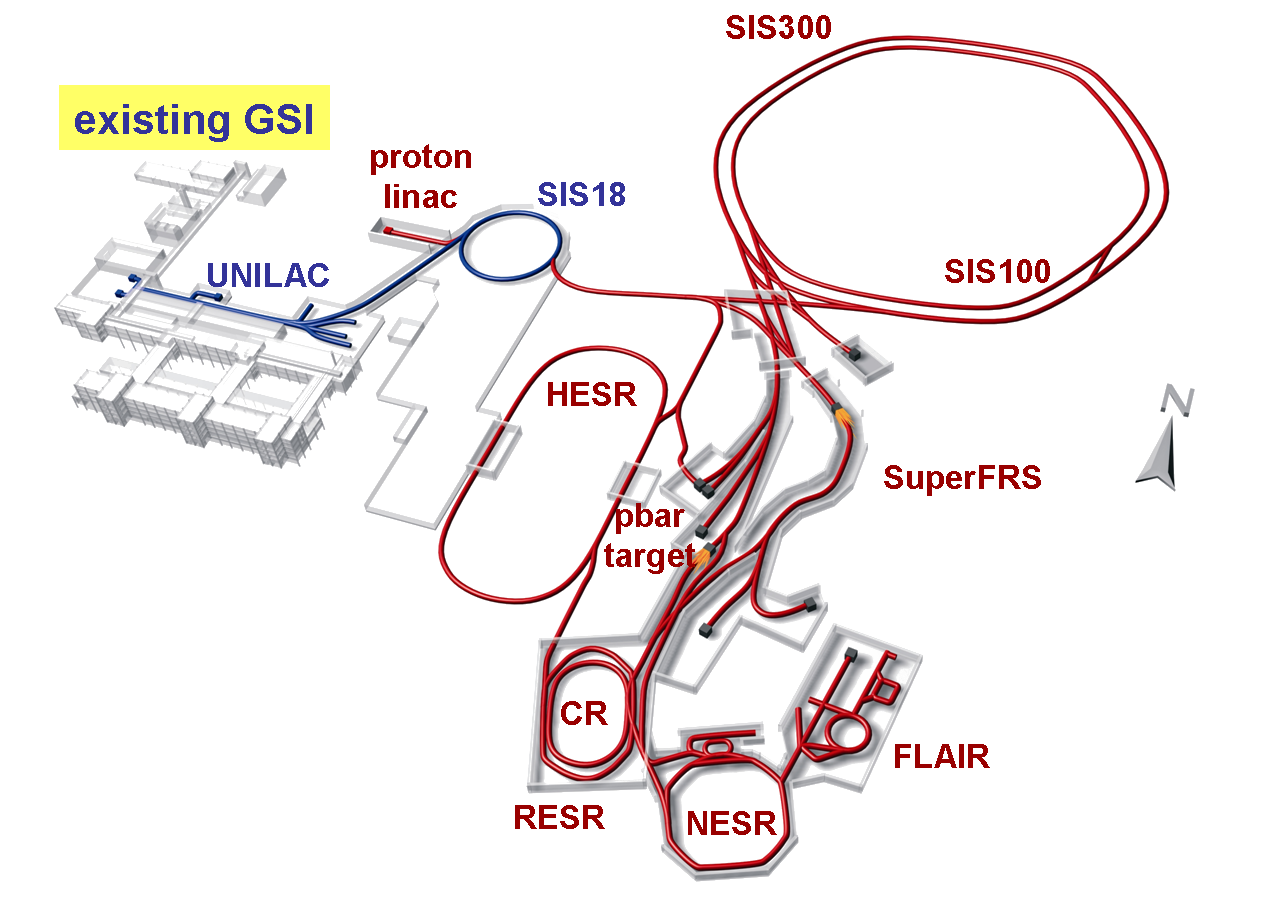}
\caption {Layout of the future FAIR facility at GSI.}
\label{fig:panda:fair}
\end{center}
\end{figure}

The future FAIR facility will be built profiting from the existing GSI facility. Its layout is presented in \Reffig{fig:panda:fair}. FAIR will provide good quality antiproton beam whose production and delivery chain is briefly described below. The proton linac will inject 70 mA of protons at 70 \mev into the existing heavy ion synchrotron SIS18. The SIS18 will be used to boost the proton energy to 2 \gev for injection into the superconducting synchrotron SIS100. The SIS100 in turn is employed to accelerate protons to the envisaged antiproton production energy of 29 \gev. Up to $4 \cdot 10^{13}$ protons per SIS100-cycle can be accelerated and before fast ejection to the antiproton production target the proton beam circulating in SIS100 has to be compressed to a single bunch of approximately 7.5 m length (~25 ns duration). With a bunch containing $4 \cdot 10^{13}$ primary protons at 29 \gev approximately $2 \cdot 10^{8}$ antiprotons are expected from the antiproton production target within the phase space acceptance of the following separator and the Collector Ring (CR). It has been demonstrated at the antiproton-facilities at CERN and FNAL, that collection and pre-cooling of antiproton beams on the one hand and accumulation on the other hand are optimized by means of two differently designed rings serving different purposes. In the CR antiprotons with a mean momentum of 3.8 \gevc are stochastically cooled and rebunched to be transferred to RESR. In RESR accumulation of antiprotons takes place. There is stochastic cooling applied during the whole accumulation process so that up to $7 \cdot 10^{10}$ antiprotons per hour should be obtained. Finally the antiproton beam is rebunched into one single bunch and transferred to the HESR.

\begin{figure}
\begin{center}
\includegraphics[width=1.0\swidth]{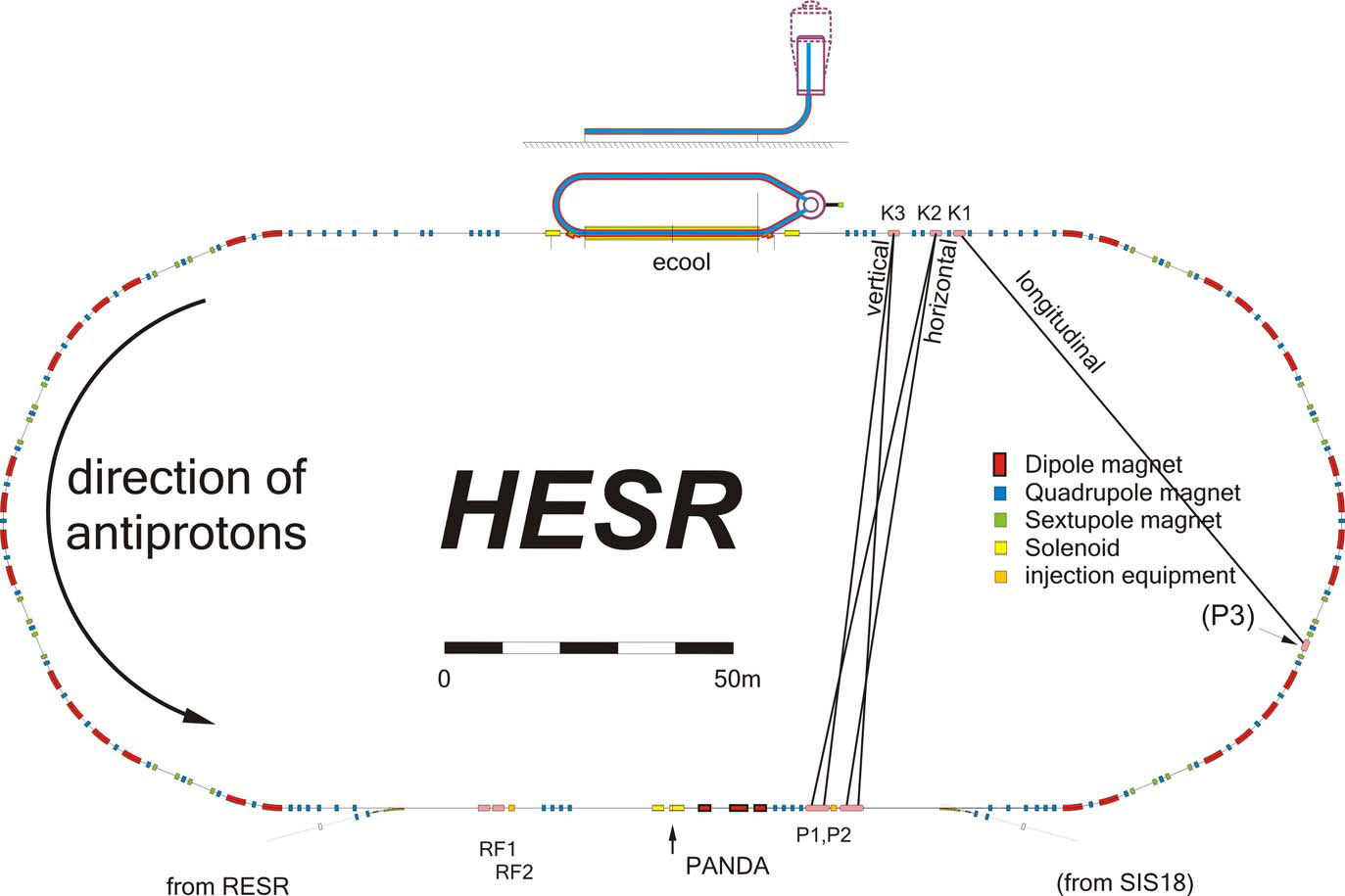}
\caption {Schematic view of the HESR with the indicated position of antiproton injection from RESR, cooling devices and experimental installations. Direct injection of protons from SIS18, which is useful for tuning of PANDA is foreseen in the later stage of construction of the system.}
\label{fig:panda:hesr}
\end{center}
\end{figure}

The synchrotron HESR (\Reffig{fig:panda:hesr}) accelerates or decelerates antiprotons to the desired momentum and serves as a storage ring for the internal target experiment PANDA. Stochastic and electron cooling of the antiproton beam are foreseen to satisfy requirements of the planned experiments. HESR is designed as a racetrack shaped ring with a circumference of 574 m, including the 132 m long straight sections. Within one of these sections the PANDA experiment will be installed. The opposite side is nearly completely used by the electron cooler. At both ends of the electron cooler compensation solenoids are foreseen. The equipment for stochastic cooling is installed in both straight sections. Two operation modes of the HESR are foreseen: high-luminosity and high-resolution with the parameters presented in the \Reftbl{tab:panda:HESRparam}.

\begin{table}
\begin{center}
\begin{tabular}{|l|l|}
\hline
\multirow{3}{*}{High resolution (HR)}
& Luminosity of ${2\cdot 10^{31}}{cm^{-2} s^{-1}}$ for $10^{10} \, \overline{p}$, \\
& RMS momentum spread $\sigma_p / p \leq 4\cdot 10^{-5}$, \\
& 1.5 to 9 \gevc, electron cooling  up to 9 \gevc \\
\hline
\multirow{3}{*}{High luminosity (HL)}
& Luminosity of ${2\cdot 10^{32}}{cm^{-2} s^{-1}}$ for $10^{11} \, \overline{p}$ \\
& RMS momentum spread $\sigma_p / p \sim 10^{-4}$, \\
& 1.5 to 15 \gevc , stochastic cooling above 1.5 \gevc \\
\hline
\end{tabular}
\caption{Parameters of the HESR operation modes.}
\label{tab:panda:HESRparam}
\end{center}
\end{table}

\section{PANDA detector overview}
\begin{figure}
\begin{center}
\includegraphics[width=1.0\swidth]{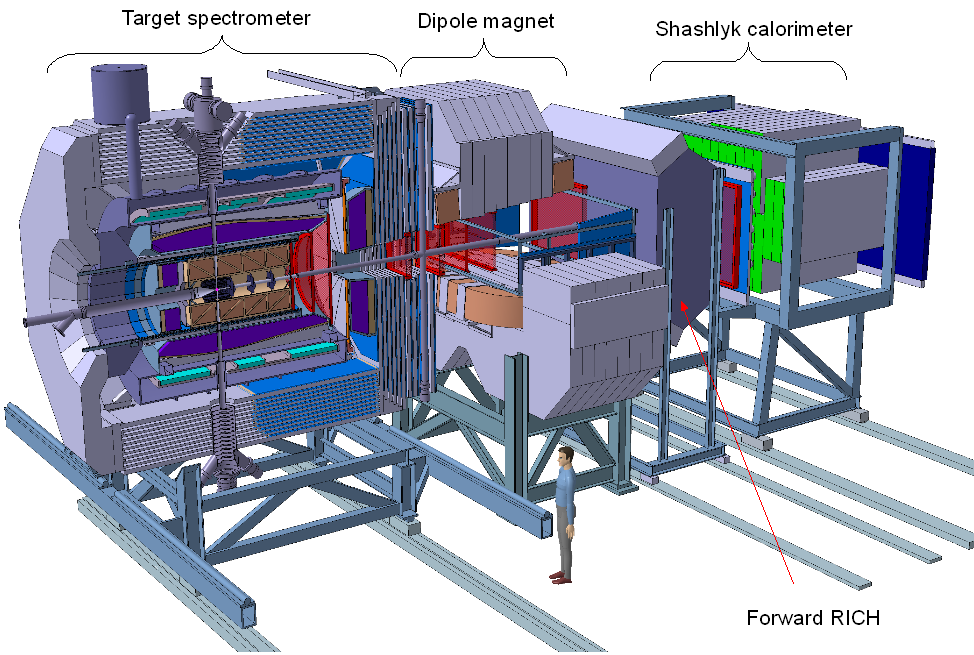}
\caption {Artistic view of the PANDA detector.}
\label{fig:panda:panda}
\end{center}
\end{figure}

The main design requirements for the PANDA detector, which are determined by the physics program, are: $4\pi$ acceptance, high resolution for tracking and calorimetry, particle identification and high rate capability. The detector is divided into two parts, the Target Spectrometer (TS) with the superconducting solenoid magnet and the Forward Spectrometer (FS) based on the dipole magnet. The main components of the PANDA detector (\Reffig{fig:panda:panda}) are briefly described below according to the PANDA Technical Progress Report \cite{bib:PANDA_TPR:2005} and the PANDA Physics Performance Report \cite{bib:PANDA_PB:2009}.

\subsection{Target Spectrometer}\label{sec:panda:target_sp}
\begin{figure}
\begin{center}
\includegraphics[width=1.0\swidth]{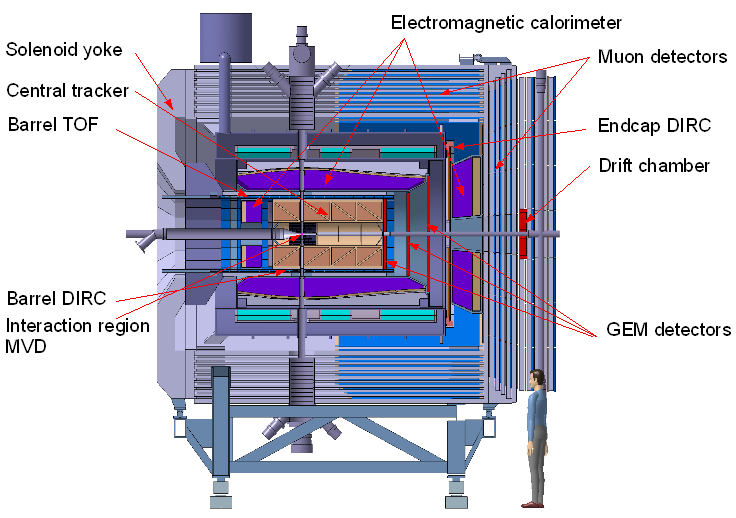}
\caption {Section of the target spectrometer showing the elements listed in \Refsec{sec:panda:target_sp}.}
\label{fig:panda:panda_ts}
\end{center}
\end{figure}

The target spectrometer, surrounding the interaction point, is split into the barrel part which covers polar angles above $22^{\circ}$ and the two end-caps, forward and backward. The forward end-cap is covering angles down to 5\degrees in the vertical and 10\degrees in the horizontal plane. The main design requirement to the target spectrometer is its compactness to minimize the cost of the magnet; this requirement imposes limitations on the size of the inner layers of the detector. The target spectrometer (\Reffig{fig:panda:panda_ts}) will contain the micro vertex detector (MVD), the tracking device with two possible alternatives presently considered, straw tube tracker (STT) or time projection chamber (TPC), forward GEM detectors, the PID detectors (Barrel DIRC, Forward Endcap DIRC and Barrel Time-of-Flight), electromagnetic calorimeter (EMC) and muon detectors. All the subsystems of the target spectrometer are described below except for EMC, which will be covered in a separate section (\Refsec{sec:panda:EMC}).

\subsubsection{Target}
The requirements to the target are dictated by two factors, the limited space available due to the desired minimal
distance from the interaction point to the vertex tracker and the need to provide about $4\cdot 10^{15}$ hydrogen atoms per $cm^2$ in order to achieve the designed luminosity of $2\cdot 10^{32}$ $cm^{-2}s^{-1}$ with $10^{11}$ antiprotons stored in HESR. To satisfy these requirements two complementary techniques have been proposed: the cluster-jet target and the pellet target.

The cluster-jet target is a homogeneous, variable density stream of hydrogen clusters formed by an expansion of pressurized cold hydrogen gas into vacuum through a Laval-type nozzle. This leads to condensation of hydrogen molecules into a narrow jet of hydrogen clusters with the size ranging from $10^{3}$ to $10^{6}$ molecules. However, the size of the interaction zone determined by the lateral spread of clusters is estimated at $\phi$ 15 mm at present, which exceeds the required. The desired density has not yet been reached with the $8 \cdot 10^{14}$ atoms/$cm^{2}$ achieved with the target prototype, therefore R\&D work of the participating teams is continuing to reach the goal.

The pellet target provides a stream of frozen hydrogen micro-spheres, called pellets, which traverse the interaction point at the velocity of about\,60 m/s with the typical pellet size of 25 - 40\,$\mu m$. Such target type is used presently at the WASA detector at the COSY machine in FZ-J\"{u}lich. It provides the density of $5\cdot 10^{15}$ hydrogen atoms per $cm^2$ required by PANDA, however the pellet stream is far from being uniform with the average distance between pellets of 2 mm. Anyway, the pellet target is an alternative of inferior quality since it provides reaction products in the pulse regime. Every time a pellet crosses the circulating antiproton beam there is a pulse of products.

\subsubsection{Solenoid Magnet}
The magnetic field of 2 T will be generated by a superconducting solenoid coil with the inner radius of 90\,cm and the
length of 2.8\,m. The field homogeneity, which is a critical parameter for the tracking devices, is foreseen to be better than 2\percent over the volume of the vertex detector and the central tracker.

\subsubsection{Microvertex Detector}
The purpose of the microvertex detector (MVD) within PANDA is to detect secondary vertices from the $D$ meson and hyperon decay and to improve the transverse momentum resolution of tracking. MVD will consist of the barrel part and eight disks perpendicular to the beam axis in the forward direction. The barrel part will consist of four layers, with the two inner layers of radiation hard silicon pixel detectors and the two outer layers of double sided silicon strip detectors. An inner radius of the barrel MVD is 2.5\,cm and an outer radius is 13\,cm. The first two of the forward disks of MVD are made entirely of pixel detectors, the following four are a combination of strip detectors on the outer radius and pixel detectors closer to the beam pipe. The last two disks, made entirely of silicon strip detectors, are placed further downstream to achieve a better acceptance for the hyperon cascades. The readout of pixel detectors via bump-bonded wafers with ASICs is foreseen as a default solution, while the pixel readout chip based on the 0.13\,$\mu m$ CMOS technology is under development for PANDA.

\subsubsection{Central Tracker}
The central tracker for PANDA will consist of two subsystems. For the barrel part around MVD two possible options are considered at present, Straw Tube Tracker (STT) or Time Projection Chamber (TPC). The second subsystem consists of three sets of GEM trackers covering forward angles. The main requirements addressed to the tracking devices are a good momentum resolution at the percent level and the ability to handle high count rates, which correspond to the designed luminosity of $2 \cdot 10^{32}$ $cm^{-2}s^{-1}$.

The STT consists of the aluminised mylar tubes, called straws, 10 $mm$ in diameter. The straws will be filled with a gas mixture of Argon and CO$_2$ used as a quencher at 1 $bar$ overpressure. They are arranged in 24 planar layers mounted in a hexagonal shape with 4200 straws in total. The resolution in $x$ and $y$ coordinates of about 150$\,\mu m$ is expected. The 8 central layers of the straws are tilted to achieve the acceptable resolution of 3\,mm also in $z$. A material budget is estimated at 1.3\percent of the radiation length.

The main advantages of TPC as an alternative to STT are low material budget and an additional particle identification capability through energy loss measurements. The TPC consists of two large gas-filled half-cylinders enclosing the target and beam pipe and surrounding the MVD. An electric field along the cylinder axis separates positive gas ions from electrons created by ionizing particles traversing the gas volume. The electrons drift with constant velocity towards the anode at the upstream end face and create avalanches detected at the pad readout plane yielding
information on two coordinates. The third coordinate of the track comes from the measurement of the drift time of each primary electron cluster. Traditionally in TPCs an amplification with multi-wire proportional chambers gated with an external trigger is used to avoid a continuous backflow of positive ions into the drift volume. The high counting rate expected with PANDA does not allow for application of the external trigger and a novel readout scheme was proposed based on GEM foils as amplification stages. Additional challenge for the TPC operation comes from the slow drift, therefore the necessity to deal with a large number of tracks coming from different interaction events at the same time. For the assignment of a track to the specific interaction the tracklet reconstruction has to take place close to the readout electronics with the following matching by time correlations with other detectors.

To perform the tracking at angles below 22\degrees in the forward direction, the gaseous micropattern detectors based on GEM foils as amplification stages are proposed. These detectors have rate capabilities three orders of magnitude higher than drift chambers and therefore can handle the particle flux of 3$\cdot10^{4}\,$cm$^{-2}$s$^{-1}$ expected in the vicinity of the 5\,cm diameter beam pipe. Current design of PANDA foresees three double planes of GEM detectors placed at 1.1, 1.4 and 1.9\,m downstream of the target. The readout is foreseen with the same front-end chips as are planned to be used for the silicon microstrips.

\subsubsection{Particle Identification}
Several systems are foreseen for the charged particle identification over a wide range of energies and momenta to fulfil the physics objectives of PANDA. The main part of the momentum spectrum above 1\,\gevc will be covered by Cherenkov detectors. For the low momenta the energy loss measurements can be performed with tracking detectors. In addition a time-of-flight barrel can be used for the identification of slow particles.

The Cherenkov detectors are based on the principle that a charged particle propagating in the medium with an index of refraction $n$, with velocity $\beta c\,>\,\mbox{1}/n$, emits a cone of light with an opening angle $\Theta_C\,=\,\arccos(\mbox{1}/n\beta$) to the direction of its propagation. The velocity information determined from $\Theta_C$, in combination with the momentum information
from the tracking detectors determines the mass of the particle. For the polar angles between 22\degrees and 140\degrees the barrel DIRC (Detection of Internally Reflected Cherenkov light) detector based on fused
silica (synthetic quartz) with a refractive index of 1.47 will be used. The 1.7$\,$cm thick perfectly polished quartz slabs will surround the beam line at a radial distance of 45 - 54 $\,$cm. The light emerging from each slab will be focused with lenses onto micro-channel plate photomultiplier tubes (MCP PMTs) which are insensitive to magnetic field.

The polar angles between 5\degrees and 22\degrees will be covered with the Forward Endcap disk DIRC based on a similar concept. The fused silica radiator will be arranged in shape of a disk with the focussing done by mirroring quartz elements reflecting onto MCP PMTs. The disk will be 2 cm thick and will have an outer radius of 110 cm.

For slow particles at large polar angles particle identification will be provided by a time-of-flight detector. In the target spectrometer the flight path is only of the order of 50 - 100\,cm. Therefore the detector must have a very good time resolution between 50 and 100\,ps. As detector candidates scintillator bars and strips or pads of
multi-gap resistive plate chambers (RPC) are considered.

\subsubsection{Muon Detectors}
The range tracking system will be implemented in the yoke of the solenoid magnet for the proper separation of primary muons from pions and decay muons. A fine segmentation of the solenoid yoke, playing the role of an absorber with aluminium drift tubes used as interleaved tracking detectors is foreseen (see \Reffig{fig:panda:panda_ts}). In the barrel region the yoke is segmented in such way that a first layer of 6\,cm iron is followed by 12 layers of 3\,cm thickness with 3\,cm wide gaps for the detectors. In the forward end-cap more material is needed, therefore, six detection layers will be placed around five iron layers each 6\,cm thick within the door, and a
removable muon filter with the additional five layers of 6\,cm iron is installed in the space between the solenoid and the dipole.

\subsection{Forward Spectrometer}
The Forward Spectrometer (FS) (see \Reffig{fig:panda:panda}), based on a wide-aperture dipole magnet, consists of the following subsystems: forward tracker (drift chambers or straw tubes), particle identification, represented by the RICH detector and time-of-flight wall, forward electromagnetic calorimeter and forward muon detector.

\subsubsection{Dipole Magnet}
The purpose of the dipole magnet with a maximum bending power of 2\,Tm is momentum analysis of charged
particles emitted in the forward direction. In the current design, the magnet yoke will occupy about 2.5$\,$m in the beam direction starting at 3.5$\,$m downstream of the target. Thus, it covers the entire angular acceptance of the forward spectrometer of $\pm$10\degrees{} in the horizontal and $\pm$5\degrees{} and in the vertical direction,
respectively.

\subsubsection{Forward Trackers}
The forward tracker, consisting of the set of six wire chambers (two located in front, two within and two behind the dipole magnet), will define the trajectories of charged particles deflected with the dipole magnet. The drift chambers with small-size cells and straw tubes are considered as possible alternatives for the forward tracker. Each set will contain three pairs of detection planes, one pair with vertical wires and two pairs with wires inclined by +10\degrees{} and
-10\degrees{} to reconstruct tracks in each chamber separately. The expected momentum resolution of the
system for 3$\,\gevc$ protons is $\delta p/p\,=\,0.2$\percent{}.

\subsubsection{Forward Particle Identification}
The RICH detector is proposed for the $\pi$/$K$ and $K$/p separation at the highest momenta with a dual radiator design similar to the one used at HERMES. Using two radiators, silica aerogel and C$_4$F$_{10}$ gas, makes it possible to achieve the $\pi$/$K$/p separation in a broad momentum range 2--15$\,\gevc$. A lightweight mirror focusing the Cherenkov light on an array of photo tubes placed outside the active volume will be used to reduce the thickness of the detector.

For the charged particle identification in the forward direction at momenta below 4-5 \gevc the time-of-flight technique can be applied. Time-of-flight wall made of plastic scintillators, read out on both ends by fast photo tubes, will be placed at 7\,m from the target. In addition to the main TOF wall, two side walls will be installed inside the dipole magnet gap, near the left and right wall of the magnet yoke, in order to register low momentum charged particles bent in the dipole field towards the yoke. The relative time-of-flight between two charged tracks will be measured for particles reaching any of the time-of-flight detectors including the barrel TOF. With the expected time resolution of $\sigma\,=\,50\,$ps $\pi$/$K$ and $K$/p separation on a 3$\,\sigma$ level will be possible up to momenta of 2.8$\,\gevc$ and 4.7$\,\gevc$, respectively.

\subsubsection{Forward Electromagnetic Calorimeter}
Detection of photons and electrons in the forward direction will be performed with the Shashlyk-type calorimeter. The modules of Shashlyk calorimeter form a sandwich of intercalating plates of lead and plastic scintillator. The readout of the module is performed with wave-length shifting fibres passing through the block and coupled to photomultipliers. The detector will consists of 26 rows and 54 columns, 1404 modules in total, with the size of front face of the module $55\times55$ mm. With the Shashlyk modules constructed for the KOPIO experiment at BNL \cite{bib:KOPIO:1999} the energy resolution of $4\percent/\sqrt{E}$ has been achieved.

\subsubsection{Forward Muon Detectors}
The range tracking system for discrimination of pions from muons in the very forward direction similar to the muon system of the target spectrometer has been designed.

\subsection{Data Acquisition}
The high data rates of $2\cdot 10^{7}$\,events/s expected at PANDA and the complexity of the experiment required the data acquisition concept different from the traditional two layer hierarchical approach. Traditionally a subset of specially instrumented detectors is used to evaluate a first level
trigger condition. For the accepted events, the full information of all detectors is then transported to the next higher trigger level or to storage.
The data acquisition (DAQ) concept of the PANDA assumes that each subdetector will be self-triggering, i.e. signals are detected autonomously by the sub-systems and are preprocessed. This requires hit-detection, noise-suppression and clustering at the readout level.  The data related to a particle hit, with a substantially reduced rate in the preprocessing step, are marked with a precise time stamp and buffered for further processing. The trigger selection finally occurs in computing nodes which access the buffers via a high-bandwidth network fabric thus each of sub-detectors can contribute to the trigger decision on the same level.

\section{Electromagnetic calorimeter}\label{sec:panda:EMC}
The detection of photons and electrons in a wide range of energy from several \mev up to 15 \gev is required in the PANDA experiment within the full solid angle. This task will be performed by an electromagnetic calorimeter (EMC).

Calorimeters, in contrast to tracking detectors, degrade the initial energy of the incident particle sharing it among the shower products, which are measured to determine the primary information on the particle. The important feature of calorimeters is that they are sensitive not only to charged particles but also to neutral particles. The electromagnetic shower propagation proceeds in the way that e.g. the initial electron creates a photon via bremsstrahlung process and in turn the photon provokes pair production, an electron and a positron. The shower develops resulting in increasing number of particles until the average energy of products decreases below a critical energy at which the main process of energy loss is ionisation. The longitudinal size of the shower is determined by the radiation length $X_{0}$ of the material where the shower propagates. The transverse dimension of the shower is characterised by
a Moli\`{e}re radius, which, by definition, is the radius of a cylinder containing on average 90\percent of the shower's energy deposition.

Two types of electromagnetic calorimeters are used: homogeneous and heterogeneous (or sampling). In the homogeneous one the same medium is used to cause the shower development and to detect the produced particles, whereas sampling calorimeters have sandwich structure which consists of passive layers of converter with high Z alternated with active detection layers.

There are several general requirements, which determine the design of the electromagnetic calorimeter in PANDA.
The structure of EMC is determined by the design of the PANDA Target Spectrometer, which assumes that EMC is placed inside the yoke of the solenoid. From the point of view of cost the size of the yoke is limited which in turns requires the calorimeter to be built from a material with small radiation length. On the other hand operation of EMC in strong magnetic field of the solenoid makes it impossible to use conventional PMTs for light readout. The fast response of EMC is necessary because of high event rates up to $2 \cdot 10^{7}$ events/s expected with PANDA. The requirements of good energy resolution over a wide energy range and high granularity are determined by the physics program of PANDA, which requires reconstruction of electrons and photons in the energy range from the low energy threshold at 10 \mev up to 15 \gev. Last but not least is the requirement of radiation hardness of the used components. More specific design issues are determined from the studies of possible registration in the particular physics channels.

Combination of these requirements led to a design of the homogeneous electromagnetic calorimeter for the PANDA detector with PWO-II scintillator material and readout with Large Area Avalanche Photodiodes (LAAPD) and vacuum phototriodes. It will be shown below how the proposed design fulfils the requirements.

The requirement of compactness determines to a significant extent the choice of PANDA EMC as a homogeneous calorimeter because a typical lateral size of the sampling calorimeter modules needed to cover the maximum energy of photons is much larger than the lateral size of the crystal for a homogeneous calorimeter. Additional arguments in favor of the homogeneous calorimeter are: typically worse energy resolution of sampling calorimeters insufficient for the study of many physics channels and an elevated value of the low energy threshold for the detection of photons.

\subsection{Scintillator material (PWO)} \label{sec:panda:PWO}
The choice of lead tungstate, $PbWO_4$ (PWO), as a scintillator material for PANDA EMC is motivated by the requirements of compactness, fast response and radiation hardness. Its density $\rho$=8.28 $g/cm^3$ and radiation length, $X_0\,=0.89\, cm$, ensure that with this choice the most compact design will be achieved as compared with any other scintillator used in high energy physics.
By its physical properties $PbWO_4$ is a birefringent, tetragonal, scheelite-type crystal belonging to the space group of monoclinic raspite, colourless and transparent in the visible spectrum. Its scintillation mechanism, optical properties and light yield are discussed below.

The two most extensive recent applications of PWO are in the electromagnetic calorimeters of the CMS (ECAL) \cite{bib:cms_ecal_tdr:1997kj} and ALICE (PHOS) detectors \cite{bib:alice_phos_tdr:1999kd} at the CERN LHC. High quality of scintillators in mass production for these detectors has been achieved, however further optimisation for the application in PANDA appeared mandatory, because of the required much lower energy range. In PANDA the detection of 10-20 \mev photons is required which asks for the increase of light yield in comparison with 9-11 phe/MeV (measured with a PMT with bi-alkali photocathode at room temperature) typical for crystals used in these two experiments.

\subsubsection{Production}
The world largest producer of PWO crystals is the Bogoroditsk Techno-Chemical Plant (BTCP) near Tula (Russia). It is considered as the main producer of crystals for PANDA. The Czochralski method is used for crystal growth from a platinum crucible. To grow high quality crystals an additional precrystallization is required. The polycrystalline $PbWO_4$ obtained during precrystallization is sintered and used as a starting material for the final crystal growth. Up to seven crystals can be grown from the same initial load of the crucible by adding some new material between the crystallizations. Ingots of 250 mm length and of slightly elliptical cross-section, up to 45 mm in diameter became available. Crystal growth along the direction of the $a$ axis appears to be the best for producing long
samples. An alternative producer of PWO crystals, the Shanghai Institute of Ceramics of the Chinese Academy of Sciences (SICCAS) uses the modified Bridgman-Stockbarger technique in closed platinum foils for the crystal growth. This technique produces simultaneously several ingots.

\subsubsection{Scintillation mechanism}
The scintillation mechanism of PWO is determined by the electronic structures of the conduction and valence
bands in this material, which are described in details in \cite{bib:PhysRevB.64.245109}. Luminescence appears in PWO due to charge-transfer transitions in anionic molecular complexes. Both regular $WO_4^{2-}$ and irregular $WO_3$ tungstate groups are the luminescence centers. Because the $WO_4^{2-}$ complex has a $T_d$ point crystalline field symmetry, the final configuration of the energy terms is found to be $^3T_1$, $^3T_2$ and $^1T_1$, $^1T_2$ with the $^1A_1$ ground state.
The $^1A_1 \rightarrow ^1T_1 \,, ^1T_2$ transitions are responsible for the blue luminescence (420nm) of the lead tungstate crystal. When an anion vacancy appears in a $WO_4^{2-}$ anionic complex the local symmetry of the new
$WO_3$ complex is reduced to $C_{3\nu}$. Luminescence in green (490 nm) is ascribed to this center.

\subsubsection{Thermal quenching in PWO}
Most of the crystals of the tungstate family have an intense but slow emission in the millisecond range. $PbWO_4$ has a rather weak but fast emission because of a strong quenching of the scintillation process. This quenching is rather complex in origin but to a large extent associated with high-temperature charge transfer processes and thermal decomposition of the excited states. As a result, the light yield of PWO is strongly temperature dependent. In the range between $+20\degrees$ and $-25\degrees$ the dependence of light yield on the temperature is almost linear with a gradient of about $-2\percent/\degrees C$.

\subsubsection{PWO-II}
\begin{figure}
\begin{center}
\includegraphics[width=0.5\swidth]{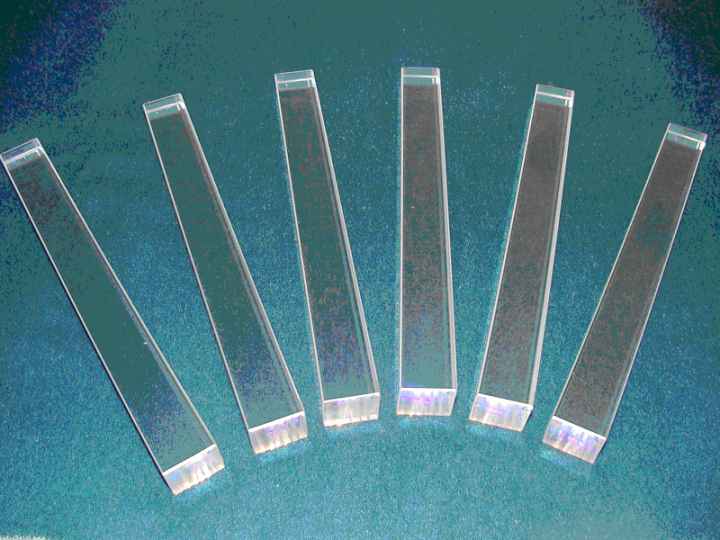}
\caption {Machined and optically polished PWO-II crystals (200mm length) for the PANDA electromagnetic calorimeter \cite{bib:emctdr}.}
\label{fig:panda:scint_pwo}
\end{center}
\end{figure}

Detailed studies have been performed to improve the light yield of PWO scintillators. During the initial development of PWO scintillators it was established that an additional doping with molybdenum (Mo) and lanthanum (La) at concentrations $< 100 \, ppm$ increases the light output, however it introduces the slow scintillation components with $\tau=1-4 \, \mu s$. The further research was in direction of increasing the structural perfection of the crystals and reduction of the concentration of point structure defects by doping with yttrium (Y), lanthanum (La) or lutetium (Lu) ions, which suppress oxygen and cation vacancies in the crystal matrix. Structural perfection of the crystal has been reached at BTCP by an improved control of the stoichiometric composition of the melt. These crystals contain at least
two times less Frenkel type defects leading to an increase of the light yield by 80\percent compared to the average CMS quality. The new generation of crystals is called PWO-II (second generation) according to the accepted terminology. The scintillators produced following this technology will be used for the construction of PANDA EMC. An additional property of PWO-II crystals is their larger average temperature gradient of the light yield $-3\percent/\degrees C$, which is 50\percent higher than for the standard PWO crystals. A sample of the machined and optically polished PWO-II crystals produced for the PANDA electromagnetic calorimeter is presented in \Reffig{fig:panda:scint_pwo}.

\subsubsection{Optical properties}\label{sec:panda:optical_properties}
The scintillation emission spectrum of PWO results from the superposition of two broad and complex emission bands at 420 nm and 500 nm respectively. The optimization of crystals leads to a Gaussian-shaped spectrum (140 nm FWHM) peaking at about 440 nm with a range from 360 nm to 570 nm at 10\percent of the maximum. The scintillation kinetics is fast at room temperature and is dominated by an exponential decay with a constant of $\tau$=6.5ns. The possible presence of slow component was verified for PWO-II crystals \cite{bib:emctdr} at $+25\degrees C$ and $-25\degrees C$. The measurements demonstrated that 95\percent of the scintillation light is collected within a time window of 300-400 ns.

Crystalline material of lead tungstate is transparent in the visible spectral range and fully colorless. An optical transmission of the crystals has been steadily improved, in particular in the region of scintillation, between 360 and 570 nm. The optical transmission of $PbWO_4$ crystals can be limited by the presence of macroscopic defects like inclusions, precipitates or veils which scatter light in all directions, or by the existence of traps which induce absorption bands. A better control of the raw material preparation and of the growth and annealing conditions, as well as the introduction of dopants, have led to a considerable improvement in the optical transmission of the full-size
$PbWO_4$ crystals.

As a result of the improved technology of the PWO crystals production, the typical values of light yield of 17-20 phe/MeV are observed for full size (200mm length) PWO-II crystals measured at room temperature with a photomultiplier with bi-alkali photocathode (quantum efficiency $\sim 20 \percent$). Reducing the thermal quenching of the luminescence process by cooling the PWO-II crystals from $+25\degrees C$ to $-25\degrees C$ leads to an increase of light yield by a factor $\approx$4.

\subsubsection{Radiation hardness}
The important characteristics of scinitallators in application for calorimetry is their behavior in radiation environment. The dose load to the crystals from the hadronic probes during the operation of the PANDA experiment was estimated \cite{bib:emctdr} with a Dual Parton Model (DPM) event generator. With the assumption of event rate of $10^7$/sec expected for PANDA the absorbed energy rates of 27 mGy/h, 1.5 mGy/h, 0.16 mGy/h and 0.03 mGy/h were calculated for the polar angles of 5\degrees, 25\degrees, 90\degrees and 135\degrees, respectively.

It has been established previously \cite{bib:Lecoq:1994yr}, that the radiation load does not damage the scintillation mechanism of PWO. The main effect is the loss of optical transmission due to creation of color centers with absorption bands in the visible spectral region. The deterioration of the optical transmission in the experiment is caused by the interplay of damage and recovery mechanisms. The latter are fast at room temperature and keep the loss of light transmission at a moderate level. The operation of the PANDA EMC foreseen at $-25\degrees C$ required additional studies of the optical transmission upon irradiation with $\gamma$-rays at this low temperature. The two irradiation facilities at IHEP (Protvino, Russia) and JLU (Giessen, Germany) have been adapted for that purpose.
The measurements \cite{bib:Semenov:2009zz} have shown that due to a slow relaxation of the color centers in the cooled crystal one deals, asymptotically after a prolonged irradiation depositing 30-50 Gy, with a reduction of 20\percent - 30\percent in the scintillation response. This reduction is compensated with a significant margin by the factor of $\approx$4 increase in light yield due to cooling (\Refsec{sec:panda:optical_properties}) relative to room temperature.

\subsection{Photosensors (APDs)}
The operation of the PANDA EMC inside the 2 T magnetic field of the solenoid makes it impossible to use conventional photomultipliers for light readout. Two types of photosensors insensitive to magnetic field have been chosen for the PANDA EMC, silicon avalanche photodiods (APD) for the barrel part and vacuum phototriodes (VPT) for the forward endcap. The current section is mainly concentrated on the description of the operation principle and properties of APDs since the experimental part of the thesis presented in \Refchap{sec:measurements} is devoted to a study of this type of readout.

\begin{figure}
\begin{center}
\includegraphics[width=0.5\swidth]{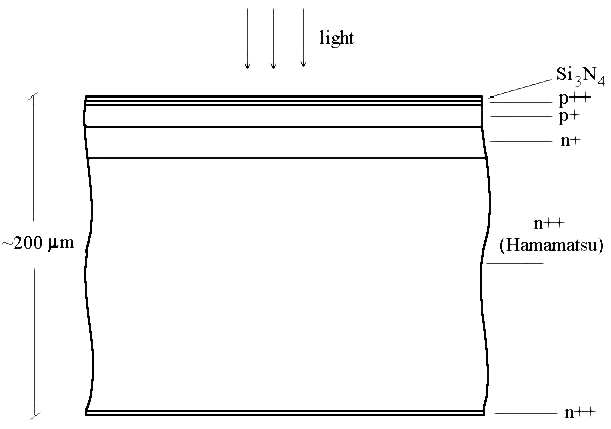}
\caption {Schematic view of an APD with reverse structure \cite{bib:cms_ecal_tdr:1997kj}.}
\label{fig:panda:apd_structre}
\end{center}
\end{figure}

Besides insensitivity to magnetic field APDs have additional properties, which motivate their application for the PANDA EMC: internal signal amplification and weak response to ionizing radiation. An APD optimised to detect the scintillation light from the lead tungstate crystals was developed for the CMS detector in collaboration with the Hamamatsu Photonics \cite{bib:cms_ecal_tdr:1997kj} factory in Japan. These APDs have the following features: compactness, with the overall thickness of less than 2 mm; fast rise time of the order of 2 ns; a quantum efficiency of 70 - 80\percent; insensitivity to magnetic field and low cost in mass production. In contrast to most APDs used for detection of red and infrared light, the inverted structure has been applied for the CMS APDs to overcome such drawbacks as a reduced gain for wavelengths shorter than 500 nm and high sensitivity to ionizing particles. Schematic view of an APD with an inverted structure is presented in \Reffig{fig:panda:apd_structre}. In these APDs light enters via the $p^{++}$ layer and is absorbed in the $p^+$ layer lying below, where electron-hole pairs are generated. The electrons then drift in the electric field towards the p–n junction. They are amplified there by an impact ionization and drift in the n- or p-material towards the $n^{++}$ electrode where the charge is collected. In front of the $p^{++}$ layer is a passivation layer which both protects the wafer and provides an anti-reflection coating. This passivation layer is made of either silicon dioxide or silicon nitride as in \Reffig{fig:panda:apd_structre}.
The amplification or gain of the APD is largest for wavelengths which are completely absorbed in the $p^+$ layer, in the present case for the wavelengths shorter than 550 nm. In APDs with the reverse structure the response to ionizing radiation is much reduced and to first order it is proportional to the thickness of the $p^+$ layer in front of the amplification region.

The main disadvantage of the S8664-55 APDs in application for the PANDA EMC is their small active area $5\times 5$ $mm^2$ in comparison with the surface of the crystal exit face which is $21.4\times 21.4$ $mm^2$. The PANDA collaboration invested into a collaborative effort with Hamamatsu Photonics, which resulted in development of large area APDs (LAAPDs) with an active area of $10\times 10$ $mm^2$ and the same reverse internal structure as in the CMS APDs. At present, there is an APD of the third generation in development whose active area $10\times 20$ mm will permit to cover completely with two APDs of this type the exit face of a PWO-II scintillator. Having two APDs on each scintillator gives an additional advantage of having the possibility to cover different dynamic ranges of the registered photons. Moreover, one has a back-up solution in case one of them fails due to radiation damage.

\subsubsection{APD gain}
One of the key parameters which define the operation of an APD is its gain. The APD gain depends on the electric field applied across the p-n junction. The higher the reverse voltage, the higher will be the gain. However, if the reverse voltage is increased above the limiting value, a break-down of the p-n junction occurs. The current flowing through the device series resistance and the load resistance cause the voltage drop and protect from permanent damage of the device \cite{bib:APD_Hamamatsu}. Moreover, the gain of an APD is temperature dependent. At a fixed reverse voltage the gain decreases with the increasing temperature. This happens because of the stronger excitation of the lattice vibrations - phonons, provoking more frequent collisions with the charge carriers and preventing them from reaching sufficient energy to trigger ionization. The temperature dependence of gain has a value of - 2.2\percent/\degrees{}C at an internal gain of M = 50 for a CMS-APD \cite{bib:Renker:2000}. This dictates the requirement of stability of the operating temperature for the PANDA EMC at the level $\Delta T = \pm 0.1\degrees{}C$.

\subsubsection{APD noise and excess noise factor}
Since an APD is designed to operate under a reverse bias, its sensitivity at low light levels will be limited by the shot noise and the APD's leakage current. Shot noise derives from the random statistical Poissonian fluctuations of the dark current, $I_d$. The dark current $I_d$ of an APD can be divided into the bulk current, $I_b$, and the surface current, $I_s$. While the surface current is independent of the applied gain, $M$, the value of the bulk current increases with increasing gain, hence the total dark current can be defined as:
\begin{equation}
I_d=I_b \cdot M + I_s.
\end{equation}

In addition, avalanche photodiodes generate excess noise due to the statistical nature of the avalanche process. These statistical fluctuations of the APD gain are characterised by an excess noise factor, $F$, and have its origin in inhomogeneities in the avalanche region and in hole multiplication. $\sqrt{F}$ is the factor by which the statistical noise of APD current (equal to the sum of the multiplied photocurrent plus the multiplied APD bulk dark current) exceeds that which would be expected from a noiseless multiplier on the basis of Poissonian statistics (shot noise) alone. The total spectral noise current for an APD is given by:
\begin{equation}
i_n=[2q(I_s+I_b \cdot M^2 \cdot F)B]^{1/2},
\end{equation}
where $q$ is the electron charge and $B$ is the bandwidth.

The excess noise factor is a function of the carrier ionization ratio, $k$, where $k$ is usually defined as the ratio of hole to electron ionization probabilities and is given by the following equation:
\begin{equation}
F=k \cdot M + (2-k)(1-\frac{1}{M}),
\end{equation}
and therefore the lower are $M$ and $k$, the lower is excess noise factor.

The excess noise factor measured for the APD S8664-1010 is $F=1.38$ at the gain $M=50$ and $F=1.57$ at the gain $M=100$ \cite{bib:emctdr}.

\subsubsection{Radiation hardness}
The effect of radiation on APD's performance is a critical issue for their operation in electromagnetic calorimeters. Two main damage mechanisms can take place. The first one is a bulk damage, due to displacement of Si atoms from their lattice sites, causing an increase in dark current. The second one is a surface damage, or the creation of defects in the front layer, which may lead to an increase of the surface current and a decrease of the quantum efficiency. The detailed studies of APDs radiation hardness have been performed with different sources of radiation at $T=-25\degrees{}C$, which corresponds to the foreseen operation conditions of the PANDA detector. The change of surface and bulk dark current induced by irradiation was studied with the proton beam at KVI, with an intense $^{60}Co$ $\gamma$-source at Giessen and with an $^{241}Am-\alpha-Be$ neutron source at Frankfurt. It was shown that the proton induced damages are mainly located inside the conversion layer and at the end of the avalanche region.
The irradiation of APDs with an integrated fluence of protons $1.1\cdot 10^{13} \, p$ which is comparable with the dose expected in 10 years of PANDA operation leads to a decrease of the APD gain by 50 - 60\percent \cite{bib:emctdr}. Measurements with photons and neutrons give no indication for any kind of induced bulk damage and it is concluded that an increase of dark current is caused by the surface damage only.

\subsubsection{Vacuum phototriodes (VPT)}
The photon detection in the forward endcap EMC has to deal with rates up to 500 kHz and magnetic fields up to 1.2 T. Therefore vacuum phototriodes (VPT) were chosen with a diameter of the photocathode of 22 mm as photon sensors for this part of the EMC. The main reasons for this choice are rate capability, radiation hardness, absence of nuclear counter effect and absence of temperature
dependence of the efficiency. In contrast to the barrel region, the magnetic field is oriented in the axial direction of the VPTs and
thus makes it feasible to use VPTs for the endcap. Vacuum phototriode is essentially a photomultiplier tube with only one dynode and a weak dependence of the gain on the magnetic field. The VPTs with standard quantum efficiency of 15\percent and gain of about 10 are available from Hamamatsu and the Research Institute Electron (RIE). However, Photonis is developing phototriodes with high quantum efficiency photocathodes and high gain $>$40 at 1000 V, which are more suitable for PANDA.

\subsection{Coverage of the EMC. Mechanical design}
\begin{figure}
\begin{center}
\includegraphics[width=0.9\swidth]{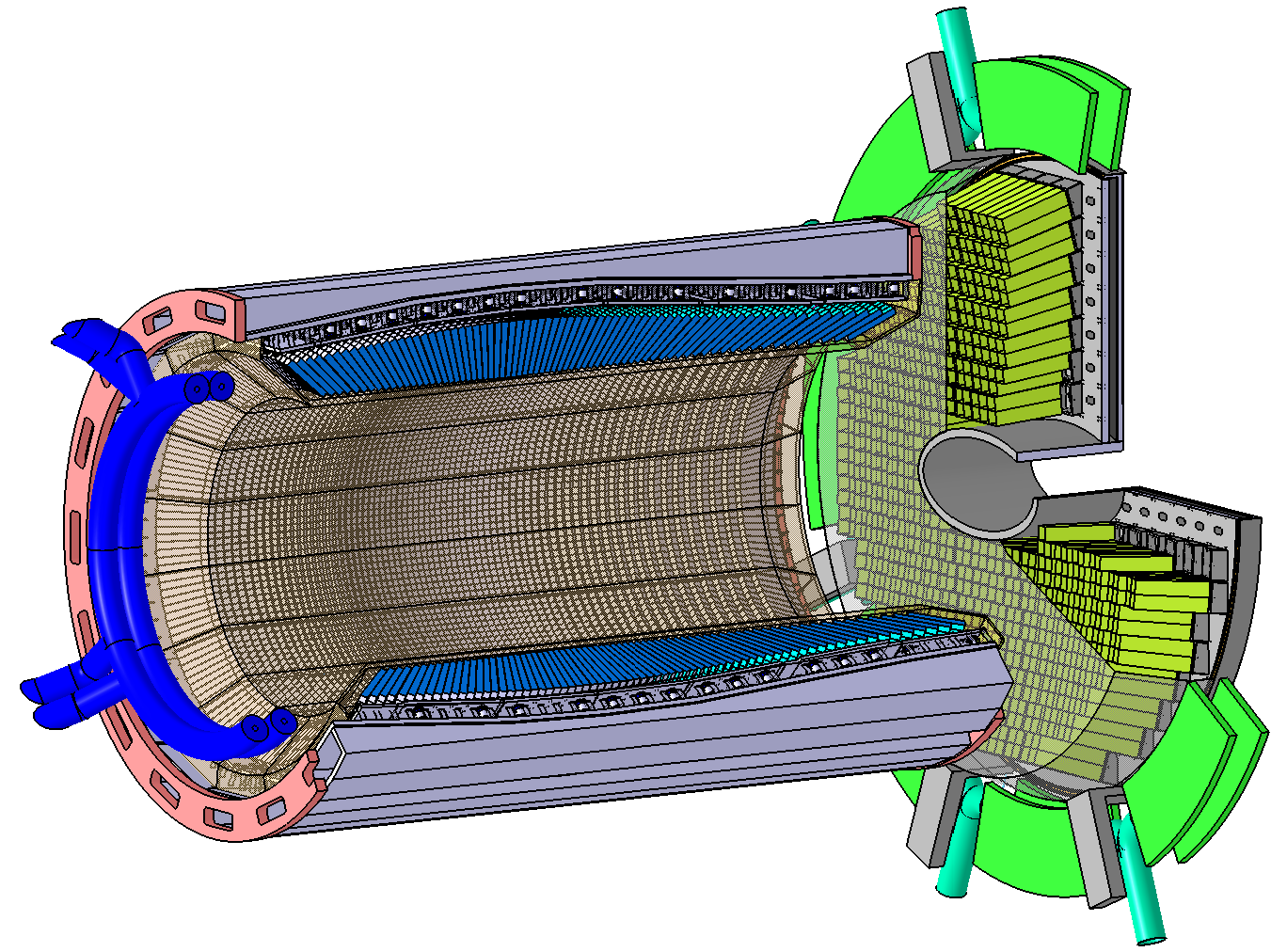}
\caption {The barrel and forward endcap of the PANDA EMC \cite{bib:emctdr}. The antiproton beam is arriving from the left-hand side of the drawing along the symmetry axis of the barrel EMC.}
\label{fig:panda:emc}
\end{center}
\end{figure}

The maximal coverage of the solid angle will be satisfied by dividing EMC into three parts: barrel and two endcaps. The barrel EMC, consisting of 11360 crystals, will cover polar angles from $22\degrees$ till $140\degrees$ and will have an inner radius of 57 cm and an outer radius of 94 cm. The forward endcap, located at 2.05 m downstream from the target, will consist of 3600 crystals and will cover polar angles from 23.6\degrees{} down to 5\degrees{} in the vertical and 10\degrees{} in the horizontal direction. The backward endcap in the current design consists of 592 crystals, covers polar angles from 151.4\degrees{} till 169.7\degrees{}, and is located at 55 cm upstream from the target. The barrel and the forward endcap of the PANDA EMC are presented in \Reffig{fig:panda:emc} following the Technical Design Report for the PANDA Electromagnetic Calorimeter \cite{bib:emctdr}.

\begin{figure}
\begin{center}
\includegraphics[width=1.0\swidth]{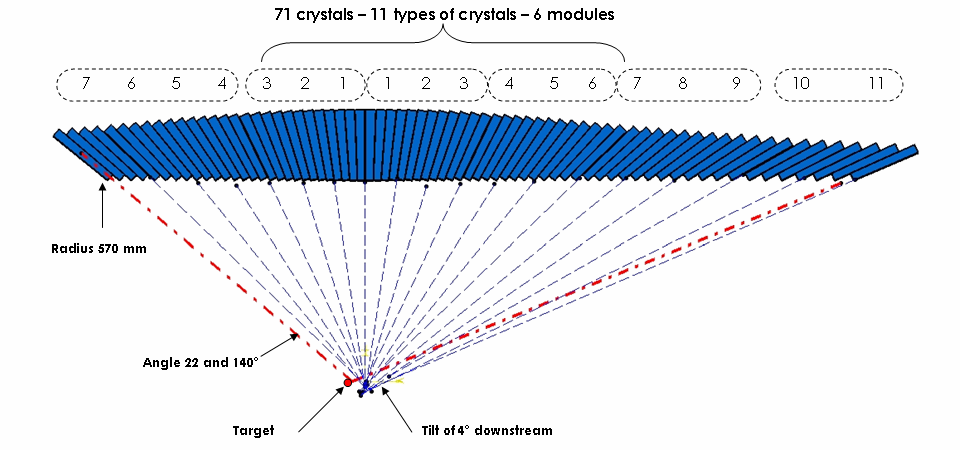}
\caption {Geometrical arrangement of the crystals of the barrel in a cut along the beam axis \cite{bib:emctdr}. Numbers indicate different crystal types.}
\label{fig:panda:emc_mech_Barrel}
\end{center}
\end{figure}

The individual shape of a crystal in the barrel part is a tapered parallelepiped, with dimensions depending on the position along the beam axis. The front face of an individual crystal is close to the square with the 20 mm side (21.18-22.02 mm), which corresponds to Moli\`{e}re radius of $PbWO_4$.
The barrel part is divided into 16 identical sectors, each subtending 22.5\degrees of the azimuthal angle, called slices. The crystals are grouped into packs of $4\times 10$ along the azimuth.
Along its length the barrel contains 71 crystals of 11 types depending on the position, as indicated in \Reffig{fig:panda:emc_mech_Barrel}. In one slice of 710 crystals, 3 or 5 alveole packs are grouped together into 6 modules: 4 modules of 120 crystals each, 1 module with 70 and 1 with 160 crystals. The off-pointing geometry of EMC is foreseen, i.e. the crystals are not pointing towards the target position. This ensures, that tracks originating at the target never pass through gaps between the crystals, but always cross a significant part of a crystal.

The dimensions of the crystals in the forward endcap are determined by the requirement to accommodate the VPT readout on the rear surface of the crystal. Currently available VPTs with the diameter of 22 mm can be accommodated on a crystal with a rear cross section of $26 \times 26 \, mm^2$. The off-pointing geometry of the forward endcap EMC, i.e. an orientation of the crystals towards a point on the beam axis 950 mm upstream the target, determines that each crystal has a front-face of $24.4\times 24.4\,mm^2$. A cluster of 4 crystals will form a mini-unit of trapezoidal cross section, mounted in a single carbon-fiber alveole. Four mini-units of crystals will be combined to form a 16-crystal subunit of ca. 19 kg that can be attached individually to the 30 mm thick aluminum mounting plate.

To optimize light collection crystals will be wrapped with a reflective material considered as Radiant Mirror Film ESR from 3M in the current design.

The cooling of crystals down to -25\degrees{}C is foreseen with stability at the level of $\pm 0.1\degrees{}C$, which will be satisfied by a cooling system in combination with a thermal shielding. The requirement of cooling power is estimated at $\approx$3300 W.

\subsection{Electronics}
According to the general scheme of the Data AcQuisition (DAQ) for the PANDA experiment, the readout of EMC assumes continuous digitization of the amplified and shaped signals from APD and VPT photosensors. The readout chain will include low-noise front\-end electronics, digitizer modules and data multiplexers.

The requirements to frontend electronics are determined by several factors. The detection of photons with energies from the maximum 12 \gev per crystal down to 3 \mev, which is the single crystal threshold with the electronics noise level of 1 \mev requires the dynamic range of 12000. Expected event rates up to 500 kHz for the forward endcap and up to 100 kHz for the barrel require short shaping times. Operation of the frontend electronics inside the cooled volume implies strict conditions on power consumption. Two complementary low-noise low-power charge sensitive preamplifier circuits have been developed. The first one, based on discrete components with very good noise performance using signal shaping with a peaking time of 650 ns, is foreseen for readout of the forward endcap EMC, where event rates up to 500 kHz are expected. A CMOS ASIC, based on 350 nm technology, will be used for the barrel EMC. It has a similar noise performance with a shorter peaking time of 250 ns and is designed for event rates up to 350 kHz.

The digitizer modules will be located at a distance of 20-30 cm and 90-100 cm for the barrel EMC and the forward endcap EMC, respectively, away from the analogue circuits and outside of the cold volume. The digitizers consist of high-frequency, low-power pipelined ADC chips, which continuously sample the amplified and shaped signals. The sampling is followed by the digital logic, which processes time-discrete digital values, detects hits and forwards hit-related
information to the multiplexer module via optical fiber links. At this step the detection of clusters of energy deposition can be efficiently implemented. The multiplexer modules will be located in the DAQ hut and they will perform advanced signal processing to extract amplitude and signal-time information.

\subsection{Performance}
The main task of the electromagnetic calorimeter is reconstruction of photons and leptons with high efficiency. EMC measures the deposited energy of a particle, $E$, and provides information on its direction of propagation by determining the point of impact ($\theta$, $\phi$). Therefore, the energy and spatial resolution are the main parameters which define performance of the EMC. Besides the energy resolution, the energy coverage is an important characteristics of the EMC and especially the low energy threshold. In addition EMC can provide the time information and therefore the time resolution of the order of 1 ns achievable with the PWO scintillators can be used for event separation. EMC can also provide $e^{\pm}$ separation from the $\pi^{\pm}$ background by an analysis of the lateral shower shape. Such separation is possible because hadronic showers have different lateral shower shape defined by the difference in energy loss per interaction and the elementary statistics of the underlying processes.

\subsubsection{Energy threshold}
In study of radiative decay of the charmonium states produced as resonances in $\overline{p}-p$ annihilation the final states with $\pi^{0}$'s and $\eta$'s produced non-resonantly are a significant source of background photons for the direct photons from the transitions in charmonium. If one of the photons from $\pi^0$ (or $\eta$) decay has energy below the threshold, the $\pi^0$ (or $\eta$) is not reconstructed. The second photon will be erroneously classified as coming from the resonance decay thus adding to the background. Therefore the probability of $\pi^0$ reconstruction depends on the the minimum photon energy $E_{thres}$ accessible with the EMC; $E_{thres}$=10 \mev is a design goal for PANDA EMC. A Monte Carlo based study of this problem is presented in \Refchap{chap:phys:MonteCarlo}.

\subsubsection{Energy resolution}
The energy dependence of the Gaussian energy resolution $\sigma_{E}/E$ can be parameterized in the following way:
\begin{equation}
\frac{\sigma_{E}}{E}=a\oplus \frac{b}{\sqrt{E(\gev)}},
\end{equation}
where $a$ is a constant term and $b$ is a stochastic term.
The stochastic term is mainly defined by photostatistics and can be expressed as
\begin{equation}
b=\sqrt{\frac{F}{N_{pe}}},
\end{equation}
where $N_{pe}$ is the number of electron-hole pairs released in the photodetector per \gev energy deposited in the scintillator and $F$ is an excess noise factor, which parametrises fluctuations of the gain process in an APD.

The constant term has contributions from leakage of energy out of the scintilator, non-uniformity of light collection and intercalibration errors. The design requirements for PANDA EMC are $a <$ 1 \percent and $b <$ 2 \percent \cite{bib:emctdr}.

\subsubsection{Position resolution}
Position resolution is mainly defined by granularity of the electromagnetic calorimeter. The reconstruction of an impact point of a particle is performed by defining the cluster, i.e. a group of neighboring crystals with the deposited energy above the single cluster threshold of $E_{cluster}=3\, \mev$, and the following weighting of the position of the crystal with the deposited energy. Different algorithms of position reconstruction can be applied, which take into account also the lateral shower shape. The good space resolution is crucial for the splitting of clusters created by overlapping photons, for example from $\pi^{0}$ with a small opening angle. For such a splitting the local maxima produced by two photons should be separated by at least two crystal widths.

%% file: crystal_apd.tex
\chapter{Scintillator and APD test measurements}\label{sec:measurements}

\section{Introductory remarks}
Interest of our equipe in development of the electromagnetic calorimeter was motivated by the proposed experiment related to $\gamma$-ray detection in the $h_c \rightarrow \eta_c$ transition in charmonium \cite{bib:PANDA_PB:2009} and our former experience of work with cooled APDs \cite{bib:Mykulyak:2004}.

\Refsec{sec:panda:EMC} outlined the present status of the EMC correlated in time with writing of this Thesis, whereas when the work of its author just started (2004) many of the details considered now as self-evident, had to be still developed. We had realized that low-energy response of scintillators will be of interest for the future PANDA-EMC. The tools available at SINS for such a project were: a C-30 compact cyclotron and a Van-de-Graaff accelerator. The C-30 cyclotron accelerates protons up to about 26 \mev. Having a source of quasi-monoenergetic particles close to the desired energy, it seemed natural to start the project using the latter tool. It was obvious, however, that proximity of the target room to the cyclotron vault might adversely affect measurements, due to the cyclotron RF noise pick-up. It was decided, therefore, to readout a scintillator from both exit surfaces in coincidence, in expectation that in this way the potential source of pick-up noise will be suppressed. Moreover, this mode of operation makes it possible to study the time response of a scintillator and readout chain.

The calorimeters existing at that stage suggested PWO \cite{bib:alice_phos_tdr:1999kd}, \cite{bib:cms_ecal_tdr:1997kj} and BGO \cite{bib:Sumner:1988cu} as the scintillators for possible application in the PANDA EMC, while LYSO was considered as a candidate for the EMC in the future detector at the ILC \cite{bib:Chen:2005daa}. An APD S8664-55 of Hamamatsu used for readout in the CMS and ALICE detectors at CERN had by far too small sensitive area of $5\times 5$ mm to satisfy the low detection threshold required for PANDA. We have, therefore, started our measurements with APDs of Advanced Photonix Inc. (USA) which had sensitive area of $\approx 2 \, cm^2$; R\&D work on an APD with the area of 1 $cm^2$ started almost simultaneously in a collaborative effort of PANDA with Hamamatsu Photonics Inc. We have tested several combinations scintillator-APD in the measurements with 25 \mev protons reported below in \Refsec{sec:measurements:protons}.

\section{Measurements with $\approx$ 25 \mev protons at the C-30 cyclotron}\label{sec:measurements:protons}
\subsection{Experimental setup}
\subsubsection{The C-30 cyclotron}
The proton beam used in the measurements was produced with the C-30 cyclotron at SINS, Swierk. The C-30 cyclotron is a compact, fixed magnetic field, fixed frequency isochronous machine. Proton beams are obtained by accelerating $H^{-}$ ions from the ion source followed by electron stripping in an aluminium foil, which leads to beam extraction. The 52 MHz radiofrequency electric field is used to accelerate protons. The extraction geometry and the magnetic field of 1.7 Tesla of the cyclotron magnet give the possibility to obtain a range of proton velocities extending from 0.2c up to 0.24c. This corresponds to proton beam energies from 19 MeV to 28 MeV, which is determined by a non intrusive time-of-flight system. An average external beam intensity up to 10 $\mu A$ can be achieved.

\subsubsection{Details of the experimental setup}
\begin{figure}
\begin{center}
\includegraphics[width=1.0\swidth]{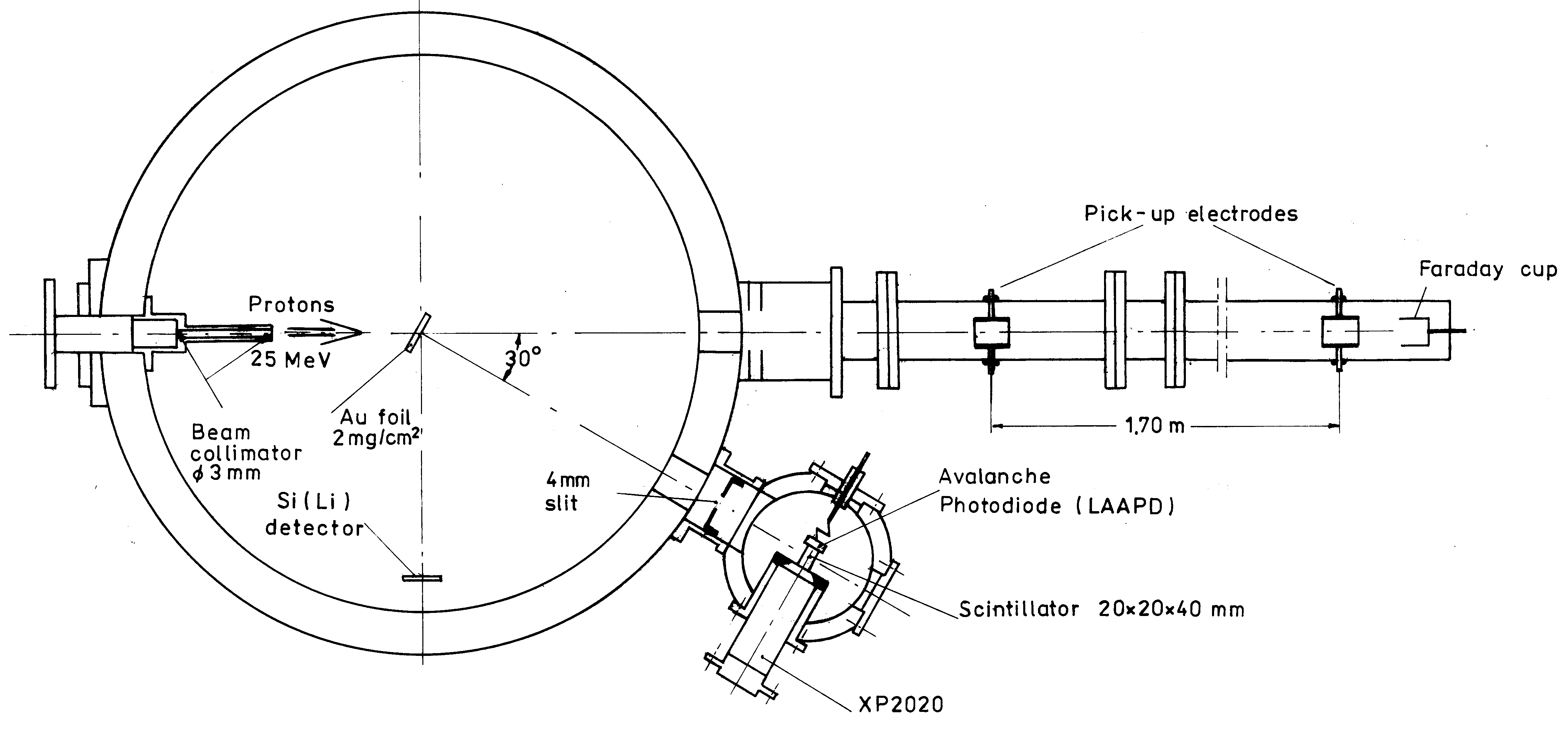}
\caption{Horizontal section of the experimental setup for measurements of energy resolution of scintillators using protons accelerated with the C-30 cyclotron.}
\label{fig:phys:c30setup1}
\end{center}
\end{figure}

\begin{figure}
\begin{center}
\includegraphics[width=0.8\swidth]{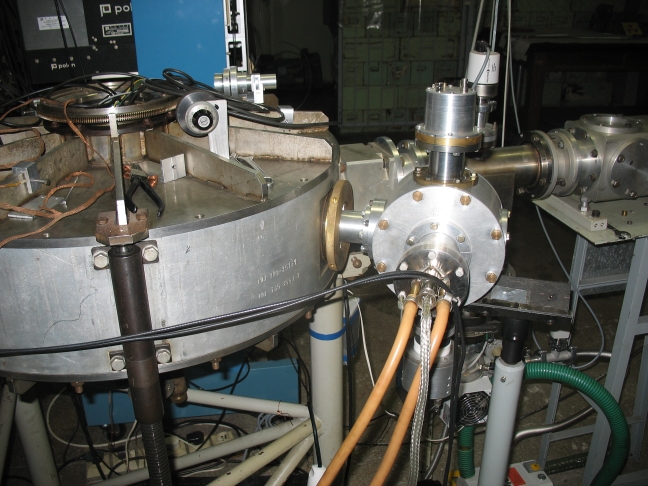}
\caption{The experimental setup seen from the side containing the cooling elements (compare with \Reffig{fig:phys:c30setup2}). Seen are the two plastic tubes providing the cooling water inlet and outlet from the U-shaped copper tube (9); whereas in between them are seen the cables of power supply to the Peltier elements in a flexible metal screen.}
\label{fig:phys:c30photo}
\end{center}
\end{figure}

The experimental setup consists of a $\phi 70$ cm diameter vacuum chamber with a 2 $mg\cdot cm^{-2}$ Au target on a thin C backing in its center and a small chamber of $\phi 15$ cm inner diameter fixed to the $30^{\circ}$ exit port of the former. The small chamber houses the tested scintillator in the form of parallelepiped $20 \times 20 \times 40 \, mm^3$. The scintillator is readout from both $20\times 20\, mm^2$ end surfaces with a PMT XP2020 and an APD. The cathode pin of an APD is soldered with a thin covar wire to minimize heat leakage to the SHV male connector installed in the body of the small chamber. The connector provides signal path to and HV from a preamp. To enhance light transport from its origin in the scintillator volume to the APD sensitive area the scintillators were coated with a few layers of a teflon tape. The horizontal section of the setup is presented in \Reffig{fig:phys:c30setup1}. In this figure the small chamber is shown rotated by $90^{\circ}$, i.e. the PMT axis which is actually directed towards the floor (see \Reffig{fig:phys:c30photo}) is drawn rotated into the plane of \Reffig{fig:phys:c30setup1} to present the entrance slit and the scintillator location.

\begin{figure}
\begin{center}
\includegraphics[width=1.0\swidth]{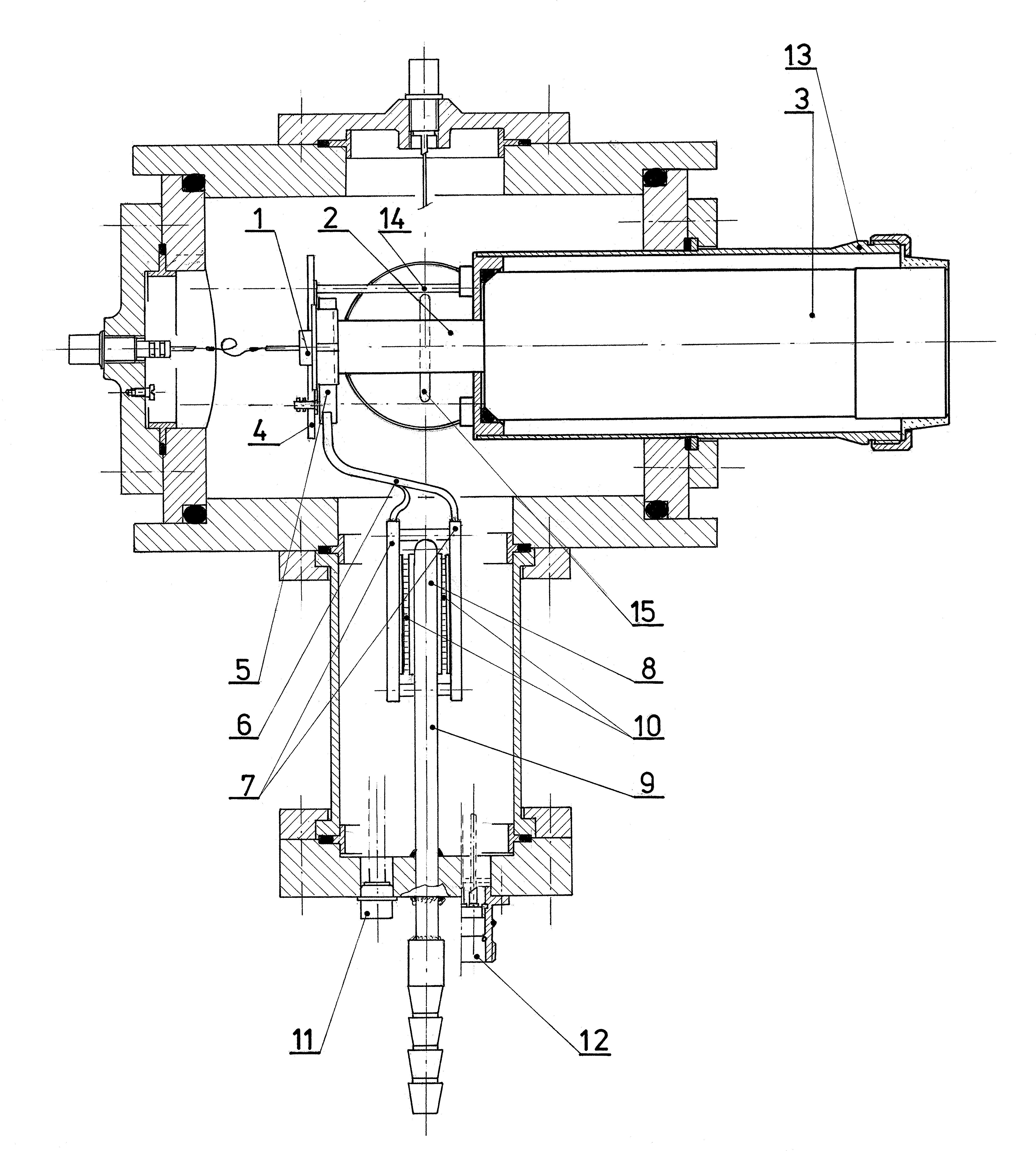}
\caption{Vertical section of the small chamber of the experimental setup, which houses the scintillator. The constituent elements are: (1)
APD, (2) scintillator, (3) PMT XP2020, (4)
Teflon plate, (5) LAAPD holder, (6) Flexible copper joint (cold-finger), (7)
Side copper plates, (8) Central plate, (9) Water-cooled U-shaped copper tube,
(10) Peltier elements DT12-4, (11) Vacuum feedthrough connecting to Pt-100
thermometers, (12) Vacuum feedthrough for power supply to DT12-4s, (13)
Soft-iron housing of the PMT, (14) $\gamma$ or $\beta$-ray source,
(15) Pumping port to the chamber with a defining slit.}
\label{fig:phys:c30setup2}
\end{center}
\end{figure}

The APD is cooled using two thermoelectric modules (Peltier elements DT12-4\footnote{Marlow Industries,
Inc., 10451 Vista Park Road, Dallas, TX 75238-1645, USA}) (see \Reffig{fig:phys:c30setup2} for details). The Peltier elements are fixed between the
two side copper plates and the central plate, which is silver-soldered in between the arms of an U-shaped, water-cooled copper tube, presenting a heat-sink for the "hot" sides of the Peltier elements. The side plates are in
thermal contact with the Peltiers' "cold" sides, which provide cooling of the LAAPD-holder via flexible copper joints (cold-fingers). In order to improve the thermal contact in vacuum, the contacting surfaces are polished and
covered with a thin layer of thermoconducting (but electroisolating) white paste P12\footnote{Wacker Chemie GmbH, Germany}. The temperature control is provided by Pt-100 resistive thermometers installed in two points: the first on at the
center of the right contact plate and the second one on the LAAPD holder. The small chamber is pumped down to $1.-2.\cdot 10^{-5}\, mbar$ together with the large one; vacuum effectively isolates all cooled elements from ambient and prevents from moisture deposition.

\subsubsection{Tested APDs and scintillators}
The main purpose of the study was to measure energy and time resolution for PWO scintillators with an APD readout. However, at the moment when these measurements have been performed BGO was considered as an alternative option for the PANDA electromagnetic calorimeter. It has an advantage of higher light yield, but has much slower fluorescence decay time in comparison with PWO. The plastic BC-408 is a very fast scintillator with large light output well matched to the maximum of spectral sensitivity of many APDs. It has been used for tuning fast timing in our experiment. In addition several measurements with a LYSO scintillator have been performed. LYSO is characterised by a larger radiation length than PWO, however it has an advantage of much higher light yield having at the same time a comparable speed. Its high price does not permit its application in calorimetry at present, however LYSO is considered as a perspective material for future applications, e.g. in the EMC at ILC \cite{bib:Chen:2005daa}. The main parameters of the mentioned scintillators, such as density, radiation length, light output and decay time are summarized in \Reftbl{tab:meas:scintillators}. For PWO both fast and slow components of the fluorescence decay time are listed. The two generations of PWO are introduced in the table; PWO-II is characterized by about twice higher light yield, it is discussed in more detail in \Refsec{sec:panda:PWO}. Only PWO-II has been used in our measurements. Light yield of a scintillator is the parameter which may vary from crystal to crystal, therefore the values listed in \Reftbl{tab:meas:scintillators} are only indicative.

\begin{table}
\begin{center}
\begin{tabular}{|l|c|c|c|c|}
\hline
Material& Density & Radiation length & LY & Decay time\\
& ($g\cdot cm^{-3}$)& ($cm$) & \% (LY NaI) & ($ns$) \\
\hline
PWO-I & 8.3 & 0.89 & 0.3 & 40/10\\
PWO-II & 8.3 & 0.89 & 0.6 & 40/10\\
BC-408 & 1.03& 43 & 24 & 2.1\\
BGO & 7.1& 1.12 & 9 & 300\\
LYSO & 7.1 & 1.2 & 75& 40\\
\hline
\end{tabular}
\caption{Main parameters of the scintillation materials used in the measurements with 25 \mev protons.}
\label{tab:meas:scintillators}
\end{center}
\end{table}

\begin{figure}
\begin{center}
\includegraphics[width=0.8\swidth]{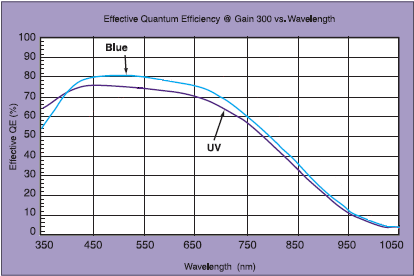}
\caption{Dependence of the APD quantum efficiency on the wavelength for the two types of Advanced Photonix APDs 630-70-74-510 ("blue") and 630-70-73-510 ("UV").}
\label{fig:phys:apd_efficiency}
\end{center}
\end{figure}

Three different types of APDs were used during our measurements; two produced by Advanced Photonix Inc. (API): 630-70-74-510 and 630-70-73-510 and one produced by Hamamatsu Photonics: S8664-1010 . The Advanced Photonix APDs differ from each other by the quantum efficiency dependence on the wave-length. The 630-70-74-510 has maximum sensitivity in the blue part of light spectrum while 630-70-73-510 in ultraviolet, see \Reffig{fig:phys:apd_efficiency}. The APDs by Advanced Photonix are fabricated with three diameters of the sensitive area: 5, 10 and 16 mm. For our measurements the largest diameter type 630-70-74 has been selected which has an area of $\approx$2 cm$^2$. This area is one half of the cross section area of the scintillator, thus about 50\% of the light emerging from its exit face can be intercepted and processed. The sensitive area of Hamamatsu S8664-1010 is a square with 10 mm side. The studied products of these two companies also differ strongly in the required working bias voltage. A typical working bias for the Advanced Photonix APDs is 1760 - 1870 V, with a gain of $\approx$50 at the lower and $\approx$300 at the upper limit of the interval, whereas Hamamatsu APDs work at a significantly lower bias of $\approx$ 400 V.

\subsubsection{Readout electronics}
\begin{figure}
\begin{center}
\includegraphics[width=1.0\swidth]{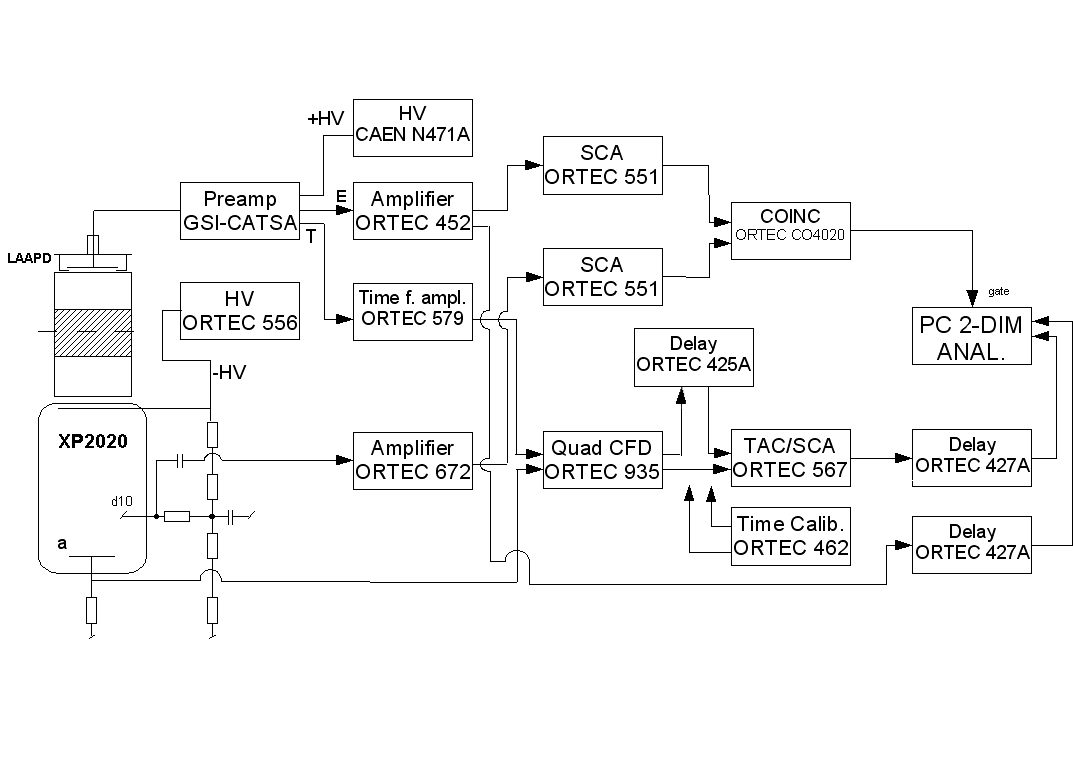}
\caption{Block-scheme of electronics used in the measurements with 25 \mev protons. The hatched area indicates the region of $\sim 4 \, mm$ width illuminated with protons (not to scale with the 40 mm length of the scintillator).}
\label{fig:phys:c30electronics}
\end{center}
\end{figure}

\begin{figure}
\begin{center}
\includegraphics[width=0.8\swidth]{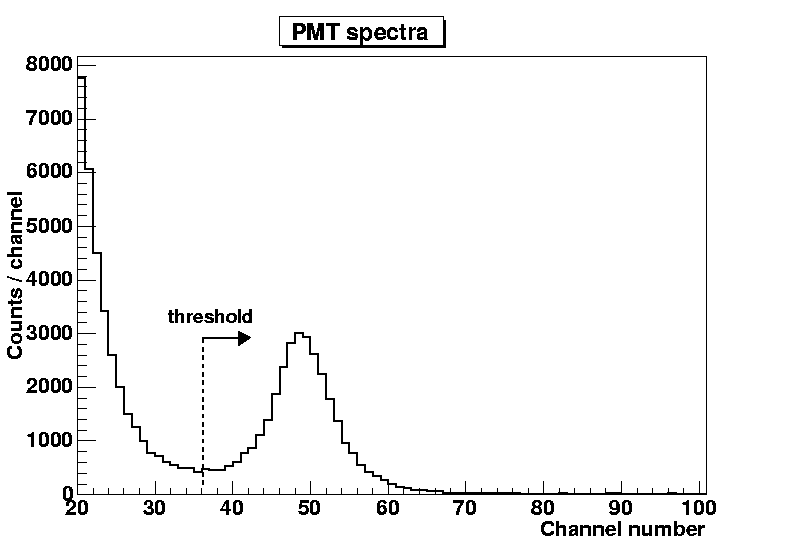}
\caption[PMT threshold for coincidence measurements.]
{PMT threshold for coincidence measurements.}
\label{fig:phys:pmt_threshold}
\end{center}
\end{figure}

The readout electronics was a fast-slow coincidence system assembled using mostly the standard ORTEC NIM modules. The complete block-scheme is presented in \Reffig{fig:phys:c30electronics}. Two signals from the PMT have been employed, the signal from the $10^{th}$ dynode was amplified with an ORTEC 672 amplifier and after activating the Single Channel Analyser (SCA) was directed to the coincidence module together with a signal from the APD to produce a gate signal for the two-dimensional analyser. The second signal from PMT was collected from its anode and was used, after shaping with the aid of the constant-fraction discriminator ORTEC 935, as a start signal for time distribution measurements with the Time-to-Amplitude Convertor (TAC) ORTEC 567. The stop signal for the TAC was derived from the fast timing branch (denoted "T") of the GSI CATSA preamplifier. The spectroscopic output from the preamplifier (denoted "E" in \Reffig{fig:phys:c30electronics}) was amplified with the standard ORTEC 452 amplifier, whose one output activated the SCA. If a coincidence occured with the "slow" PMT signal, the gate of the two-dimensional analyser is opened to accept the APD amplitude and time difference signals. The timing signal from the preamplifier after an additional amplification with the Timing-Filter-Amplifier was used as a stop signal for the TAC. The signal from TAC was fed into the second input of the analyser. To optimize time performance the timing signals from both APD and PMT were fed into the Constant Fraction Discriminators (CFD) with optimised settings of zero crossing and threshold level. As it is clear from the foregoing presentation, the registered events had two coordinates: the amplitude of an APD signal and the time difference between PMT and APD signal arrivals. This permitted to judge, neglecting the time dispersion of XP2020, on dependence of the APD signal time dispersion on its amplitude. Small signals, which might have been caused by background effects, were suppressed by setting appropriate thresholds on the signal amplitudes. A typical PMT spectrum from the 10th dynode of XP2020, with the level of threshold marked with a vertical broken line, is shown in \Reffig{fig:phys:pmt_threshold}. The delay line amplifiers were used wherever it was necessary to synchronize the signals.

\subsection{Results}
The measurements were started with the plastic scintillator BC-408. Due to its high light yield it allowed to estimate the contribution of an APD to energy and time resolution of the combination scintillator plus APD from an APD only, i.e. in this case the number of photons should not be the limiting factor. Before the measurements with a proton beam the settings of the CFDs were adjusted using radioactive sources for optimum time resolution. The measurement with a plastic scintillator BC-408 has been performed with only one type of an APD, Advanced Photonix 630-70-74-510, which has maximum efficiency in the blue part of light spectrum.

The time calibration of the time-to-amplitude converter has been performed with an ORTEC time calibrator, which produces start and stop signals separated by an integer multiple of a fixed time period, which has been selected 10 ns for our purposes.

\begin{figure}
\begin{center}
\includegraphics[width=0.8\swidth]{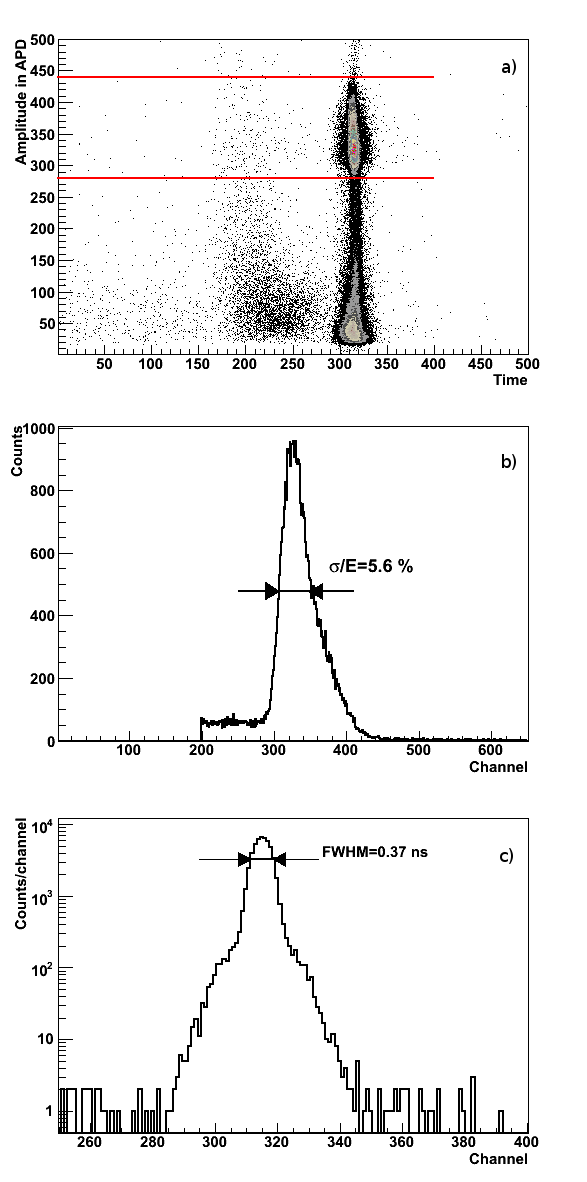}
\caption{The two-dimensional spectrum from a BC-408 scintillator with an APD readout obtained with the 26 MeV proton beam (a) with its projections on the APD amplitude axis with $\sigma/E$ determined from the peak FWHM (b) and on the time axis limited to APD amplitudes 280-440 (c).}
\label{fig:phys:bc408_all}
\end{center}
\end{figure}

With the 26 MeV proton beam the coincidence 2-dimensional spectra have been collected. A sample 2-dimensional spectrum is shown in \Reffig{fig:phys:bc408_all} (a) and in \Reffig{fig:phys:bc408_all} (b) its projection on the APD amplitude axis. Because of its asymmetric shape the energy dispersion $\sigma$ is determined as the full width at half maximum (FWHM) divided by 2.35.

The two-dimensional spectrum is presented in \Reffig{fig:phys:bc408_all} (a). The time resolution has been estimated by projecting that part of the two-dimensional spectrum which corresponds to the signal from the proton beam, i.e. having APD amplitudes between the channels 280 and 440.

The projection of \Reffig{fig:phys:bc408_all} (a) on the time axis, limited to APD amplitudes 280-440, containing clean proton hits, is presented in \Reffig{fig:phys:bc408_all} (c). The time resolution, defined as FWHM is equal 0.37 ns. The two numbers extracted from the results of \Reffig{fig:phys:bc408_all} (a) are representative for a fast scintillator with large light output. One may conclude that under these conditions it is feasible to reach subnanosecond timing using an APD. These results will be further compared with those obtained under less favorable conditions.

\begin{figure}
\begin{center}
\includegraphics[width=0.8\swidth]{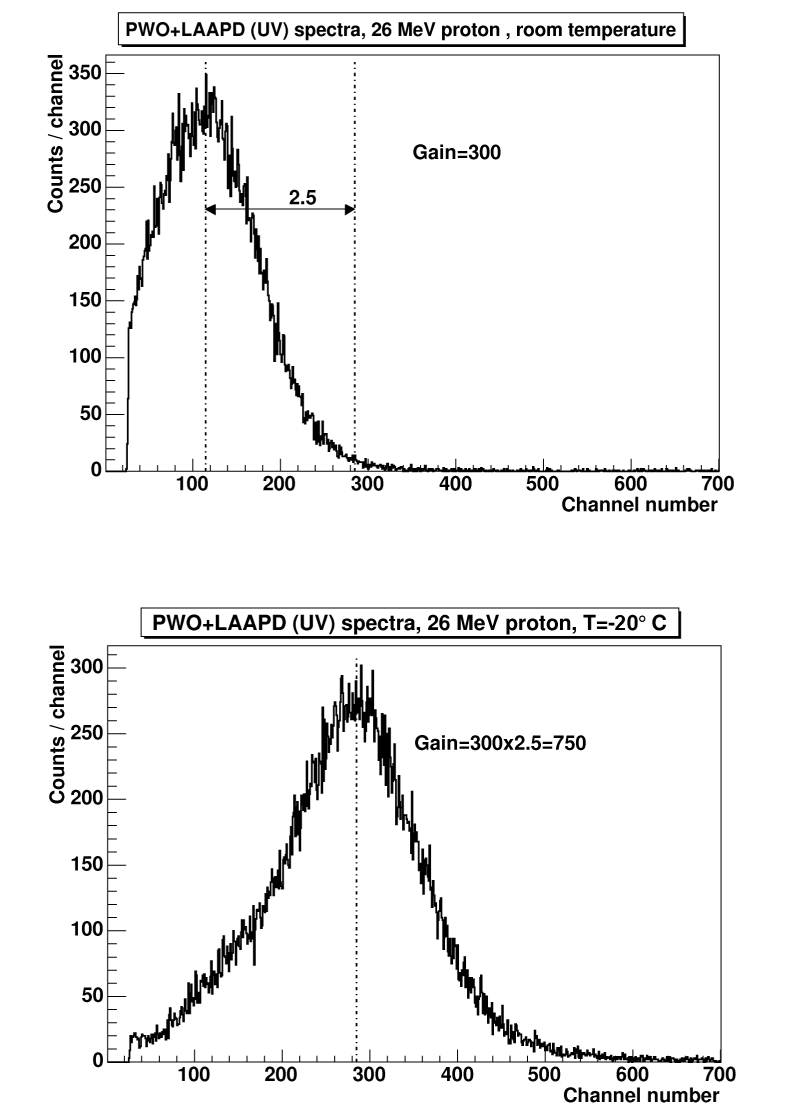}
\caption{The amplitude spectra from PWO+APD at room temperature (upper spectrum) and cooled down to $-20^{\circ}C$ (lower spectrum).}
\label{fig:phys:apd_proton_cooling}
\end{center}
\end{figure}

The measurements with PWO-II scintillator which was the main purpose of this study have been performed with all the three types of APDs. An additional aspect of the measurements was a study of the effect of APD cooling on the APD gain. In \Reffig{fig:phys:apd_proton_cooling} the two spectra from PWO plus Advanced Photonix APD are presented. The upper one was measured with an APD at room temperature and the lower one with APD at $-20^{\circ}C$. We observe the APD gain increase due to cooling by a factor of 2.5. The dark current of the APD was controlled during the measurements and its decrease from 50-60 nA at room temperature down to 3-4 nA at $-20 \degrees C$ is observed. This indicates that for amplitude and time resolution an APD contribution to electronic noise must have also decreased.

\begin{figure}
\begin{center}
\includegraphics[width=0.8\swidth]{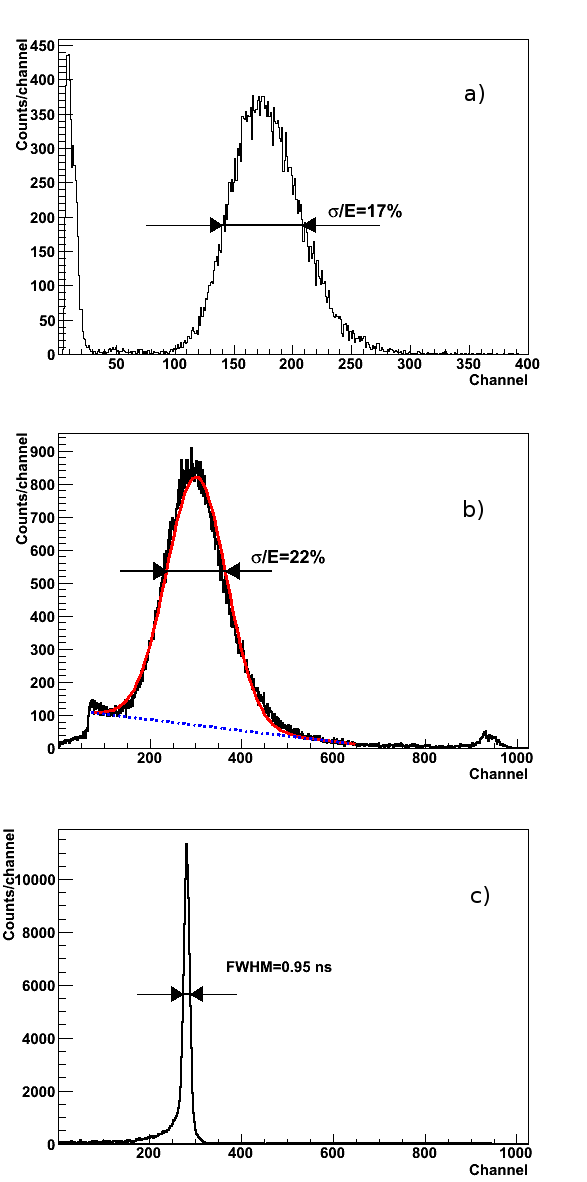}
\caption
{The response of PWO scintillator to the proton beam, a) energy spectrum taken with an API APD "blue", b) energy spectrum with an API APD "UV", c) time spectrum measured with an API APD "UV".}
\label{fig:phys:pwo_API_all}
\end{center}
\end{figure}

The APD amplitude spectra from the two types of Advanced Photonix Inc. (API) APDs (blue and UV) are presented in \Reffig{fig:phys:pwo_API_all} a) and b). These two measurements have been performed with slightly different energies of the proton beam, 26 MeV with API "blue" and 22 MeV with API "UV". The obtained amplitude dispersions are determined as FWHM/2.35 and $\sigma/E$ values are equal 17 \% and 22 \%, respectively. For the spectrum in \Reffig{fig:phys:pwo_API_all} b) the fit assuming a gaussian peak plus a linear background has been performed. The difference in resolution seems reasonable since the worse energy resolution for APD "UV" is expected due to the lower quantum efficiency of the latter over PWO emission spectrum (see \Reffig{fig:phys:apd_efficiency}). \Reffig{fig:phys:pwo_API_all} c) presents the time spectrum. The width at half maximum is equal 0.95 ns, a loss in resolution in comparison with \Reffig{fig:phys:bc408_all} c) not unexpected in view of the decreased e-h statistics.

\begin{figure}
\begin{center}
\includegraphics[width=0.8\swidth]{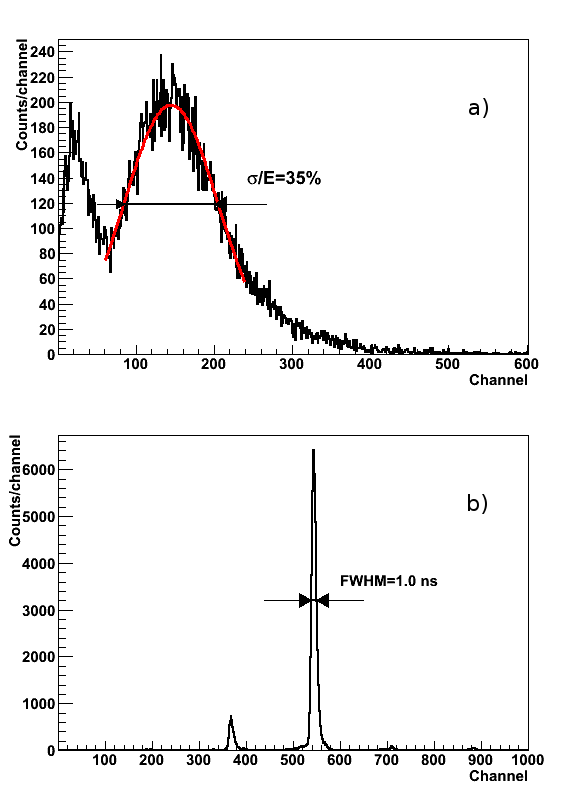}
\caption{Response of the PWO scintillator readout with the Hamamatsu S8664-1010 APD to 26 \mev protons, reflected in a) energy spectrum and b) time spectrum.}
\label{fig:phys:pwo_ham_all}
\end{center}
\end{figure}

The second series of measurement has been performed with an APD produced by Hamamatsu Photonics Inc. This APD possessed $10 \times 10$ mm sensitive area and was specially designed for the PANDA needs. However its area is twice smaller than that of API APDs which affects the measured energy resolution due to inferior light collection. In \Reffig{fig:phys:pwo_ham_all} a the amplitude spectrum with the extracted energy resolution is presented and the time spectrum with its resolution is presented in \Reffig{fig:phys:pwo_ham_all}(b).

The energy resolution worsened in comparison with \Reffig{fig:phys:pwo_API_all} to $\sigma/E = 35 \, \%$. A comment on this observation will be made in the next subsection. On the other hand the time resolution of 1.0 ns does not differ very much from the one previously obtained with the API APD.

\begin{figure}
\begin{center}
\includegraphics[width=0.8\swidth]{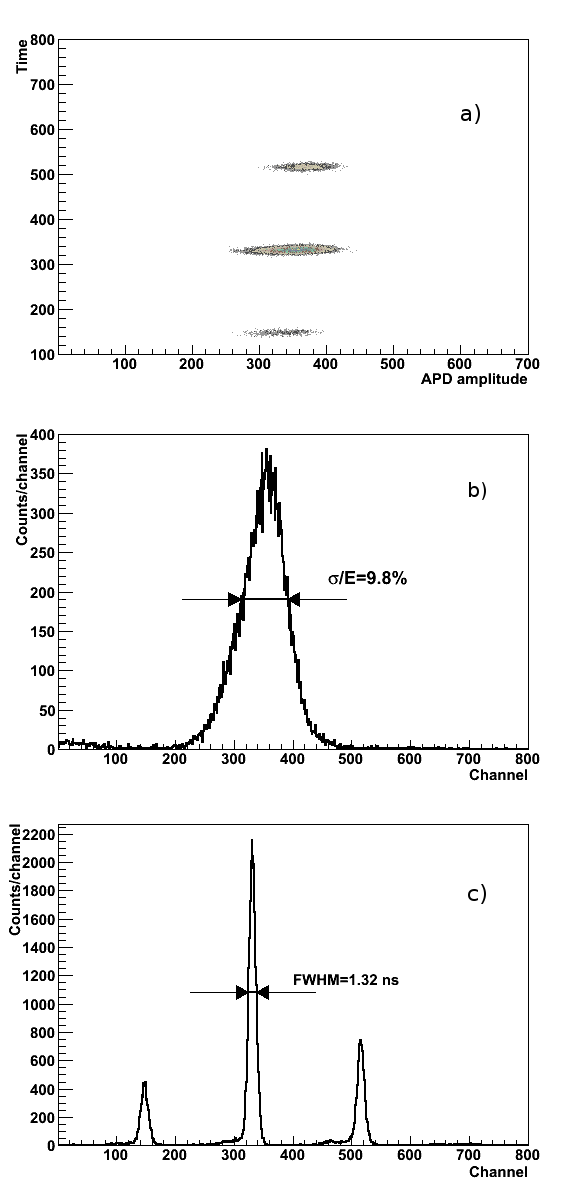}
\caption
{The response of BGO scintillator to the proton beam, a) 2-dimensional time-energy spectrum, b) energy spectrum with Hamamatsu APD, c) time spectrum.}
\label{fig:phys:bgo_apd_all}
\end{center}
\end{figure}

BGO and LYSO were the two additional tested scintillators. In both cases the Hamamatsu S8664-1010 APD has been used. \Reffig{fig:phys:bgo_apd_all} a) presents the two-dimensional time-energy spectrum obtained in the measurements with BGO. \Reffig{fig:phys:bgo_apd_all} b) and c) represent energy and time spectra correspondingly which are the projections of this 2-dimensional spectra. the events are grouped in the three areas with the extreme top and bottom area equidistant from the central one. The time distance corresponds to the period of the cyclotron high frequency. Apparently, the TAC start and stop signals are derived from different cyclotron bunches in these extreme groups of events. The estimated amplitude resolution is based only on the events in the central group. This is equal to 5.6 \% and the time resolution is 1.32 ns.

\begin{figure}
\begin{center}
\includegraphics[width=0.8\swidth]{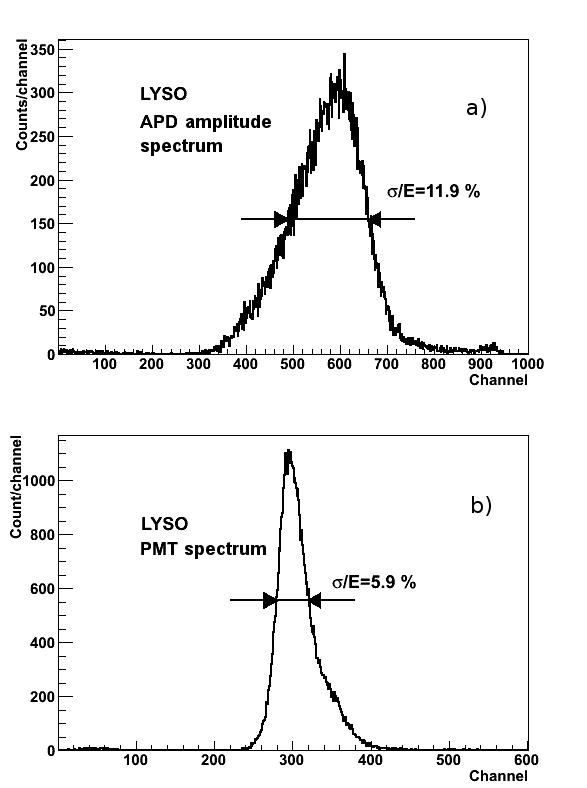}
\caption
{The response of LYSO scintillator to the proton beam, a) energy spectrum with Hamamatsu APD, b) energy spectrum with PMT.}
\label{fig:phys:lyso_all}
\end{center}
\end{figure}

\Reffig{fig:phys:lyso_all} presents the spectra from the APD (a) and from the PMT (b) for the LYSO scintillator. The energy resolution for the APD is 11.9 \% and for the PMT is 5.9\%.

The summary of the measured values for energy resolution is presented in \Reftbl{tab:meas:energy_res_proton} and for time resolution in \Reftbl{tab:meas:time_res_proton}.
\begin{table}
\begin{center}
\begin{tabular}{|c|c|c|c|}
\hline
& API("Blue") & API("UV") & Hamamatsu\\
\hline
PWO-II & $17 \pm 2\, \%$ & $22 \pm 3 \%$ & $35 \pm 3\, \%$\\
BC-408 & $5.6 \pm 0.7\, \%$& - & - \\
LYSO   & - & - & $11.9 \pm 0.9\, \%$\\
BGO    & - & - & $9.8 \pm 0.8\,\%$ \\
\hline
\end{tabular}
\caption{Summary of the measured energy resolutions for scintillators with (cooled) APD readout for the different combinations scintillator-APD.}
\label{tab:meas:energy_res_proton}
\end{center}
\end{table}

\begin{table}
\begin{center}
\begin{tabular}{|c|c|c|c|}
\hline
& API("Blue") & API("UV") & Hamamatsu\\
\hline
PWO-II & $0.95\pm0.01\, ns$ & - & $1.0\pm0.01\, ns$ \\
BC-408 & $0.37\pm0.01\, ns$ & - & - \\
LYSO   & - & - & - \\
BGO    & - & - & $1.32\pm0.01\, ns$\\
\hline
\end{tabular}
\caption{Summary of the measured time resolutions for scintillators with (cooled) APD readout for the different combinations scintillators-APD.}
\label{tab:meas:time_res_proton}
\end{center}
\end{table}

\subsection{Interpretation of the results. Simulation of the light collection efficiency}
The measured value of the time resolution for PWO scintillator with an APD readout 0.95-1.0 ns appears very promising for the purpose of the PANDA experiment where the time resolution below 1 ns is required. The quantity measured in our experiment is the dispersion of signal arrivals in time between APD and PMT. The PMT XP2020 has negligible signal dispersion in comparison with an APD. The time dispersion, $\sigma_{\rm time}$, of an APD is proportional to the ratio of two factors:
\begin{equation}
\sigma_{\rm time}\sim\sigma_{\rm noise}/(dV/dt),
\end{equation}
where $\sigma_{\rm noise}$ is the rms voltage noise and $dV/dt$ is the signal slope at threshold of the timing discriminator. In view of the rather low gain of $\approx$300 of the applied APD in contrast to
$\approx$$10^6$ typical for PMTs the contribution of APD to the measured value of time resolution is dominating over the PMT.

The measurements of the time resolution of full-size $200 \times 20 \times 20 \, mm^3$ PWO-II scintillator readout by Hamamatsu APD and with 10 bit 80 MHz sampling ADC used as a digitizer performed recently at the University of Groningen \cite{bib:emctdr} gives the number of the order of 1 ns, which is comparable with our results.

On the other hand the energy resolution for the PWO scintillator with Hamamatsu APD readout appears worse than we expected. The quenching effect was verified as a possible source of this discrepancy. Measurements of the energy resolution with low energy proton beam, which we have performed, can be compared to energy resolution for $\gamma$-rays on the basis of Birks' theory \cite{bib:Birks:1951}. This theory empirically describes the effect of quenching of light yield for highly ionising particles. The light yield $dL$ over the element $dx$ of path-length is related to the energy loss $dE$ by the following equation:
\begin{equation}
\label{eqn:meas:birks}
dL/dx=S\frac{dE/dx}{1+kB\frac{dE}{dx}},
\end{equation}
where $kB$ is a Birks' constant and $S$ is a scintillation efficiency. The total light yield produced by a particle is obtained by integration of \Refeq{eqn:meas:birks} from its initial energy down to zero. The Birks' formula \Refeq{eqn:meas:birks} is based on the assumption that the incident particle excites the number of molecular structures in scintillator, proportional to the stopping power, damaging a fraction $B dE/dx$ of them. The latter are assumed to act as light-quenching centers with an efficiency $k$. The quenched fraction goes to zero in the limit of low ionization density, therefore the effect should be largest for incident ions with low energy. Although the high energy ions eventually slow down, the average $dE/dx$ along the particle track is less for high energy incident particles. The factor $kB$ is material dependent and should be extracted from measurements for each scintillator and particle type. For the PWO scintillator Ref. \cite{bib:Hoek:2002ss} presents the results of measurements with 85 MeV protons. A comparison with response to cosmic muons, which suffer a negligible quenching, yields the quenching factor $Q_{F}=0.92$. For the 25 MeV proton the quenching factor has been rescaled with the Birks' formula via its numerical integration over the energy range of interest using the tabulated values of $dE/dx$. This estimate gives $Q_{F}= 0.9$ for 25 \mev protons in PWO. This number tells that the quenching effect alone cannot be considered as the main source of poor energy resolution in our experiments.

It has been argued in \Refsec{sec:panda:EMC} that the two main contributions to energy resolution are electronic noise and photon statistics. One of the factors which influence photon statistics is light collection efficiency. We have performed light transport simulations for our setup with the Litrani program \cite{bib:Gentit:2001ky} to determine light collection efficiency.

The setup implemented in simulations, consisted of a crystal, a photomultiplier and an APD. The $20\times 20 \times 40 \, mm^3$ crystal was oriented with its long axis along the OZ axis. The PMT was placed under the crystal, as in \Reffig{fig:phys:c30photo}, and an optical contact with its entrance window was via an optical glue with a thickness of 0.03 mm. Its refractive index was not known, therefore for the simulations $n=1.704$ was assumed as possessed by Meltmount, the optical coupling material implemented in Litrani. It should be stressed however, that changing the refractive index in the range 1.6-1.7 does not change the result of simulations significantly. The bottom side of the crystal was completely covered by the PMT entrance window which is $2^{\verb+"+}$ in diameter. The APD with square sensitive domain of $1 \times 1 \, cm^2$ was located above the crystal not having an optical contact with it. Because of the limited cooling power of the used Peltier elements, such a contact would immediately deteriorate the minimum temperature reached by our APD. The size of the gap was of the paper card thickness and was assumed 0.3 mm in simulations. The teflon coating of the crystal was treated as a perfect diffusor in simulations.

An area illuminated with protons extended over the band of 4 mm width along the length of the scintillator at its midplane and over its entire width, being defined by the collimator at the entrance to the small chamber. The depth of proton penetration was 1.6 mm, determined by the stopping power of $PbWO_4$. The fluorescence light was generated with a wavelength of 420 nm, which is one of the two main components of $PbWO_4$ scintillation. The absorption length was assumed the same as for the CMS PWO crystals, which is implemented within the program and is equal 18.5 cm at the 420 nm wavelength. The quantum efficiencies of PMT and APD were taken as 20\percent and 70\percent, respectively.

The main output of the simulations with Litrani is the calculated ratio of the number of produced photoelectrons in PMT and e-h pairs in APD. This ratio is determined by the light collection efficiency for given geometry and quantum efficiencies of the given photodetector. It is equal 6:1 for PMT vs. APD, despite almost equal product of quantum efficiency and covered area. This is the effect of a lack of optical contact between the scintillator and the APD. To estimate the statistical term of the energy resolution the number of registered photoelectrons in APD should be calculated. The standard characteristics of the light yield of a scintillator is the number of photoelectrons per \mev measured with the PMT with bi-alkali photocathode with quantum efficiency of 20\percent. For PWO-II crystals a typical value is 20 photoelectrons per \mev. This number of photoelectrons can be assumed to be equal to what is measured with PMT in our setup due to similar conditions of measurements. The number of photoelectrons registered in APD can be estimated as a number of photoelectrons registered in PMT multiplied by the ratio obtained from the Litrani simulation. One thus obtains 3.3 e-h pairs per \mev energy deposited in the scintillator and emitted as light.

The statistical term in energy resolution is:
\begin{equation}
\frac{\sigma_{E}}{E}=\sqrt{\frac{F}{N_{e-h}}}\cdot \frac{1}{\sqrt{E(MeV)}}.
\end{equation}

With the excess noise factor $F=1.23$ and the energy 22.5 \mev emitted as light, which corresponds to 25 \mev energy of the proton with 0.9 quenching factor, one obtains that the statistical term in energy resolution is equal $\sigma/E=13.2\percent$. This contribution is rather high, however it does not explain completely the measured resolution of 35\percent. One of the sources of uncertainty is that the PWO-II light yield can differ from crystal to crystal and that our sample has inferior properties than assumed as the mean. The rest of the difference should be attributed to electronic noise, which could not be estimated with sufficient certainty. One should add on this occasion that we had to modify our setup shown in \Reffig{fig:phys:c30setup1} (and in \Reffig{fig:phys:c30setup2}) when going from APDs with 2 $cm^2$ sensitive area (Advanced Photonix Inc.) to APDs with 1 $cm^2$ (Hamamatsu Photonics). The modifications, consisted in replacement of the large-sized preamplifier (GSI-CATSA), which was kept outside the small chamber and contacted with the APD via a short ($\sim$ 10 cm) cable, with a preamp of very small size which was installed within the extension shown on top of the small chamber (see \Reffig{fig:phys:c30photo}). The length of the contact wire was 2 cm in this case. Moreover, this way the whole system was enclosed in a Faraday cage formed by the massive walls of the chamber and the extension. This reduced the noise pick-up significantly and made it possible to perform measurements with an APD of reduced area and a scintillator with low light yield (PWO).

One should stress, however, that our team acquired a lot of practical experience in scintillator readout with APDs in the course of experiments at the cyclotron C-30. This experience was essential to perform the work described in \Refsec{sec:measurements:VDG}. The results of \Refsec{sec:measurements:protons} are partly summarized in a conference presentation \cite{bib:Melnychuk2006}.

\subsection{Outlook}
The conditions under which the measurements of \Refsec{sec:measurements:protons} with 25 \mev protons have been performed were rather remote from those that are expected to be met with the PANDA EMC. This conclusion emerged from the measurements with a $3\times 3$ matrix of full-length PWO-II scintillators performed at the MAMI microtron in the J. Gutenberg University in Mainz \cite{bib:Makonyi2006}. Our team was invited to participate in this experiment by the equipe led by Prof. Rainer Novotny from the University of Giessen, the spokesperson of PANDA EMC. Quasi-monochromatic gamma-rays were produced by bremsstrahlung of 855 \mev electrons in a thin $C$ target with a simultaneous momentum analysis of the electron in a magnetic analyser. The latter has a position sensitive detector along its focal plane. By selecting a narrow band of electron momenta one obtains a collimated beam of photons in the direction of the primary electron with well defined energy. The matrix was equipped with PMT readout and was kept at -25\degrees{}C in a thermally isolated box whose interior was cooled with dry and cool nitrogen gas from a powerful cooler. The 9 scintillators were inserted one-by-one on the axis of the collimator to perform energy calibration of the whole matrix. This way it was feasible to determine eventwise the distribution of energies of the electromagnetic shower in the central crystal and its nearest neighbors when the central crystal was irradiated. By adding all these contributions it was possible to restore the line shape as a function of the photon energy. The line shape appeared close to Gaussian with the following dependence of the dispersion, $\sigma(E)$, on photon energy:
\begin{equation}
\label{eq:sigma_e:pmt}
\frac{\sigma (E)}{E}=\frac{0.95\,\%}{\sqrt{E}}+0.91\,\% ,
\end{equation}
where energies $E$ are expressed in \gev, so that $\sigma(E)/E$ is dimensionless.

This experiment was a milestone of the PANDA-EMC development and its results constituted a reference point for the development of our own instrumentation. A special small-size charge-sensitive preamplifier has been developed by the group of electronic engineers from the University of Basel \cite{bib:emctdr} for R\&D work on the PANDA EMC. The above $3\times 3$ matrix was equipped with Hamamatsu's S8664-1010 APDs and the Basel preamplifiers. The experiment was repeated with sixteen tagged photon energies in the range 40.9- 674.5 \mev, which were used to determine the energy dependence of the response function \cite{bib:Novotny:2008zz}. The matrix was kept at $0\degrees{C}$. The following dependence of $\sigma (E)/E$ on $E$ has been obtained in the quoted experiment:
\begin{equation}
\label{eq:sigma_e:apd}
\frac{\sigma(E)}{E}=\frac{1.86\,\%}{\sqrt{E}}+0.65\,\%.
\end{equation}
One notes some increase in the statistical term in comparison with the PMT readout \Refeq{eq:sigma_e:pmt}, however overall the resolution is excellent and a forecast for energy resolution close to 2\percent at 1 \gev obtained.

It is clear that the experiments with bremsstrahlung $\gamma$-rays performed at MAMI leave a wide gap of about 0-50 \mev in which measurements are necessary in order to verify whether the dependence of $\sigma(E)/E$ on the photon energy suggested by \Refeq{eq:sigma_e:pmt} or \Refeq{eq:sigma_e:apd} is also obeyed at these lower energies. It is necessary, however, that similar conditions are preserved:
\begin{itemize}
\item use of monochromatic $\gamma$-rays,
\item use of full-size $20\times 20 \times 200 \, mm$ PWO-II scintillators,
\item use of the same reflector in tight contact with the scintillator, achieved with the aid of a thermoshrinking plastic tubing,
\item use of an Hamamatsu S8664-1010 APD in contact with the scintillator via an optical oil of the same or a similar kind,
\item use of the preamplifier of the same type,
\item application of cooling of the scintillator and an APD, preferably down to $-25\degrees{C}$.
\end{itemize}
In the following section \Refsec{sec:measurements:VDG} an apparatus will be described which incorporates the above features.

\section{Measurements with low energy $\gamma$-rays at the Van-de-Graaff accelerator}\label{sec:measurements:VDG}
The measurements of PWO energy resolution with APD readout performed with the 25 \mev proton beam, which were described above, had several drawbacks. The limited cooling capacity of the Peltier elements did not allow to cool scintillators but only APDs. In case of PWO its cooling from room temperature down to $-25 ^{\circ}C$ should give fourfold increase in light output, that is the reason why a measurement at this temperature presents a lot of interest. The second drawback of the performed measurements with the proton beam was the distance between the APD and scintillator which had to be left to keep APD thermally isolated. Unfortunately, this worsens significantly the light collection which leads to high statistical term in energy resolution, with the measured values worse than expected at the given energy.

The experimental setup has been constructed for measurements of energy resolution of the cooled PWO with APD readout to detect low energy $\gamma$-rays produced in radiative proton capture reactions. The setup was designed to incorporate single crystals of the size $20\times20\times 200$ mm, which is close to the final shape of the crystals foreseen in the PANDA experiment \cite{bib:emctdr}.

\subsection{Production of photons in radiative capture reactions}\label{sec:measurements:reactions}
To determine the possible source of low energy $\gamma$-rays the following limitations as to the choice of the reaction had to be taken into account:
\begin{enumerate}
\item
the beam energy accessible with the Van-de-Graaff generator is limited to 2 MeV for protons and deuterons,
\item
the possibility to produce a target with reasonable thickness,
\item
internal spread of $\gamma$-ray spectra,
\item
the reaction cross-section, which defines the expected event rate.
\end{enumerate}

After considering the above mentioned limitations the following reaction was selected as the possible source of $\gamma$-rays:
\begin{equation}
^{11}B(p,\gamma)^{12}C.
\end{equation}

This reaction was studied in detail in the 50-ties \cite{bib:Huus:1953} and the 60-ties \cite{bib:Allas:1964}. The study of excitation curve for this reaction has shown the existence of three resonances at 0.163 MeV, 0.675 MeV and 1.388 MeV; the latter rather broad with $\Gamma \approx$1.15 \mev. The $\gamma$-rays are emitted in a direct transition to the ground state of $^{12}C$ as well as in a cascade transition through the 4.44 MeV level of $^{12}C$. For the ground state transition the angular distribution is peaked at $90^{\circ}$ whereas for the excited state transition it is nearly isotropic for the energy range $4 \,\mev\leq E_p \leq 14\, \mev$, where the data are available \cite{bib:Allas:1964}. This favours measurements at $90^{\circ}$ to the proton beam momentum. \Reftbl{tab:meas:egamma_11B} presents $\gamma$-ray energies for all the three resonances.

\begin{table}
\begin{center}
\vspace{-10pt}
\begin{tabular}{|c|c|c|c|}
\hline
$E_p (MeV)$& \multicolumn{3}{|c|}{$E_{\gamma}$ (MeV)}\\
\hline
0.163 & 4.44 & 11.6 & 16.1\\
0.675 & 4.44 & 12.2 & 16.6\\
1.388 & 4.44 & 12.8 & 17.2\\
\hline
\end{tabular}
\caption{Gamma ray energies at the three resonances in $^{11}B(p,\gamma)^{12}C$ reaction at 90\degrees to the beam direction.}
\label{tab:meas:egamma_11B}
\vspace{-15pt}
\end{center}
\end{table}

For the proton energy $E_p=$ 0.675 MeV the intensity of the transition to the ground state is $\approx$ 20 times lower than that to the 4.44 \mev state which yields the correspondingly smaller intensity of the 16.6 MeV line in comparison with the 12.2 MeV one. For the 1.388 MeV resonance the intensities of both lines are comparable at 90\degrees to the beam direction.

It has been found in the performed measurements that our B targets are contaminated with $^{19}F$, which produces $\gamma$-rays with the energy 6.13 MeV from the reaction $^{19}F(p, \alpha \gamma)^{16}O$. This reaction has a strong resonance at $E_p$=1.374 MeV with cross-section much higher than the reactions with $^{11}B$ \cite{bib:Hellborg:1972}. Subsequently, this reaction was used as the source of calibration $\gamma$-rays with energy 6.13 MeV.

A complementary reaction to those mentioned above is $T(p,\gamma)^{4}He$, producing a mono\-chromatic $\gamma$-ray line with the energy 20 \mev. However, the low concentration of $T$ in the available target did not allow to perform measurements in a reasonable time, therefore all the results presented below are related to the $^{11}B$ target.

\subsection{Concept of the measurements. Monte Carlo simulation of the experimental setup.}\label{sec:measurements:VdG_simulations}
\begin{figure}
\begin{center}
\vspace{-15pt}
\includegraphics[width=0.8\swidth]{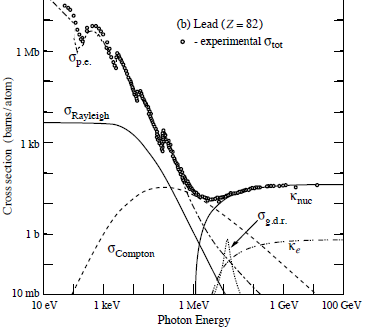}
\caption
{Photon total cross section as a function of energy in lead, showing contribution of different processes: $\sigma_{p.e}$ - atomic photoelectric effect, $\sigma_{Rayleigh}$ - Rayleigh coherent scattering, $\sigma_{Compton}$ - incoherent scattering, $\kappa_{nuc}$ - pair production (nuclear field), $\kappa_{e}$ - pair production (electron field) \cite{bib:pdg}.}
\label{fig:phys:gamma_crossection}
\vspace{-15pt}
\end{center}
\end{figure}

The three main processes of $\gamma$-ray interaction with matter in the energy range 0-20 \mev are: photoelectric effect, Compton scattering and pair production in the nuclear electric field. The energy dependence of these processes for lead is presented in \Reffig{fig:phys:gamma_crossection}. The dominant process of interaction above some energy is electron-positron pair formation. For $PbWO_4$ material this energy is around 10 MeV, as follows from \Reffig{fig:phys:gamma_crossection} and photons produced in radiative capture reactions in $^{11}B$ lie above this energy. It was assumed that for a significant part of events the first interaction of the detected $\gamma$ with the PWO scintillator is an electron-positron pair creation, however the Compton scattering is not negligible. The produced positron annihilates with emission of two 511 \kev photons. Both of them can be absorbed in the scintillator but there is a high probability that one or even two of them escape. The coincidence setup has been proposed with the PWO scintillator surrounded by two plastic scintillators, which present two halves of a cylinder with the PWO scintillator located along its axis. The requirement of no signal in both plastic scintillators could be an indication that all the energy of the primary $\gamma$-ray is deposited in PWO, so that the recorded spectrum gives its proper resolution at the incident energy.

The proposed setup might be termed "three crystal pair spectrometer". Such spectrometers have been used since the sixties for the study of radiative capture nuclear reactions \cite{bib:Ask:1961}.

To prove its feasibility in the current application detailed Geant 4 Monte-Carlo simulations have been performed. Simulations were done under the following assumptions:
\begin{itemize}
\item
\textbf{Geometry}.
3 active volumes were implemented - PWO crystal and 2 plastic scintillators. No support structure was taken into account. The PWO crystal was implemented as a G4Box with dimensions $20 \times 2 \times 2 \,cm^3$ and the 2 plastic scintillators were implemented as G4Tubes, i.e. sections of a tube with the internal radius 2.6 cm, external radius 5.5 cm and 13.4 cm height, covering PWO over almost the full azimuthal angle since a gap between the 2 plastics of 2 mm is implemented. The PWO crystal is oriented along the OZ axis and its center is at 40 cm from the origin of coordinates, at which the source of $\gamma$-rays is assumed to be located.
\item
\textbf{Generated events}.
The $\gamma$'s with energies 6.13 MeV, 12.8 MeV and 17.2 MeV were generated propagating in the OZ direction from a disk of 4 mm radius in the XOY plane with its center at the origin of coordinates. 100 000 events for each energy were generated.
\item
\textbf{Physics list}.
The standard electromagnetic physics list was used, with the default cut of 1 mm for the range of $e^{+}, e^{-}$ and $\gamma$ .
\end{itemize}

\begin{figure}
\begin{center}
\vspace{-15pt}
\includegraphics[width=0.8\swidth]{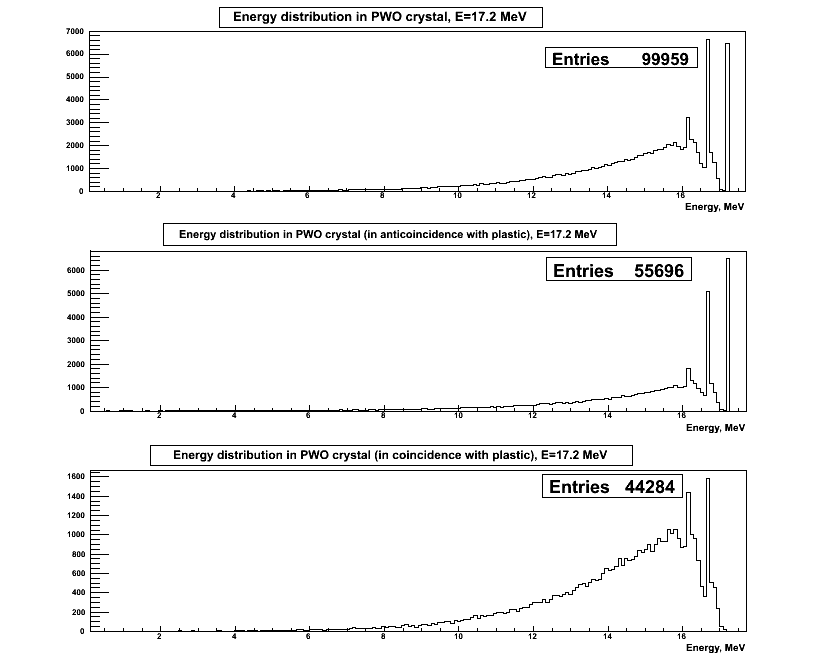}
\caption{Energy distributions in PWO for the 17.2 MeV $\gamma$, resulting from the Monte-Carlo simulations described in \Refsec{sec:measurements:VdG_simulations}. The three panels correspond to: the spectrum of all events registered in PWO (upper panel), the spectrum in anti-coincidence with both plastics (central panel) and in coincidence with one of the plastic scintillators (lower panel).}
\label{pwo_energy_16_9}
\vspace{-15pt}
\end{center}
\end{figure}

The energy deposited in the PWO scintillator for the 17.2 MeV $\gamma$ is presented in Fig. \ref{pwo_energy_16_9}. The top spectrum represents the energy distribution of all recorded events. The middle and bottom spectra correspond to anti-coincidence and coincidence with one of the plastic scintillators. The three peaks can be seen in the spectra corresponding to the total energy, the one-$\gamma$ escape and the double escape of the positron annihilation photons. In the upper spectrum the single escape peak is the dominant one. It can be also seen from the central spectrum that the requirement of no signal in any of the plastic scintillators increases the significance of the full energy peak but does not suppress the single- and double-escape peaks completely. The requirement of a coincidence (lower panel) removed the full-energy peak and enhanced the continuous background of Compton scattering in PWO.

\begin{figure}
\begin{center}
\includegraphics[width=0.8\swidth]{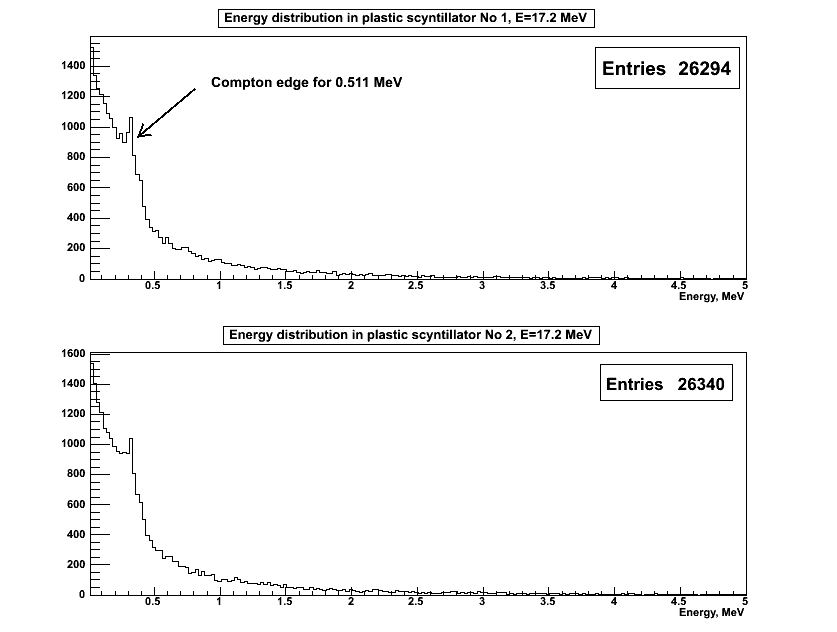}
\caption{Energy distributions in the two plastic scintillators for the 17.2 MeV $\gamma$-rays detected in PWO. The spectra result from the Monte-Carlo simulations described in \Refsec{sec:measurements:VdG_simulations}}
\label{plastic_energy_16_9}
\end{center}
\end{figure}

In \Reffig{plastic_energy_16_9} the energy deposited in plastic scintillators is presented. The peak which is seen in both spectra corresponds to the Compton edge of the 0.511 MeV $\gamma$.
According to the formula
\begin{equation}
E_{max} = E \cdot \frac{2E}{m_e c^2 +2E},
\end{equation}

the Compton edge of the 0.511 \mev $\gamma$ is at 0.341 \mev. The reader should be aware that the plastics, containing only $H$ and $C$, respond almost exclusively via Compton scattering to low energy $\gamma$-rays.

\begin{figure}
\begin{center}
\includegraphics[width=0.8\swidth]{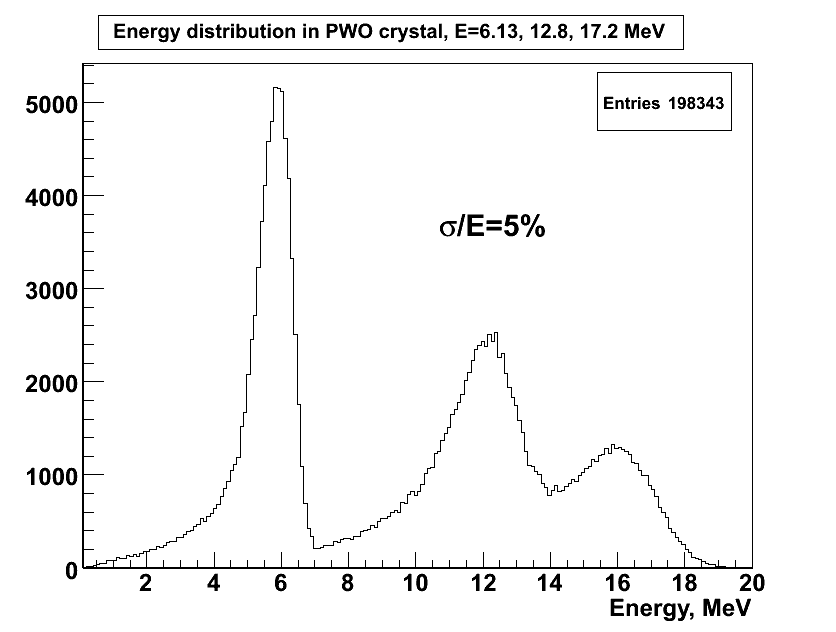}
\caption{Energy distribution in PWO for the 6.13, 12.8 and 17.2 MeV $\gamma$'s having equal intensities, detected with the energy resolution $\sigma/E=5\%$ (see \Refsec{sec:measurements:VdG_simulations}).}
\label{VDG_total_energy_5}
\end{center}
\end{figure}

In \Reffig{VDG_total_energy_5} the energy distribution in PWO for all the 3 analysed energies (6.13, 12.8 and 17.2 MeV) is presented folded with the energy resolution of $\sigma / E =5\%$. The anti-coincidence with plastic scintillators is assumed. This plot gives an orientation as to the shape of energy spectrum expected in our measurements, however the actual values of energy resolution will appear to be worse than assumed in \Reffig{VDG_total_energy_5} (see \Refsec{sec:VDG:results}).

\subsection{Experimental setup}

\subsubsection{Van de Graaff accelerator}
\begin{figure}
\begin{center}
\vspace{-15pt}
\includegraphics[width=0.6\swidth]{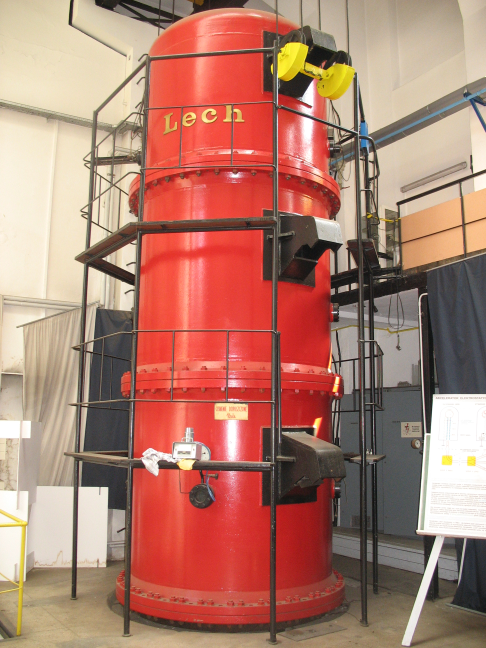}
\caption
{The SINS/UW Van-de-Graaff accelerator inside the pressure tank used to accelerate protons for production of low energy $\gamma$-rays in radiative capture reactions. In the top part a crane is shown used to lift the top section for ion-source installation and/or repair.}
\label{fig:phys:VanDeGraaf}
\vspace{-15pt}
\end{center}
\end{figure}

The low energy $\gamma$-rays for the present study were produced in radiative proton capture reactions with protons accelerated in a Van de Graaff accelerator belonging to SINS and the Warsaw University. This 3.5 MeV Van de Graff machine provides p, d, $^3He$ and $^4He$ beams with energy resolution of about 1 keV. The total accessible $p$ and $d$ beam currents on a target are up to 50 $\mu A$ and for $^3He$ and $^4He$ ions up to about 30 $\mu A$. During our measurements a typical beam current on target was about 10 $\mu A$. A photo of the accelerator located inside the closed pressure tank is presented in \Reffig{fig:phys:VanDeGraaf}. The beam is directed vertically into the target room lying underneath, protected with about 1 m of concrete. In the target room it is bent by 90\degrees with a magnetic analyser and then it may be redirected in the horizontal plane with the aid of a distributing magnet into one of the three beam pipes terminated with an experimental set-up. One of them was used for the present scintillator studies, the other two are currently used in solid-state and chemical analysis studies, the latter with the PIXE method. Tools for steering, focusing and visual observation are provided along the path of the beam.

\subsubsection{Setup for the measurements}
\begin{figure}
\begin{center}
\includegraphics[width=1.0\swidth]{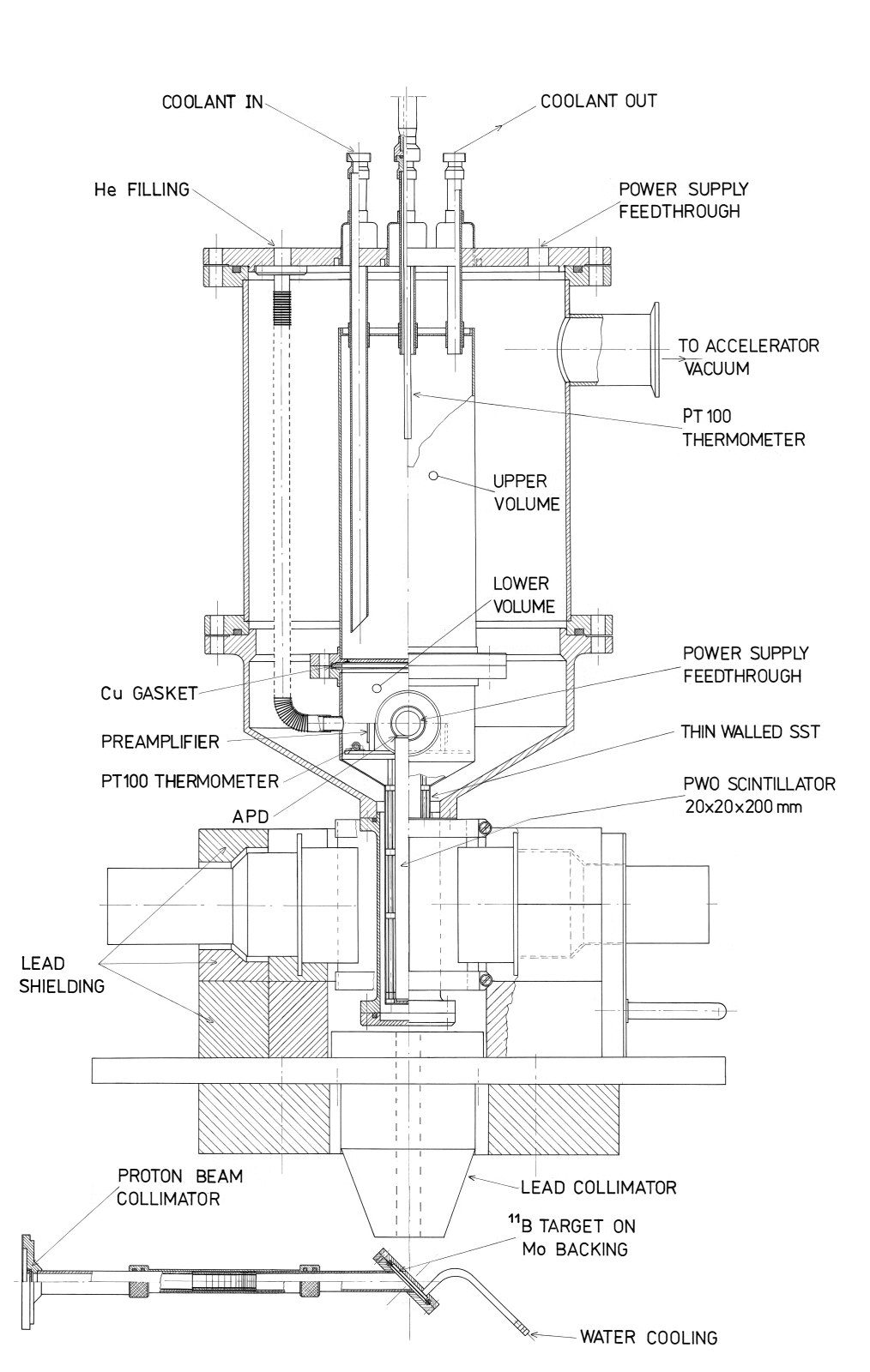}
\caption
{Vertical section of the setup used for the measurements of energy resolution of cooled PWO scintillators with low energy $\gamma$-rays from the radiative capture reactions.}
\label{fig:phys:setup_VanDeGraaf}
\end{center}
\end{figure}

The vertical section of the experimental setup is presented in \Reffig{fig:phys:setup_VanDeGraaf}. The common vacuum jacket houses two volumes separated from the ambient. The lower volume of the setup houses a PWO ($PbWO_{4}$) scintillator, an APD (Hammamatsu S8664-1010), a preamplifier for APD signals and a Pt-100 sensor to measure temperature close to the APD. The PWO-II scintillator of the size $20\times 20 \times 200$ mm, produced by BTCP, was used in the measurements.

\begin{figure}
\begin{center}
\vspace{-10pt}
\includegraphics[width=0.6\swidth]{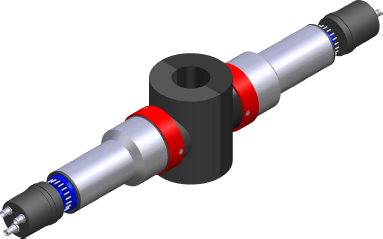}
\caption{A drawing of the plastic scintillators, which are used for coincidence measurements. EJ-200 is machined to form a cylinder cut into two halves along the vertical axis. A cylindrical plexiglass light-guide is glued at the center of each of the two half-cylinders to maximize light transmission to the photocathode of a $3^{\inch}$ diameter XP-4312. The PMs are protected with $\mu$-metal shields from stray electromagnetic fields. Transistorized voltage dividers are seen at the two PM extremes.}
\label{fig:phys:plastics}
\vspace{-15pt}
\end{center}
\end{figure}

\begin{figure}
\begin{center}
\includegraphics[width=0.6\swidth]{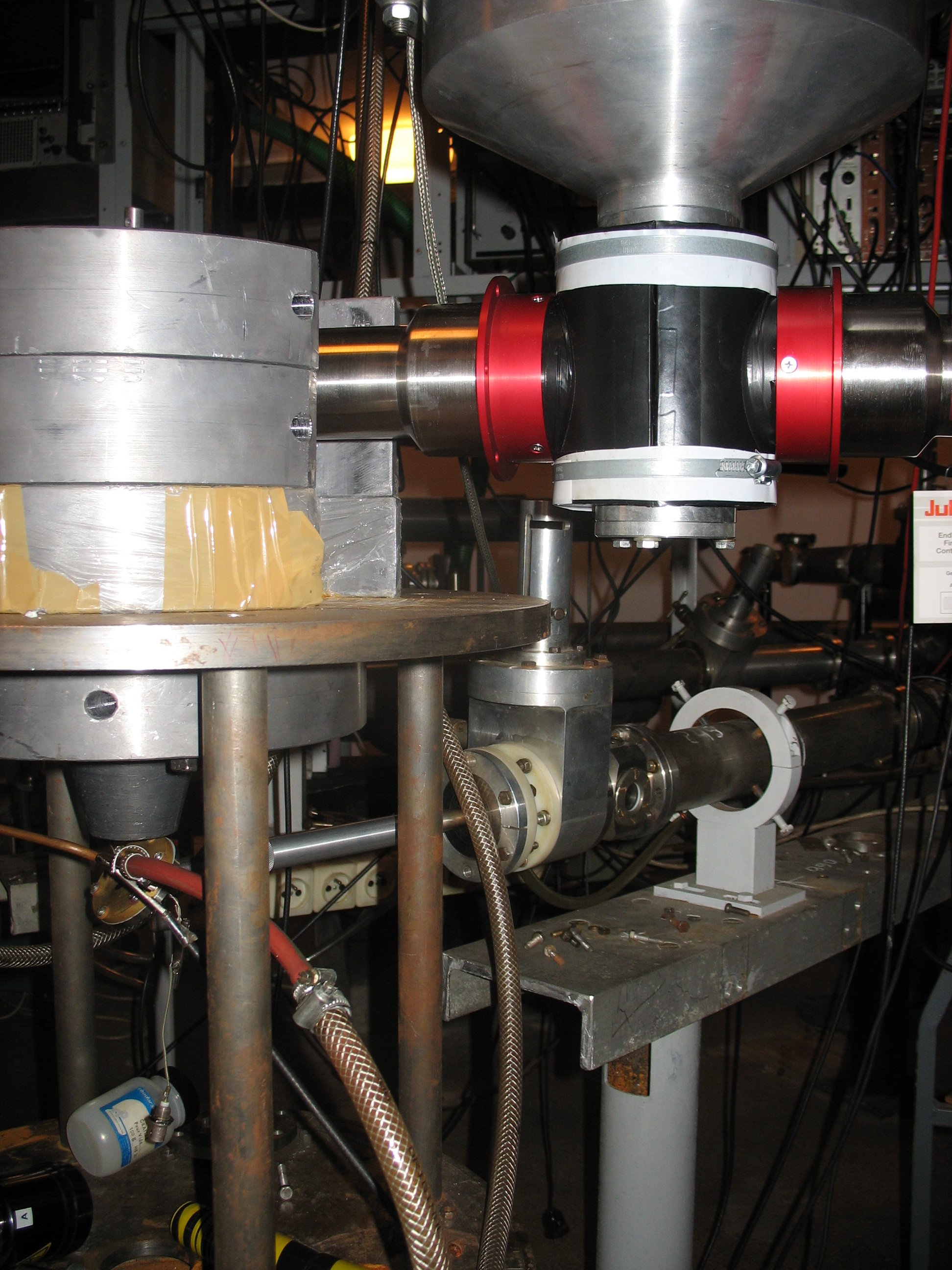}
\caption{A photograph of the setup demonstrating the fastening of plastic scintillators. The cryostat is shifted out of the heavy Pb shielding. A beam-line with a gate-valve and a small-diameter stainless-steel tube, which terminates with a water-cooled target are seen below the cryostat.}
\label{fig:phys:Setup_VDG_foto2}
\end{center}
\end{figure}

The PWO crystal is surrounded by two cylindrical plastic scintillators (EJ-200) with PMT (XP4312) readout for coincidence measurements. The length of the plastic scintillators along the vertical axis is 134 mm, the internal and external radii of the cylinder are 26 and 55 mm, respectively. The EJ-200 material is characterised by high light output - 64 \% of anthracene, long attenuation length and the emission spectra matching well the used photomultipliers. A drawing of the plastic scintillators is presented in \Reffig{fig:phys:plastics}. \Reffig{fig:phys:Setup_VDG_foto2} illustrates the way they are fastened to the aluminum tube, which is a part of the outer jacket of the cryostat. The lead collimator is placed between the target and the scintillator with a hole of $\phi$ 1.5 cm. The collimator prevents from direct photon hits of the plastic scintillators. On the other hand a rather precise relative positioning of the target, collimator and the axis of the PWO crystal is required.

\subsubsection{Cooling}
Through the upper volume of the cryostat the coolant is circulated from the JULABO F32-ME Refrigerated/Heating Circulator. The circulator is suitable for internal and external temperature application in the range of temperatures from $-35^{\circ}C$ to $+200^{\circ}C$. The nominal cooling capacity is 150 W at $-20^{\circ}C$ and 30 W at $-35 ^{\circ}C$, i.e. decreases rapidly with the decreasing temperature. Two types of coolant have been tested: JULABO "Thermal" bath fluid H5S and Galden heat transfer fluid HT-70 of Solvay Solexis. The results presented below had been taken at -20$^{\circ}$C which was achieved with the H5S coolant. It is slightly higher than $-25^{\circ}C$ planned for the PANDA EMC environment. The HT-70 coolant was also tested in an attempt to reach lower temperatures. It is typified by the lower viscosity of 1.09 $mm^2\cdot s^{-1}$ vs 4 $mm^2\cdot s^{-1}$ for the Julabo H5S coolant, so that it permits a higher velocity of the circulating liquid. However no significant improvement has been achieved with the difference at the level of $0.5^{\circ}C$. We conclude that with the given thermal losses in our system we need a more powerful refrigerator to reach the desired -25\degrees{C}.

The cooled elements are isolated from ambient by vacuum common with the accelerator vacuum system, therefore the minimum achieved temperature depends on the vacuum level. The lowest temperature was achieved after a prolongued pumping when vacuum stabilised at $6\cdot 10^{-6}$ Torr. The lower volume in \Reffig{fig:phys:setup_VanDeGraaf} is filled with He at slightly above atmospheric pressure which serves as a contact medium between the cooled bottom copper plate of the upper volume and the scintillator, APD and the preamplifier located in the lower volume. Temperature in the lower volume was monitored with a Pt-100 sensor, whose resistance was measured with a Keithley digital multimeter, which provides the possibility to store the temperature information in a PC via an USB port.

The working temperature of -20$^{\circ}$C was achieved on the average after 6 hours of cooling.

\subsubsection{Targets}\label{sec:VDG:targets}
\begin{figure}
\begin{center}
\vspace{-15pt}
\includegraphics[width=0.8\swidth]{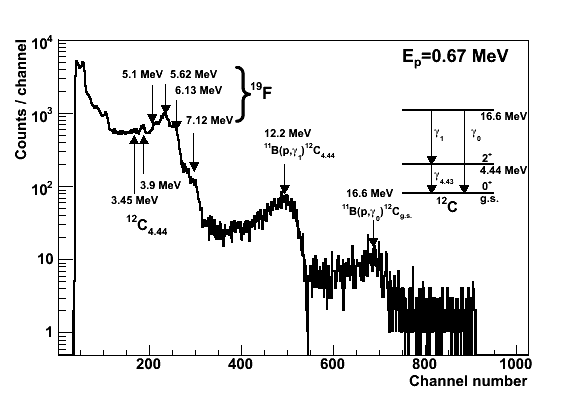}
\caption
{Energy spectrum of $\gamma$-rays produced in a 2 $\mu m$ thick boron target at $E_{p}=0.67$ MeV, detected with the $\phi \, 4^{\inch} \times 4^{\inch}$ $NaI(Tl)$ detector mentioned in \Refsec{sec:VDG:targets}.}
\label{fig:phys:nai_spectra}
\vspace{-15pt}
\end{center}
\end{figure}

The high melting point of boron ($\approx 2300^{\circ}C$) makes it difficult to prepare thin boron targets of uniform thickness by an evaporation method. The targets were prepared by evaporating natural boron on Mo backings 0.5 mm thick, using electron bombardment of a boron pellet in a Ta boat, as described in \cite{bib:Erskine1963397}. Several targets have been produced with thicknesses ranging from 2 $\mu m$, corresponding to $\sim$100 keV energy loss for 1 MeV protons, to 10 $\mu m$. The target holder is constructed so, that the back surface of the Mo backing is water cooled (see the bottom part of \Reffig{fig:phys:setup_VanDeGraaf}). It is tilted at 45\degrees, while the axis of the Pb collimator is at 90\degrees to the beam direction. The quality of the target has been studied with a large NaI(Tl) scintillator $\phi \, 4^{\inch} \times 4^{\inch}$ coupled to a $4^{\inch}$ diameter PMT. The measurements at the beam energy $E_{p}=0.67 MeV $ have been performed. The two high energy lines originating from the transitions from the capturing state to the ground state and the 4.44 \mev first excited state in $^{12}C$ are clearly seen in the spectrum of \Reffig{fig:phys:nai_spectra}. However, the peaks corresponding to the transition $4.44 MeV \rightarrow g.s.\, (ground\, state)$, which would be also useful for the PWO response studies, are obscured by the $^{19}F$ contaminant, which has high chemical affinity to $B$. It should be stressed that $^{19}F(p,\alpha)^{16}O^{*}(6.13 \, MeV)$ proceeds with a "normal nuclear" cross-section of the order of $mb$, while $^{11}B(p,\gamma)^{12}C$ is mediated by an electromagnetic interaction, therefore its intensity is relatively suppressed.

\subsubsection{Readout electronics}
\begin{figure}
\begin{center}
\vspace{-15pt}
\includegraphics[width=0.8\swidth]{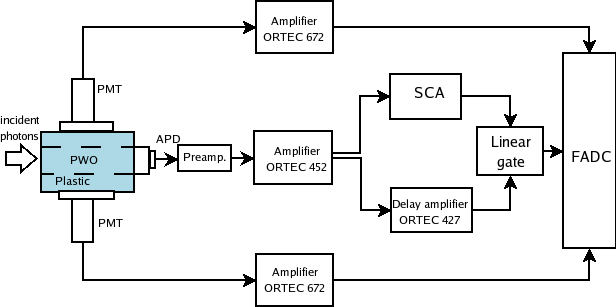}
\caption
{A scheme of readout electronics used for the measurements of energy resolution of PWO scintillators with low energy $\gamma$-rays.}
\label{fig:phys:Electronics_VDG}
\vspace{-20pt}
\end{center}
\end{figure}

The readout chain of PWO starts with a preamplifier, which has been designed and produced at the Basel University especially for PANDA EMC design studies \cite{bib:emctdr}. It is the Low Noise / Low Power (LNP) charge preamplifier with charge sensitivity 0.5 V/pC. The LNP preamplifier fits detector capacitances in the range from 50 pF to 300 pF, therefore it is well applicable to work with ($10\times10$ $mm^2$) Large Area APDs S8664-1010 produced by Hamamatsu possessing the capacitance of about 270 pF. The measured noise in combination with an APD and the postamplifier with the shaping time of 750 ns is 1900 $e_{RMS}$ at $-25^{\circ}C$ \cite{bib:emctdr}. The signal from the preamplifier goes into the standard spectroscopic amplifier ORTEC 452 with the shaping time of 1 $\mu s$. The amplified signal is sent in parallel to a Linear Gate and a Single Channel Analiser (SCA) with a threshold set to suppress the electronic noise, whose output triggers the linear gate. For both plastic scintillators the signals from PMTs are amplified with ORTEC amplifiers with the same shaping time of 1 $\mu s$. The three output signals are fed into a 4-channel Flash ADC CAEN N1728B (100 MHz, 14 bits). CAEN N1728B Flash ADC has a 10 ms wide buffer which is filled continuously with digitized amplitudes from each of the three channels. In the acquisition software for the N1728B FADC the signal coming from PWO is assigned as a trigger, i.e. when it exceeds some threshold the inputs in all the three channels are stored from the FADC buffer 10 $\mu s$ before and 20 $\mu s$ after the trigger into hard drive of a PC. A slightly simplified scheme of the readout is presented in the \Reffig{fig:phys:Electronics_VDG}.

\subsection{Results}\label{sec:VDG:results}
The measurements of energy resolution of PWO scintillators with Large Area APD readout have been performed at the proton beam energy of 1.4 \mev. It is close to the 1.390 \mev resonance in $^{11}B(p,\gamma)^{12}C$ reaction at which (see \Refsec{sec:measurements:reactions}) $\gamma$-rays with energies 12.8 \mev and 17.2 \mev should be produced with comparable cross-sections.

\begin{figure}
\begin{center}
\vspace{-15pt}
\includegraphics[width=0.8\swidth]{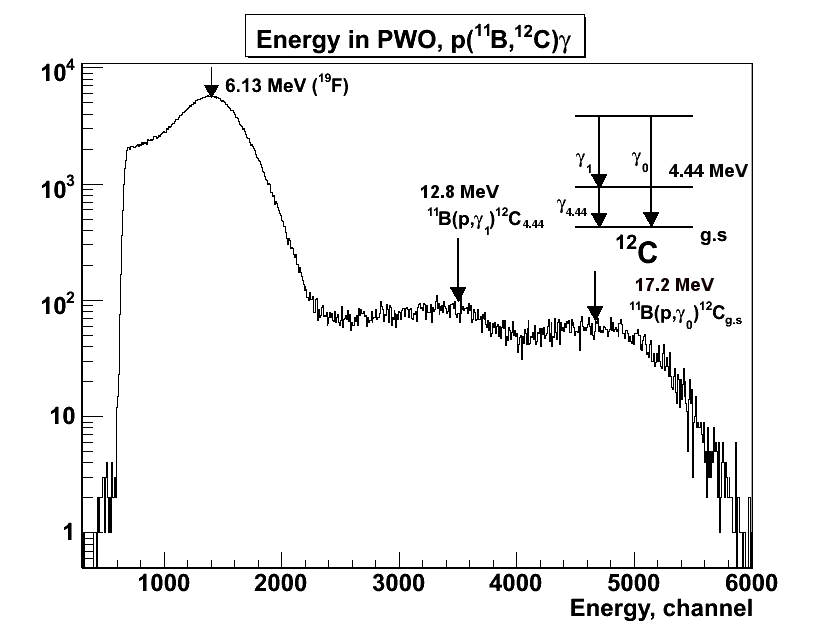}
\caption{Spectrum of $^{11}B(p,\gamma)^{12}C$ and $^{19}F(p,\alpha)^{16}O$ (target contaminant) gamma-rays emitted at $E_p$=1.39 MeV, detected with the $20\times20\times200$ mm PWO scintillator cooled down to -20\degrees{C} (logarithmic scale).}
\label{fig:phys:whole_spectra}
\vspace{-15pt}
\end{center}
\end{figure}

The cryostat was cooled during 20 hours before the measurements were started to stabilize the temperature at $-20^{\circ}C$. The bias voltage on the APD was set at +377 V and its inverse dark current was continuously monitored at an average level of 25 nA. The data have been collected for 10 hours. The collected energy spectrum is plotted in \Reffig{fig:phys:whole_spectra} in the logarithmic scale. Three energy peaks are seen: the intense peak is assigned to 6.13 \mev from the $^{19}F(p, \alpha \gamma)^{16}O$ reaction and the two weaker maxima to 12.8 \mev and 17.2 \mev lines from $^{11}B(p,\gamma)^{12}C$, following our previous experiment with the NaI(Tl) shown in \Reffig{fig:phys:nai_spectra}.

\begin{figure}
\begin{center}
\includegraphics[width=0.8\swidth]{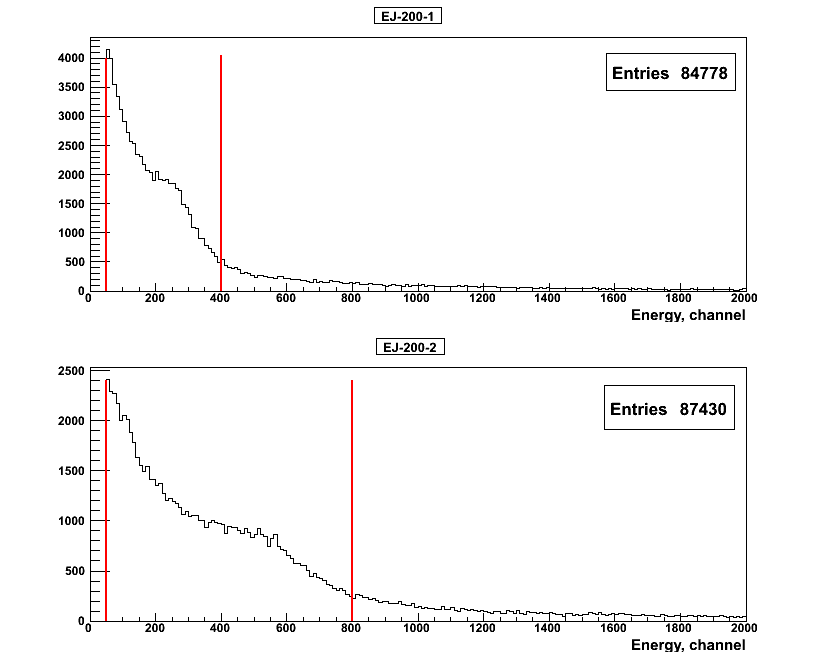}
\caption{Energy spectra in the two plastic scintillators. The range of amplitudes between the vertical lines is used for the anticoincidence requirement.}
\label{fig:phys:energy_plastic}
\end{center}
\end{figure}

\Reffig{fig:phys:energy_plastic} presents the energy spectra measured with the plastic scintillators. The edge of the Compton continuum, corresponding to the 0.511 \mev annihilation $\gamma$-ray, is clearly seen in both spectra. A single ORTEC power supply provided HV for both PMTs, so that the differences in amplitude of the Compton edge reflect the asymmetry of PMT gains at -1.1 kV of the common voltage. Vertical lines indicate the range of energies of annihilation quanta used in coincidence or anticoincidence requirements. A signal within the shown range was considered as the detection of a 0.511 \mev annihilation $\gamma$-ray, whereas its absence in both plastic scintillators served as an indication that the entire energy was deposited in the PWO scintillator. It was found that the anticoinidence requirement tends to improve the widths of the observed peaks and shifts them to higher energy since without this condition peak positions correspond to large extent to a single escape peak (see also \Refsec{sec:measurements:VdG_simulations}).

\begin{figure}
\begin{center}
\vspace{-10pt}
\includegraphics[width=0.8\swidth]{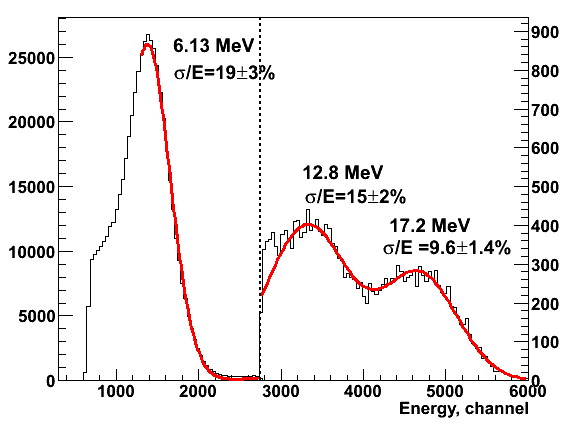}
\caption{Spectrum of $^{11}B(p,\gamma)^{12}C$ and $^{19}F(p,\alpha)^{16}O$ (target contaminant) gamma-rays emitted at $E_p$=1.39 MeV, detected with the $20\times20\times200$ mm PWO scintillator cooled to -20\degrees{C} (linear scale, note that the two parts of the plot have different scales) in anticoincidence with the plastic scintillators. The solid line is a fit to the measured spectrum using three Gaussians with the indicated relative dispersions.}
\label{fig:phys:energy}
\vspace{-10pt}
\end{center}
\end{figure}

\Reffig{fig:phys:energy} presents for convenience the energy spectrum of $\gamma$-rays in the linear scale. The left and the right parts are scaled differently. The presented spectrum corresponds to the anticoincidence requirement with the plastic scintillators. A fit with gaussians has been performed to extract energy resolutions. The presence of transitions to the higher states of $^{16}O$ with energies 6.92 \mev and 7.12 \mev was taken into account in the fit. According to \cite{bib:Hellborg:1972} the intensities of these two lines are at the level 3-8\percent of the 6.13 \mev transition depending on the proton beam energy. For the fit the intensities of each of them were fixed at the level of 5\percent of the transition $6.13 \mev \rightarrow \, g.s.$. Moreover, due to the asymmetry of the 6.13 \mev peak the fit was performed only over its right wing. The fitted Gaussian energy resolution is $19\pm 3\%$ for the 6.13 MeV, $15 \pm 2\%$ for the 12.8 MeV and $9.6 \pm 1.4\%$ for the 17.2 MeV peak.

\begin{figure}
\begin{center}
\vspace{-15pt}
\includegraphics[width=0.8\swidth]{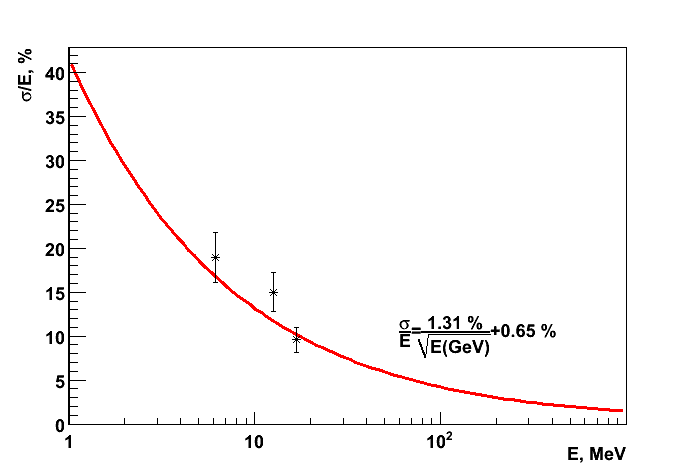}
\caption
{Measured gaussian resolutions (solid points with error bars) compared with the solid curve parameterizing the results obtained in the 40.9-675 \mev range at MAMI \cite{bib:Novotny:2008zz}.}
\label{fig:phys:sigmaE}
\vspace{-15pt}
\end{center}
\end{figure}

A comparison with the results of PWO resolution measurements using bremsstrahlung photons in the 40.9 - 675 \mev range \cite{bib:Novotny:2008zz} at MAMI is of interest. The measurements at MAMI were performed with a $3\times3$ matrix of $20 \times 20 \times 200$ $mm^3$ PWO scintillators with APD readout cooled down to $0^{\circ}C$. APD and preamplifiers of the same type as in our work have been used in \cite{bib:Novotny:2008zz}. The measured energy resolution is well described by the following dependence on the photon energy (in \gev): $\sigma/E= 1.86 \%/\sqrt{E(GeV)}+0.65 \%$. It is estimated by the authors of \cite{bib:Novotny:2008zz} that lowering the temperature from 0\degrees{C} to -25\degrees{C} should increase light output by the factor of 2 thus the statistical term in energy resolution will be reduced by the factor $\sqrt{2}$, so that the resolution to be compared with our data reads:
\begin{equation}
\label{eq:sigmaE_mami}
\sigma/E= 1.31 \%/\sqrt{E(GeV)}+0.65 \%.
\end{equation}
\Reffig{fig:phys:sigmaE} compares our experimental points and the extrapolation of the results from MAMI corrected for the difference in working temperatures (solid curve). One may conclude that our results prove that \Refeq{eq:sigmaE_mami}is a valid extrapolation of the high energy data into the lower energy range, where resolution has not yet been measured. The surplus of counts above the solid line seen in \Reffig{fig:phys:energy} between the 6.13 and 12.8 \mev peaks suggests that an assumption of pure gaussian line shapes is an oversimplification of the reality. However, the measured spectrum does not leave enough space to test more complex hypotheses (e.g. Gaussians with tails). Assuming such an eventuality would make our present estimates lower limits of the true resolution. The proper testing ground might be offered by the $T(p,\gamma)^4He$ reaction, which is expected to demonstrate the true line shape in the space between the 20 \mev and the 6.13 \mev peaks, the latter ever present in the measured spectrum. The tritium target needed for such an experiment is, however, an expensive investment.

%% file: monte_carlo.tex
\chapter{Monte-Carlo simulation of PANDA physics performance}\label{chap:phys:MonteCarlo}
This part of the PhD Thesis is devoted to demonstration of the PANDA detector physical performance in detection of the $h_c$ state in charmonium. The importance of a study of $h_c$, i.e. of the measurement of its mass, width and precise determination of its quantum numbers has been justified in \Refsec{sec:phys:hc}. The first part of this chapter (\Refsec{sec:MC:software}) is devoted to the description of software used in the analysis and the second part (Sec. \ref{sec:MC:hc_general}-\ref{sec:MC:angdist}) provides the results of simulations.

\section{Panda offline software.}\label{sec:MC:software}
The offline software developed for the PANDA physics studies has been used in this analysis and the following description is based on the Physics Performance Report for PANDA \cite{bib:PANDA_PB:2009}. The software consists of several tools adapted from other HEP experiments which include event generators (EvtGen, DPM), particle tracking with GEANT4, digitization of the signal, reconstruction and identification of charged and neutral particles and high level analysis tools providing vertex and kinematical fits.

\subsection{Event generators.}
Two event generators have been used in this analysis. EvtGen is used to produce the reaction of interest with the possibility to define decay modes of the particle and angular distributions. Dual Parton Model (DPM) can be used to produce generic background from $\overline{p}p$ annihilations.

\subsubsection{EvtGen}
EvtGen event generator \cite{Lange:2001uf} was originally developed for the Babar experiment at SLAC to study the physics related to B-mesons. The modular design of EvtGen allows to extend it by incorporating arbitrary physics channels and decay models, for these reasons it was adapted by the ATLAS and PANDA experiments. Decay amplitudes, instead of probabilities, are used for the simulation of decays. The amplitude for each node in a decay tree is used to simulate the entire decay chain, including all angular and time-dependent correlations. The implementation of each decay amplitude is independent of how the mother particle was generated or how the daughter particles are going to decay. EvtGen computes the spin density matrices and the decay probability and decay models, which implement a single node in a decay tree, must only specify the decay amplitude for each combination of the mother and daughter spin states.

\subsubsection{DPM}
In high energy proton-antiproton collisions soft processes dominate in multiparticle production and perturbative QCD is not applicable for their description. However, suitable large N expansions of QCD provide a topological classification of diagrams and a potentially useful non-perturbative approach. This topological expansion, when supplemented with generally accepted theoretical principles like duality, unitarity, Regge behavior and the parton structure of hadrons, provides the basis underlying the dual parton model (DPM) \cite{bib:sim:DPM}. It has been shown that DPM provides a complete phenomenological description of all facets of soft processes.

The main objects of the model are constituent quarks having masses $\sim$ 300-350 MeV, strings, and string junctions for baryons. It is assumed that mesons consist of a quark and an anti-quark which are coupled by color forces. The vortex lines of the field are concentrated in a small space region forming a string-like configuration. So, mesons are
considered as strings with small masses. Various processes are considered in baryon-antibaryon interactions, such as string junction annihilation and creation of three strings, quark-antiquark annihilation and string creation between
diquark and anti-diquark, quark-antiquark and string junctions annihilation and so on. The energy dependence of cross sections for these processes is parametrized to reach an agreement with experimental data. The string fragmentation is performed by the mechanism similar to that included in the LUND model \cite{bib:Andersson:1978vj}; after the string fragmentation all unstable hadrons decay with the aid of the code DECAY \cite{bib:Hanssgen:1983qh}.

\subsection{Particle Tracking and Detector Simulation.}
The detector simulation starts with propagation of the generated particles through the PANDA detector using GEANT4 transport code. It takes into account all the possible processes of particle interactions and decays. As an output GEANT produces the collection of hits, which contains information on the intersection points and energy lost by all the particles in corresponding parts of the detector. For the description of the detector geometry the approach developed by the CMS experiment is followed. The Detector Description Database (DDD) is used in which the detector is represented as a multigraph structure with a compactified description. The Detector Description Language (DDL) based on XML is used to encode the geometrical properties
of the components, their relative positions and the materials. The simulations have been performed with the complete detector setup and the Straw Tube Tracker (STT) chosen as the central tracker. The pellet target scenario was considered in order to take into account the material budget presented by the target pipe. The Gaussian spread of interaction points typified by $\sigma$=0.275 mm was taken into account in simulations.

\subsubsection{Digitization}
The output of Geant simulation is used as an input for the digitization step, which models the processing of the signal from individual subdetectors by their front-end electronics. The output of the digitization should correspond to the detector response in the real experiment.

\textbf{Readout of MVD.} MVD will use two different silicon detector types, silicon strip and pixel detectors. The readout of the silicon devices is for both types different and is treated differently in the digitization scheme.
The hit position on the sensor surface defines the channel number and the deposited energy defines the charge collected by the electronics. Strip sensors will be sensitive on both sides and the formation of digitized channels is done independently on both sides of the sensor. In the case of pixel detectors, the trajectory is projected to the surface and depending on its relative orientation, all excited pixel cells are calculated and the charge signal is shared among all pixel cells depending on the fraction of the local track.

\textbf{Straw Tube Tracker and Drift Chambers.} Digitization for the Straw Tube Tracker (STT) and the Drift Chambers (DCH) has been treated in similar way. Both devices consist of wires inside an $Ar-CO_2$ gas mixture volume. If a charged particle traverses this gas volume, the local
helix trajectory is derived from the corresponding GEANT4 intersection points. The drift time of the ionisation electrons is estimated from the smallest distance of this helix trajectory to the wire $d_{poca}$ ($poca$ - point of closest approach). The uncertainty of the drift time is taken into account by smearing $d_{poca}$ with a Gaussian distribution with a standard deviation of $\sigma$=150 $\mu m$ for the STT, and $\sigma$ = 200 $\mu m$ for the DCH devices. The energy signal of a straw tube is finally calculated by taking into account Poisson statistics.

\textbf{GEM readout.} Each GEM station consists of two detection planes. The distance between the detection planes is 1 cm. It has been assumed that each detection plane has two strip detection layers with strips having perpendicular orientation to each other. The gas amplification process and the response of the strip detector has not been simulated in detail. Instead, the entry point of a charged track into the detector plane has been taken directly from GEANT4 and smeared with a Gaussian distribution of 70 $\mu m$ width in each strip orientation direction.

\textbf{DIRC detectors.} The light propagation in the Cherenkov radiators, the signal processing in the front-end-electronics, and the reconstruction of the Cherenkov angle have been modeled in a single effective step. The resolution of the reconstructed Cherenkov angle $\sigma_{C}$ is mainly driven by the uncertainty of the single photon angle $\sigma_{C,\gamma}$ and the statistics of the relatively small number of the detected Cherenkov photons, $N_{ph}$. A single photon resolution of $\sigma_{C,\gamma}$ = 10 mrad was used, corresponding to the experience with the existing DIRC detectors. The number of detected photons was calculated from the velocity, $\beta$, of charged particles passing through the quartz radiators and the
path length, $L$, within the radiator. The sensitive wavelength interval, [$\lambda_{min};\, \lambda_{max}$], was chosen as [280nm; 350 nm], and a total efficiency of $\varepsilon$ = 7.5 \percent was used to take into account the transmission and reflectivity losses as well as the quantum efficiency
of the photo detectors. The precision of the measured Cherenkov angle obtained with the digitization and reconstruction procedure has a gaussian shape with $\sigma$=2.33 mrad.

\textbf{EMC readout.} The parameters of the EMC which were taken into account for digitization in the target spectrometer are electronics noise and photon statistics. According to the prototypes of PANDA EMC front-end electronics the noise level at $-25\degrees{}C$ is at the level of 0.5-1 \mev. The Gaussian distribution of the noise with 1 \mev width was used in simulations as a conservative estimate. For the PWO-II scintillator around 80 e-h pairs/\mev are expected with the LAAPD readout. Additional contribution to photon statistics is given by the excess noise factor of an APD, which was measured for Hamamatsu APDs at 1.38 for the gain $M=50$. For the shashlyk calorimeter in the Forward Spectrometer, consisting of lead-scintillator sandwiches only a fraction of roughly 30\percent of the incident energy is deposited in the scintillator material. Based on this energy deposit the electronics
noise with $\sigma$= 3MeV has been considered for digitization.

\subsection{Reconstruction}
\subsubsection{Charged particle track reconstruction}
\textbf{MVD cluster reconstruction.} The MVD provides very precise space point measurements as a basis for the track and vertex reconstruction.
The hit resolution of individual MVD measurements has $\sigma=21.0\,\mu m$ for hits where only one pixel contributes to the hit cluster and $\sigma=3.8\,\mu m$ for multihit clusters, where the charge weighting between pixel cells in the cluster can be used to calculate the mean position of the hit.

\textbf{Global track reconstruction.} The track object provides information about a charged particle path through space. It contains a collection of hits in the individual tracking subdetectors. An idealized pattern recognition has been used for track building based on Monte Carlo information to assign reconstructed hits to their original tracks. The tracks in the target spectrometer are fitted with the Kalman Filter algorithm, which considers not only the measurements and their corresponding resolutions but also the effect of the interaction with the detector material, i.e. multiple scattering and energy loss. A typical choice for the parametrisation of the track in a solenoidal field is a five-parameter helix along the principle field direction z. The task of the track-fitting algorithm is to determine the optimal parameter vector and its covariance matrix as a function of the
flight length $l$ in order to create a representation of the track as a piecewise helix. In the physics analysis the particle position and momentum can then be accessed through this piecewise helix representation.

\textbf{Tracking performance.} With the used reconstruction software more than 90\percent track reconstruction efficiency has been achieved for the transverse momenta $p_t>0.2 \, \gevc$ dropping down to 70\percent for $p_t=0.1 \, \gevc$. The momentum uncertainty can be characterised with $\sigma_p/p=1\percent$ for the pions of 1 \gevc momentum at a polar angle 20\degrees and the achieved vertex space resolution is $\sigma_{d0}=47.5 \, \mu m$ and $\sigma_{z0}=49.2 \, \mu m$ for the 3 \gevc pions.

\subsubsection{Photon reconstruction}
Most of the EMC reconstruction code is based on algorithms developed and applied by the Babar experiment \cite{bib:Strother:1998}. The first step in EMC reconstruction procedure is a cluster reconstruction. Electromagnetic shower during its development deposes energy in a set of neighboring crystals which is called cluster. The energy deposits in the crystals, which form the cluster, and their positions allow to determine the energy and direction of the initial photon. Cluster finding algorithm starts from the crystal with largest energy deposit. Its neighbors are then added to the cluster if the energy deposit is above a certain threshold $E_{xtl}$. This threshold, on the one hand, should be as low as possible, to detect low energy photons, on the other hand, it must be sufficiently high to separate the reconstructed photons from the readout noise. $E_{xtl}$ was set to 3 \mev, which corresponds to the energy equivalent to the $3 \sigma$ of the expected noise of the electronics. The procedure of adding new elements to the cluster is continued for the neighbors until no further module fulfils the threshold criteria. A cluster is accepted if the total energy deposit in all of its elements is above a second threshold $E_{cl}$, which was set to 10 \mev.

The next step in EMC reconstruction procedure is to search for bumps within each reconstructed cluster, which can appear if the cluster is formed by more than one particle detected at small relative angular distances. In this case the cluster has to be subdivided into regions which can be associated with the individual particles. This procedure is called bump splitting. The bump is defined by two criteria. First is that the energy deposit in one scintillator module, $E_{local}$, must be above the threshold, $E_{max}$, while all neighbor modules have smaller energies. $E_{max}=20 \, \mev$ was chosen for the Target Spectrometer. In addition the highest energy $E_{Nmax}$ of any of the N neighboring modules must fulfill the requirement $0.5(N-2.5)>E_{Nmax}/E_{local}$. The total cluster energy is then shared between the bumps, taking into account the shower shape of the cluster. For this step an iterative algorithm is used, which assigns a weight $w_i$ to each scintillator module, so that the bump energy is defined as
$E_b=\sum_i w_i E_i$. $E_i$ represents the energy deposit in the $i^{th}$ module and the sum extends over all modules within the cluster. The crystal weight for each bump is calculated with:
\begin{equation}
w_i=\frac{E_iexp(-2.5r_i/r_m)}{\sum_jE_jexp(-2.5 r_j/r_m)},
\end{equation}
where $r_m$ - Moli\`{e}re radius of the scintillator, $r_i$, $r_j$ - distances of the $i^{th}$ and $j^{th}$ module to the center of the bump. The procedure is iterated until convergence.

The spatial position of a bump is calculated via a center-of-gravity method. The radial energy distribution, originating from a photon, decreases approximately exponentially. Therefore, a logarithmic weighting with
\begin{equation}
w_i=max(0,A(E_b)+ln(E_i/E_b))
\end{equation}
was chosen, where only modules with positive weights are used. The energy dependent factor $A(E_b)$ varies between 2.1 for the lowest and 3.6 for the highest photon energies.

In case of the shashlyk calorimeter in the Forward Spectrometer the same reconstruction algorithms are used however with different thresholds: $E_{xtl}=8 \, \mev$, $E_{cl}= 15 \, \mev$ and $E_{max}=10 \, \mev$.

The sum of energies deposited in the scintillator material of the calorimeters is in general less than the energy of the incident photon. While only a few percent is lost in the TS EMC, which mainly originates from energy losses in the material between the individual crystals, a fraction of roughly 70\percent of the energy is deposited in the absorber material for the shashlyk calorimeter. The reconstructed energy of the photon in the TS
EMC is expressed as a product of the measured total energy deposit and a correction function which depends logarithmically on the energy and - due
to the layout - also on the polar angle. The correction function $f(\ln{E}, \theta)$ has been obtained from the Monte Carlo simulations with single photons. In case of shashlyk calorimeter in FS a correction has been considered depending only on the collected energy.

\subsection{Charged particle identification (PID)}
In the PANDA experiment several subdetectors will provide useful PID information for specific particle species and momenta. Energy loss measurements within the trackers provide good criteria for the distinction between the different particle types below 1 \gevc; the DIRC detector is the most suitable device for the identification of particles with momenta above the Cherenkov threshold; EMC, in combination with the tracking detectors, is the most powerful detector for an efficient and clean electron identification; the Muon detector is designed for the separation of muons from  other particle species. The best PID performance, however, can be obtained by taking into consideration all the available information from all subdetectors.

The PID software is divided in two parts. In the first stage the recognition is done for each detector individually, which results in probabilities for all five particle hypotheses ($e$, $\mu$, $\pi$, $K$ and $p$). In the second stage the global PID combines
this information by applying a standard likelihood method.

\subsubsection{Subdetector PID}
\textbf{dE/dx measurements.}
The energy loss of particles in thin layers of material directly provides an access to the dE/dx. As can be seen directly from the Bethe-Bloch formula, a given momentum particles of different types have different specific energy losses, dE/dx. This property can be used for particle identification. The method however suffers from two limitations. First of all, at the crossing points, there is no possibility to disentangle particles. Secondly,
the distribution of the specific energy loss displays a long tail which constitutes a limitation to the separation, especially when large differences exist between different particle yields. In PANDA, two detectors will give access to a dE/dx measurement, the MVD detector setup, and the central spectrometer tracking system.

\textbf{PID with DIRC.} Charged tracks are considered if they can be associated with the production of Cherenkov light in the DIRC detector. Based on the reconstructed momentum, the reconstructed path length of the particle in the quartz radiator, and the particle type hypothesis the expected Cherenkov angle and its error are estimated. Compared with the measured Cherenkov angle the likelihood and significance level for each particle species are calculated.

\textbf{Electron identification with the EMC.} While muons and hadrons in general loose only a certain fraction of their kinetic energy by ionisation processes, electrons deposit their complete energy in an electromagnetic shower. The ratio of the measured energy deposit in the calorimeter to the reconstructed track momentum ($E/p$) will be approximately unity. Furthermore, the shower shape of a cluster is helpful to distinguish between electrons, muons and hadrons. Since the chosen size of the scintillator modules corresponds to the Moli\`{e}re radius of the material, the largest fraction of an electromagnetic shower originating from an electron is contained in just a few modules. Instead, a hadronic shower with a similar energy deposit is less concentrated. These differences are reflected in the shower shape of the cluster. A neural network based algorithm was implemented for electron recognition based on different parameters characterizing cluster shape.

\textbf{PID with the muon detector.} The particle ID for the muon detector is based on propagation of the charged particles from the tracking volume outward through the neighboring detectors like DIRC and EMC and finally through the iron of the solenoid yoke, where the detection layers of the muon device are located. Then the extrapolated intersection points with the detection cells are compared with the detected muon hit positions. In case the distance between the expected and the detected hit is smaller than 12 cm, according to 4$\sigma$ of the corresponding distribution, the muon hit will be associated with the corresponding charged track. The procedure results in a good muon identification for momenta above approximately 1 \gevc. While a complete muon-electron separation is to be expected, a contamination of a few percent by pions can at best be achieved.

\subsubsection{Global PID}
The global PID, which combines the relevant information of all subdetectors associated with one track, has been accomplished with the aid of a standard likelihood
method. Based on the likelihoods obtained from each of the individual subdetectors, the probability for a track originating from a specific particle type $p(k)$ is evaluated from the likelihoods as follows:
\begin{equation}
p(k)=\frac{\prod_i p_i(k)}{\sum_j \prod_i p_i(j)},
\end{equation}
where the product with index $i$ runs over all the considered subdetectors and the sum with index $j$ over the five particle types $e$, $\mu$, $\pi$, $K$ and $p$.

Due to the variety of requirements imposed by the different characteristics of the benchmark channels various kinds of particle candidate lists depending on different selection criteria on the global likelihood are provided for the analysis. The usage of the so-called VeryLoose and Loose
candidate lists allows to achieve good efficiencies, and the Tight and VeryTight lists are optimized to obtain a good purity with efficient background
rejection.

\subsection{Physics analysis tools}
The analysis user has the choice to reconstruct decay trees, perform geometrical and kinematical fits, and to refine the event selection by using Beta, BetaTools and the fitters provided by the analysis software \cite{bib:Jacobsen:1997} directly in an application framework module, or by defining the analysis in a more abstract way using SimpleComposition tools. This TCL based high level analysis tool package provides an easy-to-learn user interface for the definition of an analysis task and the production of n-tuples, and it allows to set up analysis jobs without the need to compile any code. SimpleComposition as well as Beta, BetaTools and the geometric and kinematic fitters were taken over as well tested packages from BaBar, and adapted and slightly extended for the usage in PANDA. The last n-tuple based analysis steps were carried out utilizing the ROOT toolbox \cite{bib:Brun:1997pa}, which provides powerful, interactively usable instruments among others for cutting, histogramming, and fitting.

\subsubsection{4C-fit}
For the exclusive benchmark channels it turned out that especially a 4C-fit of the reconstructed decay tree is a powerful tool to improve the data quality and to suppress background. Kinematic fits allow to test if the measured final state is compatible with the hypothesis of the particular assumed reaction kinematics. 4C-fit uses the fact that the four-momentum of the initial $p\overline{p}$ state is precisely known and should be equal to the sum of 4-momenta of the final state particles.
Least-square fit is performed for the kinematic variables of the measured final state with 4 constraints:
\begin{equation}
\overrightarrow{p}_b=\sum_k \overrightarrow{p}_k,
\end{equation}
\begin{equation}
E_b=\sum_k E_k = \sum_k \sqrt{|\overrightarrow{p}_k|^2+m_k^2}.
\end{equation}
The constraints are taken into account based on the method of Lagrange multipliers. The fit gives for each event the probability that the considered final state is compatible with the beam energy-momentum. 4C-fits and cuts on the results can also be defined with the SimpleComposition tool.

\section{$h_c$ detection. General remarks}\label{sec:MC:hc_general}
According to theoretical predictions and previous experimental
observations \cite{bib:sim:hc_E835, bib:sim:hc_CLEO}, one of the most
promising decay modes for observation of the $h_c$ is its
electromagnetic transition to the ground charmonium state:
\begin{equation}
h_c \rightarrow \eta_{c} + \gamma,
\end{equation}
where the energy of the photon is $E_{\gamma}$ = 503 MeV. The $\eta_c$
can be detected through many exclusive decay channels, neutral
($\eta_c \rightarrow \gamma \gamma$) or hadronic. Several selected decay
modes of the $\eta_c$ with the corresponding branching ratios,
according to \cite{bib:pdg}, are listed in \Reftbl{tab:sim:etac_decays}.

\begin{table}
\begin{center}
\begin{tabular}{|l|c|}
  \hline
  Decay mode & BR \\
  \hline
  $K^{0}_{S} K^{\pm} \pi^{\mp}$ & $(1.9\pm 0.4) \cdot 10^{-2}$ \\
  $K^{0}_{L} K^{\pm} \pi^{\mp}$ & $(1.9\pm 0.4) \cdot 10^{-2}$ \\
  $K^{+} K^{-} \pi^{+} \pi^{-}$ & $(1.5\pm 0.6) \cdot 10^{-2}$ \\
  $\pi^{+} \pi^{-} \pi^{+} \pi^{-}$ & $(1.2\pm 0.3)\cdot 10^{-2}$ \\
  $K^{*}(892) \overline{K}^{*}(892)$ & $(9.2\pm 3.4)\cdot 10^{-3}$ \\
  $\phi \phi$ & $(2.7\pm 0.9) \cdot 10^{-3}$ \\
  $\gamma \gamma$ & $(2.4\pm1.1)\cdot 10^{-4}$ \\
  \hline
\end{tabular}
\caption[Decay modes of $\eta_c$ with the corresponding branching ratios \cite{bib:pdg}]
{Decay modes of $\eta_c$ with the corresponding branching ratios \cite{bib:pdg}}
\label{tab:sim:etac_decays}
\end{center}
\end{table}

The previous observation of $h_c$ was done in neutral decay mode of $\eta_c$ in
$\overline{p} p$ annihilation \cite{bib:sim:hc_E835} or in hadronic decay modes of $\eta_c$ formed along the chain $e^{+} e^{-} \rightarrow \Psi(2S) \rightarrow \pi^{0} h_c, h_c \rightarrow \eta_c \gamma$ \cite{bib:sim:hc_CLEO}. In the case of PANDA experiment, where $h_c$ is produced as a resonance in $\overline{p} p$ annihilation, the detector is capable to study
hadronic final states, hence, the $h_c$ can be observed exclusively whether in one of the hadronic
or neutral ($\eta_c \rightarrow \gamma \gamma$) decay modes of $\eta_c$ (see \Reftbl{tab:sim:etac_decays}).

The expected signal to background ratio is one of the criteria of feasibility to detect $h_c$ experimentally.
For hadronic final states the evaluation of signal to background ratio is even more important due to the fact that off-resonance production of the hadronic final states is much more significant in $\overline{p} p$ annihilation in comparison with $e^{+}e^{-}$.

In order to estimate the signal cross-section we calculate the value of
the Breit-Wigner formula at the resonance energy, $E_{R}$:
\begin{equation}
\label{eq:MC:sigma}
\sigma_{p}=\frac{3\pi}{k^{2}}B_{p\overline{p}}B_{\eta_{c}\gamma},
\end{equation}
where $k^{2}=(E_{R}^{2}-4m_{p}^{2})$ and the $B$'s represent the
branching ratios of $h_c$ into the initial and final states, respectively. The antiproton momentum at resonance $p_R=5.609 \, \gevc$ fits both low- and high-resolution ranges of the HESR synchrotron operation (see \Refsec{sec:panda:hesr}).

Using the value measured by E835 \cite{bib:sim:hc_E835}
$\Gamma_{p\overline{p}}B_{\eta_{c}\gamma}$=10 eV and assuming a value
of 0.5 MeV for the $h_c$ width \cite{bib:pdg} we obtain
$\sigma_{p}=$33 nb.

Physics performance of the PANDA detector was studied in the present Thesis for several selected decay modes of $\eta_c$ ($\gamma \gamma$, $K^0_S K^{+} \pi^{-}$, $K^{*} \overline{K}^{*}$ and $\phi \phi$).

\section{$h_c \rightarrow 3 \gamma$ decay mode}\label{sec:MC:hc3gamma}
\begin{figure}
\begin{center}
\includegraphics[width=0.8\swidth]{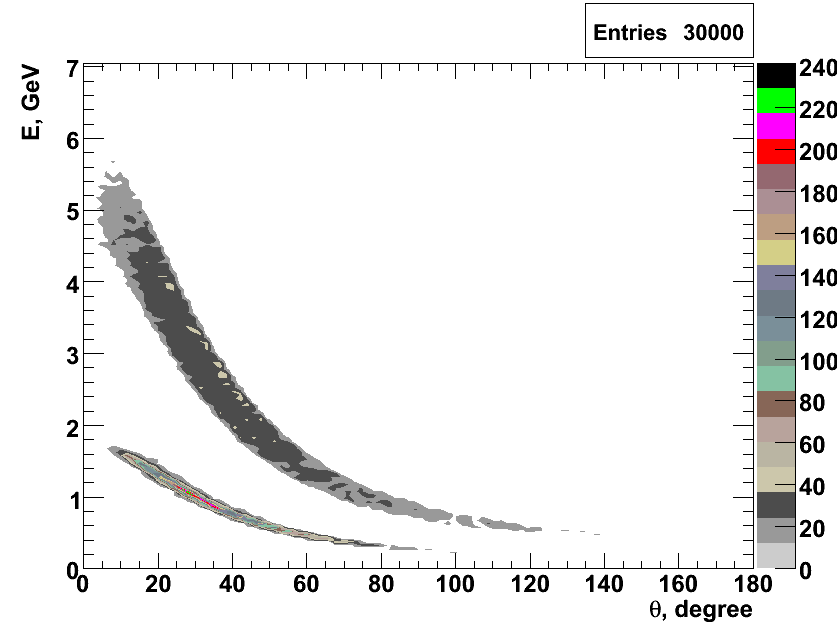}
\caption{Distribution of events on the $\gamma$-ray energy - emission
angle plane for the $h_c\rightarrow \eta_c \gamma\rightarrow 3 \gamma$
decay.}
\label{fig:sim:hc_3gamma_egamma_theta}
\end{center}
\end{figure}

This decay mode was observed at Fermilab by the experiment E835 \cite{bib:sim:hc_E835}.
It is characterized by a fairly clean final state, but the low value of
the $\eta_c\rightarrow\gamma \gamma$ branching ratio ($4.3\cdot10^{-4}$)
(see \Reftbl{tab:sim:etac_decays}) results in a relatively low event
rate in comparison with the hadronic decay modes of $\eta_c$. The
energies of $\gamma$'s produced in this decay mode are plotted in
\Reffig{fig:sim:hc_3gamma_egamma_theta} as a function of the polar angle
in the laboratory system. The lower band corresponds to $\gamma$'s
from the radiative transition $h_c\rightarrow \eta_c \gamma$. The upper
band of highly energetic photons is due to the
$\eta_c \rightarrow \gamma \gamma$ decay. One may note, that
observation of both decays, requires registration of $\gamma$'s in the
energy range from 150 MeV up to 5.5 GeV.

\subsection{Background considerations}\label{sec:mc:hc_3gamma_bkgr}
The observation of $h_c$ in 3$\gamma$ final state in proton-antiproton
collisions by E835 \cite{bib:sim:hc_E835} has demonstrated that it is
feasible to achieve sufficient background suppression for this final
state, however differences with PANDA in the detector setup require
estimation of the degree of background suppression also for the planned
experiments. The main contributors to the background for the 3$\gamma$
final state are $\gamma$'s from the $\pi^{0}$, $\eta$ and $\eta'$ decay
in $\gamma \gamma$ decay modes: the loss of one or more $\gamma$'s
outside the detector acceptance or below the energy threshold of the
electromagnetic calorimeter (EMC), will result in a 3$\gamma$ final
state. The background channels considered in this analysis are listed in
\Reftbl{tab:sim:hc_3gamma_bkgr} with the corresponding cross-sections
measured by E760 and E835 \cite{bib:sim:hc_E835} integrated over the
angular range $|\cos(\theta)|<0.6$.

\begin{table}
\begin{center}
\begin{tabular}{|l|c|}
  \hline
  Channel & Cross-section, nb \\
  \hline
  $p\overline{p}\rightarrow\pi^{0}\pi^{0}$ & 31.4 \\
  $p\overline{p}\rightarrow\pi^{0}\gamma$ & 1.4 \\
  $p\overline{p}\rightarrow\pi^{0}\eta$ & 33.6 \\
  $p\overline{p}\rightarrow\eta\eta$ & 34.0 \\
  $p\overline{p}\rightarrow \pi^{0}\eta'$ & 50.0 \\
  \hline
\end{tabular}
\caption{The main background contributors to $h_c \rightarrow 3 \gamma$
with the corresponding cross-sections \cite{bib:Andreotti:2005zr, bib:Armstrong:1997gv}
integrated over the range of angles $|\cos(\theta)|<0.6$.}
\label{tab:sim:hc_3gamma_bkgr}
\end{center}
\end{table}

\begin{figure}
\begin{center}
\includegraphics[width=0.8\columnwidth]{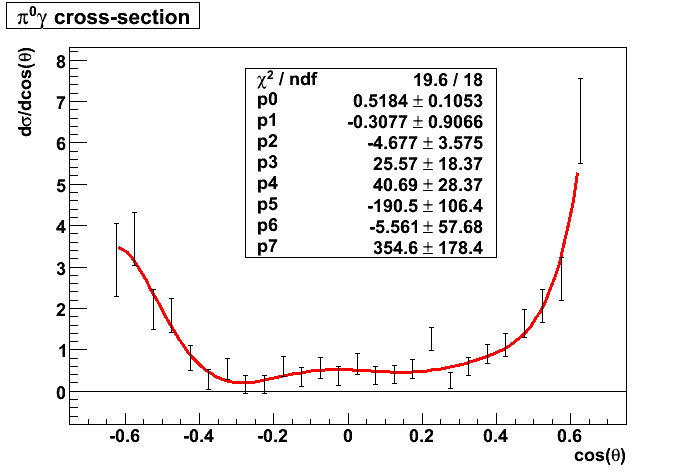}
\caption{Angular dependence of the $\pi^{0}\gamma$ cross-sections
parametrized (solid line) with a sum of powers in $\cos(\theta)$ used in
Monte-Carlo simulation. The coefficients of the fit together with their
errors are indicated in the inset.}
\label{fig:sim:hc_pi0gamma_crosssection}
\end{center}
\end{figure}

\begin{figure}
\begin{center}
\includegraphics[width=0.8\swidth]{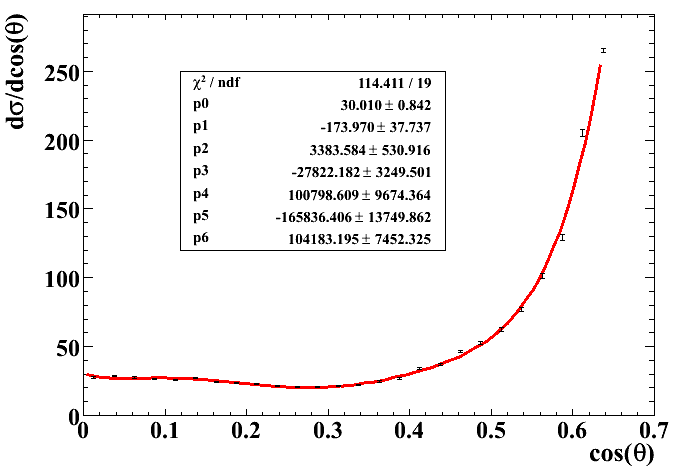}
\caption{Angular dependence of the $\pi^{0}\pi^{0}$ cross-sections
parametrized (solid line) with a sum of powers in $\cos(\theta)$ used in
Monte-Carlo simulation. The coefficients of the fit together with their
errors are indicated in the inset.}
\label{fig:sim:hc_pi0pi0_crosssection}
\end{center}
\end{figure}

The angular dependence for all the studied background channels is
strongly peaked in the forward and backward direction, which is typical
for two and three meson production in antiproton-proton annihilations at
energies of interest. For the Monte-Carlo study the angular dependence
of the cross-sections was parameterized with 6th or 7th order
polynomials in $\cos(\theta)$. As an example we show in
\Reffig{fig:sim:hc_pi0gamma_crosssection} the angular distribution for
the $\pi^{0}\gamma$ channel and in \Reffig{fig:sim:hc_pi0pi0_crosssection}
for the channel $\pi^{0}\pi^{0}$, respectively.

\begin{figure}
\begin{center}
\includegraphics[width=0.8\swidth]{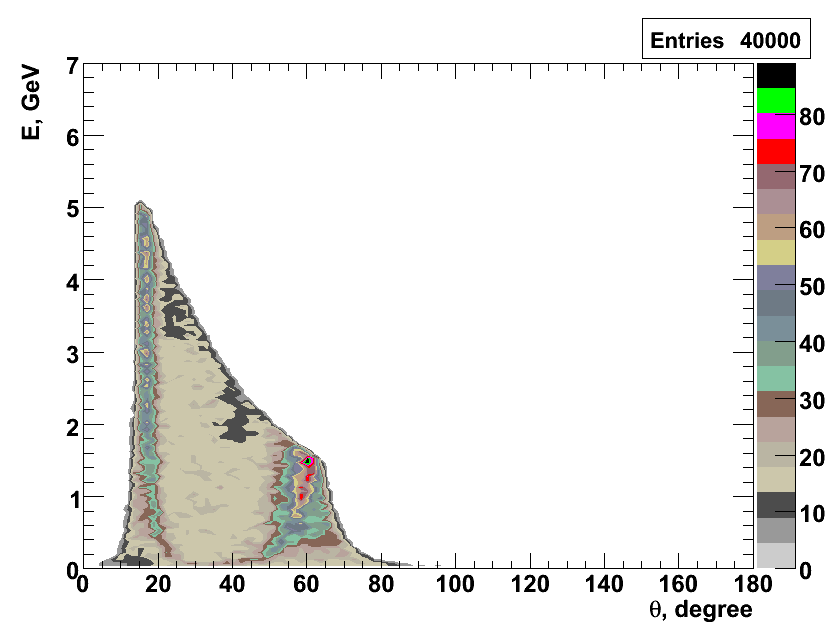}
\caption{Distribution of events on the energy-emission angle plane for
the $p \overline{p} \rightarrow \pi^{0}\pi^{0}$ background channel.}
\label{fig:sim:pi0pi0_egamma_theta}
\end{center}
\end{figure}

The distribution of events in $\gamma$-ray energy emission angle plane
for the background channel
$p \overline{p} \rightarrow \pi^{0}\pi^{0}$ is shown in
\Reffig{fig:sim:pi0pi0_egamma_theta}. In contrast to the signal (see
\Reffig{fig:sim:hc_3gamma_egamma_theta}) events cover the plane starting
from zero energy. The forward cut-off in
\Reffig{fig:sim:pi0pi0_egamma_theta} is caused by event generation
over limited range of angles dictated by the experimental data available
only in the range $|\cos(\theta)|<0.6$.

\subsection {Event selection and background suppression}\label{sec:MC:hc3gamma_eventselection}
The number of Monte-Carlo events used for this analysis for the signal and
all the background channels is collected in
\Reftbl{tab:sim:hc_3gamma_bkgr_nevts}.

\begin{table}
\begin{center}
\begin{tabular}{|l|c|}
  \hline
  Channel & Number of events\\
  \hline
  $p\overline{p}\rightarrow h_c \rightarrow 3 \gamma$ & 20 k \\
  $p\overline{p}\rightarrow\pi^{0}\pi^{0}$ & 1.3\,M \\
  $p\overline{p}\rightarrow\pi^{0}\gamma$ & 100 k \\
  $p\overline{p}\rightarrow\pi^{0}\eta$ & 1.3\,M \\
  $p\overline{p}\rightarrow\eta\eta$ & 1.3\,M \\
  $p\overline{p}\rightarrow \pi^{0}\eta'$ & 100 k \\
  \hline
\end{tabular}
\caption{Number of generated Monte-Carlo events used for different background channels in the $h_c \rightarrow 3 \gamma$
analysis.}
\label{tab:sim:hc_3gamma_bkgr_nevts}
\end{center}
\end{table}

\begin{figure}
\begin{center}
\includegraphics[width=0.8\swidth]{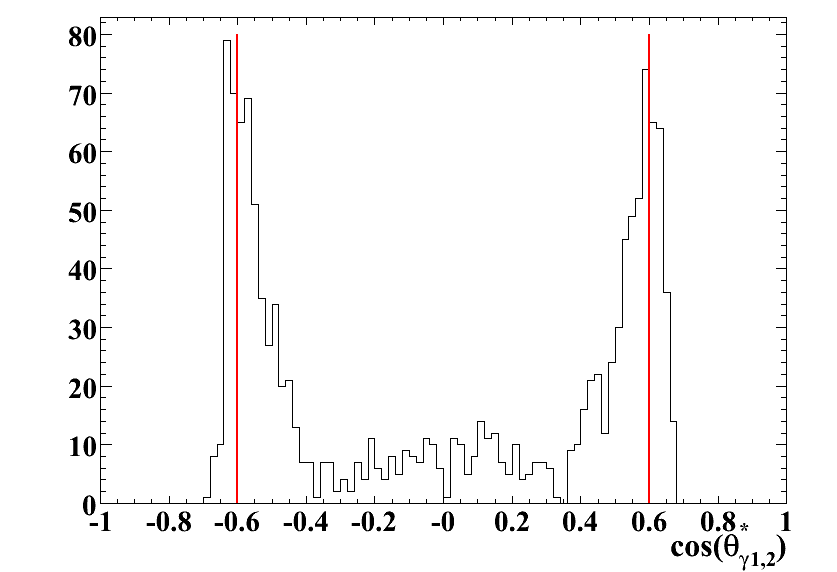}
\caption{Distribution of the reconstructed $\cos{\theta}$ of the $\gamma$'s assigned to $\eta_c\rightarrow \gamma \gamma$ decay in CM system for the $p \overline{p} \rightarrow \pi^{0}\pi^{0}$ background reaction.}
\label{fig:sim:pi0pi0_costheta12}
\end{center}
\end{figure}

\begin{figure}
\begin{center}
\includegraphics[width=0.8\swidth]{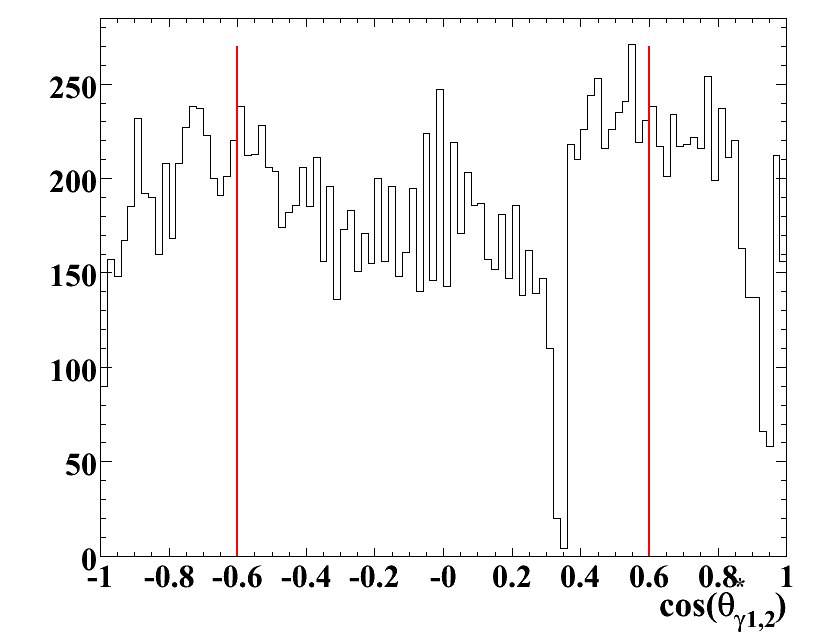}
\caption{Distribution of the reconstructed $\cos{\theta}$ of the $\gamma$'s assigned to $\eta_c\rightarrow \gamma \gamma$ in CM system for the $h_c\rightarrow \eta_c \gamma$ decay.}
\label{fig:sim:hc_3gamma_costheta12}
\end{center}
\end{figure}

The event selection is done in the following steps:
\begin{enumerate}
\item
An $\eta_c$ candidate is formed by pairing two $\gamma$'s with an
invariant mass in the window [2.6; 3.2] GeV. The third $\gamma$ is added
to this pair to form the $h_c$ candidate.
\item
A 4C-fit to beam energy-momentum is applied to the $h_c$ candidate and
the information on the $h_c$ and the updated information on the daughter
$\gamma$'s is stored into the root ntuple.
\item
The following cuts are applied at the ntuple level to suppress
background:
\begin{enumerate}
\item
Events with 3$\gamma$'s were selected. This cut keeps 47$\%$ of the
initial events,
\item
Cut on the confidence level of the 4C-fit: $CL > 10^{-4}$,
\item
Cut on the CM energy of the $\gamma$ from the
$h_c \rightarrow \eta_c \gamma$ radiative transition:
0.4 GeV $\leq E_{\gamma} \leq$ 0.6 GeV.
\item
Angular cut $|\cos(\theta)|<0.6$, to reject the background which is
strongly peaked in the forward and backward directions. The
$\cos(\theta)$ distributions for the background channel
($\pi^{0} \pi^{0}$) and for the signal are shown in
\Reffig{fig:sim:pi0pi0_costheta12} and
\Reffig{fig:sim:hc_3gamma_costheta12}, respectively,
\item
The cut on invariant mass for combinations $M(\gamma_1,\gamma_3)>$1.0
GeV and $M(\gamma_2,\gamma_3)>1.0 \;\gev$ (the value of the cut is
determined by the $\eta'$ mass).
\end{enumerate}
\end{enumerate}
An impact of the applied selection criteria on the signal and background
events is discussed below.

\begin{figure}
\begin{center}
\includegraphics[width=0.8\swidth]{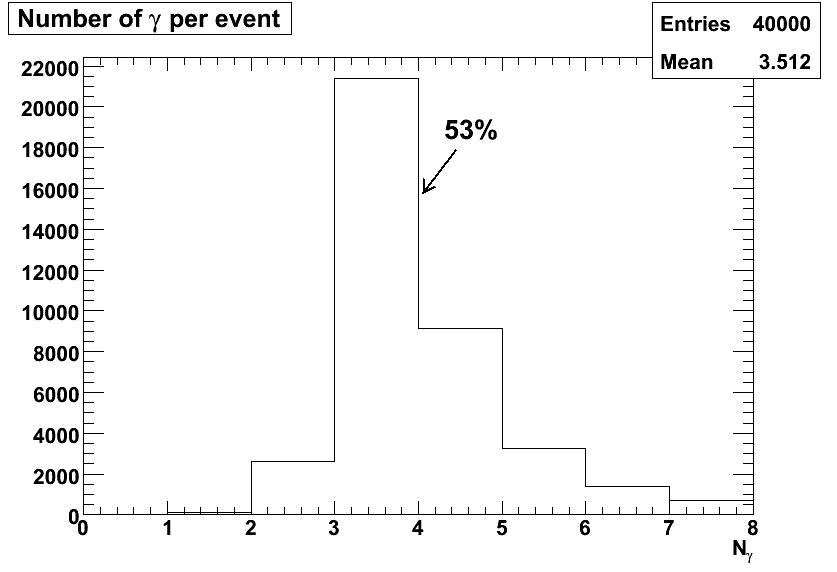}
\caption{Multiplicity of the reconstructed $\gamma$'s in the
$h_c\rightarrow \eta_c \gamma\rightarrow 3 \gamma$ decay.}
\label{fig:sim:hc_3gamma_gammamultiplicity}
\end{center}
\end{figure}

\begin{figure}
\begin{center}
\includegraphics[width=0.8\swidth]{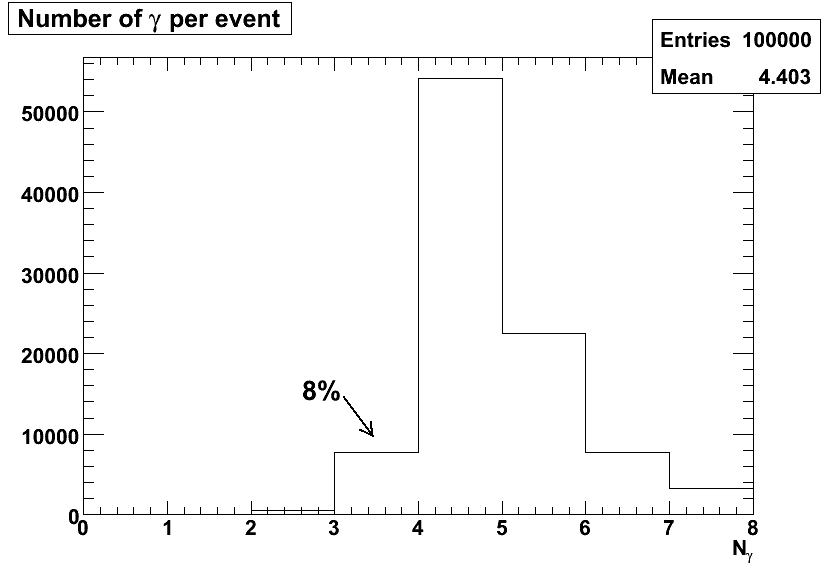}
\caption{Multiplicity of the reconstructed $\gamma$'s ($\gamma_3$) for the
$p \overline{p}\rightarrow \pi^{0} \pi^{0}$ background reaction.}
\label{fig:sim:pi0pi0_gammamultiplicity}
\end{center}
\end{figure}

\Reffig{fig:sim:hc_3gamma_gammamultiplicity} demonstrates the
distribution of $\gamma$'s in multiplicity for the signal events;
53$\%$ of the events have exactly 3 reconstructed neutral particle
candidates. A small fraction of events has less than 3 neutral
candidates due to detector acceptance. The events with $N_{\gamma}>3$
are caused by the electromagnetic split-offs. For
$p\overline{p}\rightarrow \pi^{0} \pi^{0}$ the distribution of $\gamma$'s
in multiplicity is shown in \Reffig{fig:sim:pi0pi0_gammamultiplicity};
$8\%$ of these events have three neutral candidates. One may conclude,
taking into account the ratio of signal to background cross-sections,
that the number of background events will exceed the number of expected
3$\gamma$ events for the signal. A similar situation is encountered for
all the remaining background channels listed in
\Reftbl{tab:sim:hc_3gamma_bkgr}, i.e. they have a significant fraction
of 3$\gamma$ events.

\begin{figure}
\begin{center}
\includegraphics[width=0.8\swidth]{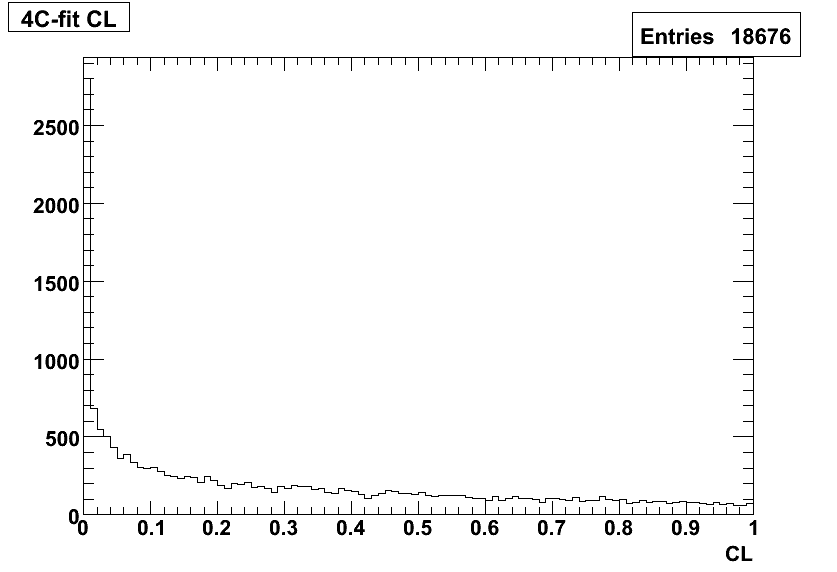}
\caption{Distribution of confidence level of 4C-fit for $h_c\rightarrow \eta_c \gamma\rightarrow 3 \gamma$.}
\label{fig:sim:hc_3gamma_CL}
\end{center}
\end{figure}

\begin{figure}
\begin{center}
\includegraphics[width=0.8\swidth]{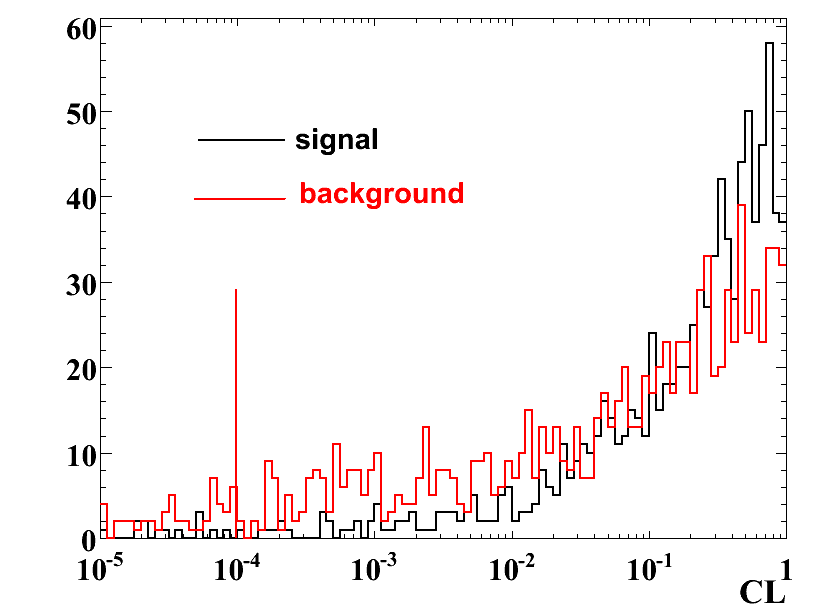}
\caption{Distribution of confidence level of the 4C-fit for $h_c\rightarrow \eta_c \gamma\rightarrow 3 \gamma$ and $p \overline{p} \rightarrow \pi^{0}\pi^{0}$
background events on a logarithmic scale of the abscissa. The vertical line indicates the assumed position of a cut on $CL>10^{-4}$ (see \Refsec{sec:MC:hc3gamma_eventselection}).}
\label{fig:sim:hc_3gamma_CL_log}
\end{center}
\end{figure}

The distribution of confidence level for the 4C-fit to beam energy-momentum
for the $h_c\rightarrow \eta_c \gamma\rightarrow 3 \gamma$ events is
presented in \Reffig{fig:sim:hc_3gamma_CL} on the linear and in
\Reffig{fig:sim:hc_3gamma_CL_log} on the logarithmic scale of the abscissa in order to expand the region with $CL\lesssim \, 0.1$. The cut is applied at the
level $CL>10^{-4}$, as indicated above in the step 3b. Lower values of
CL correspond to higher $\chi^{2}$ of the fit. The CL distribution is
almost flat in the range [0.1; 1.0] for the signal events, whereas for
the background channels the probability of higher CL values is lower in
general as it can be seen from \Reffig{fig:sim:hc_3gamma_CL_log} for the
$p \overline{p}\rightarrow \pi^{0} \pi^{0}$ background channel.
For many background events the 4C-fit does not converge. Such
events get assigned negative CL values and are removed by the
$CL>10^{-4}$ cut from subsequent analysis.

\begin{figure}
\begin{center}
\includegraphics[width=0.8\swidth]{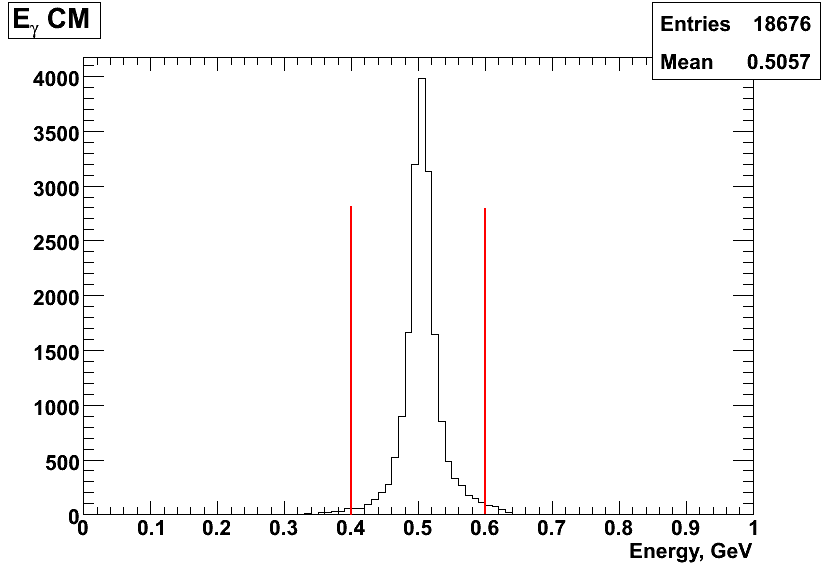}
\caption{Distribution of reconstructed energy of the $\gamma$ ($\gamma_3$) from the $h_c\rightarrow \eta_c \gamma$ radiative transition. Vertical lines indicates the window imposed to select this transition.}
\label{fig:sim:hc_3gamma_egamma}
\end{center}
\end{figure}

\begin{figure}
\begin{center}
\includegraphics[width=0.8\swidth]{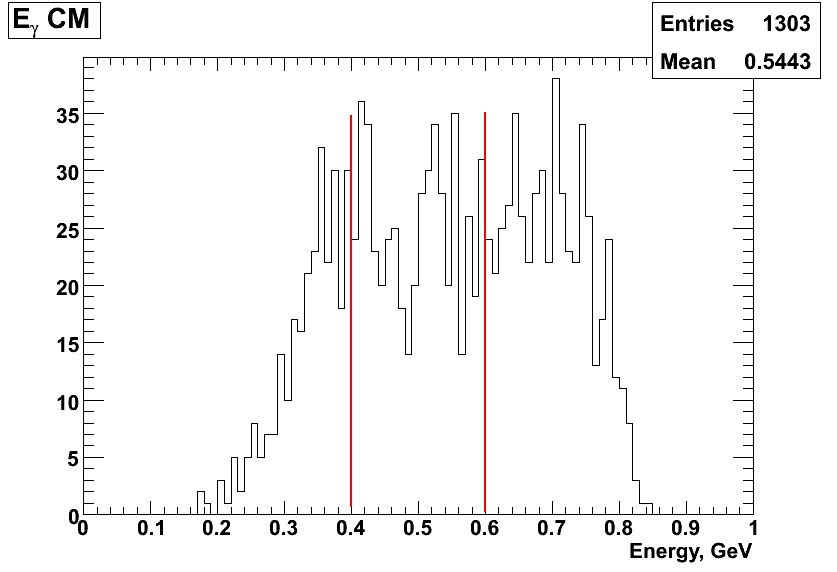}
\caption{Distribution of reconstructed energy of the $\gamma_3$ emitted in the background channel
$p \overline{p} \rightarrow \pi^{0}\pi^{0}$. Events between the vertical lines would be erroneously assigned to $h_c\rightarrow \eta_c \gamma$ if not eliminated by other selection criteria (see \Refsec{sec:MC:hc3gamma_eventselection}).}
\label{fig:sim:pi0pi0_egamma}
\end{center}
\end{figure}

The $\gamma$'s which are combined to $\eta_c$ are labeled 1 and 2,
whereas number 3 is assigned to the $\gamma$ emitted in the radiative
transition $h_c \rightarrow \eta_c \gamma$. The distribution
of reconstructed energies of the $3^{rd}$ $\gamma$ in the $p \overline{p}$
CM system is presented in \Reffig{fig:sim:hc_3gamma_egamma}. The expected
energy of the $\gamma$ emitted in the $h_c \rightarrow \eta_c \gamma$
transition should be 503 MeV according to the compilation \cite{bib:pdg}.
The reconstructed distribution is peaked close to this value. The range
[0.4; 0.6] GeV between the two vertical lines defines the limits of the
effect (cut 3c). The corresponding distribution for $\gamma$'s emitted
in $p \overline{p}\rightarrow \pi^{0} \pi^{0}$ is presented in
\Reffig{fig:sim:pi0pi0_egamma}; 40 $\%$ of the events for this
particular background channel pass the latter selection cut.

\begin{figure}
\begin{center}
\includegraphics[width=0.8\swidth]{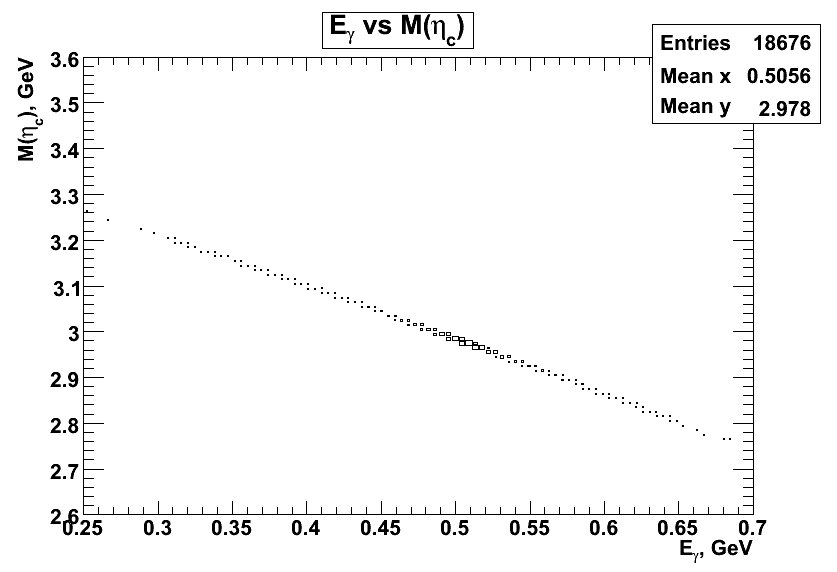}
\caption{Distribution of reconstructed energy of $\gamma$ ($\gamma_3$) versus $\eta_c$ invariant mass for $h_c\rightarrow \eta_c \gamma$.}
\label{fig:sim:hc_3gamma_egamma_metac}
\end{center}
\end{figure}

\begin{figure}
\begin{center}
\includegraphics[width=0.8\swidth]{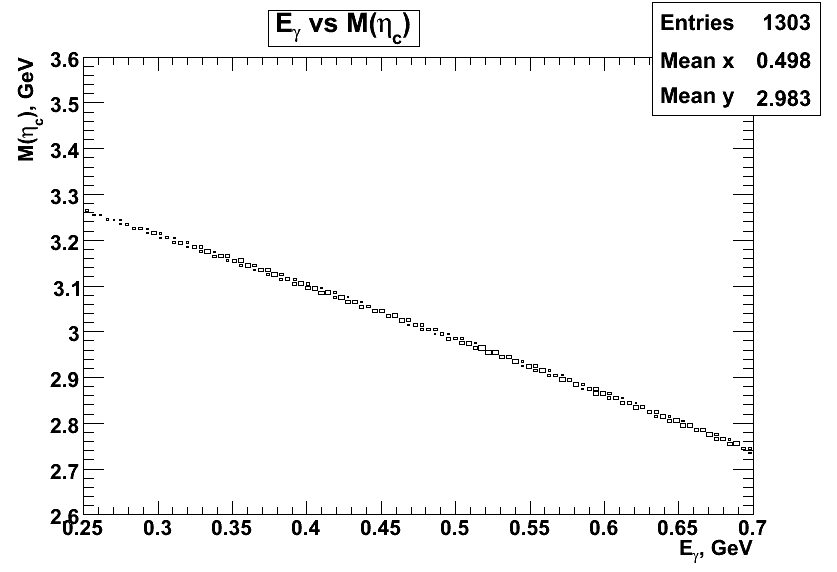}
\caption{Distribution of reconstructed energy of $\gamma$ versus $\eta_c$ invariant mass for the $p \overline{p} \rightarrow \pi^{0}\pi^{0}$ background reaction.}
\label{fig:sim:pi0pi0_egamma_metac}
\end{center}
\end{figure}

The distribution of events in the $E_{\gamma 3}$-invariant mass $m(\eta_c)$
plane is presented in \Reffig{fig:sim:hc_3gamma_egamma_metac}
for the $h_c \rightarrow \eta_c \gamma$ channel. All the events are
concentrated along a line which is defined by the applied 4C-fit,
consequently, after the fit the $E_{\gamma 3}$ and $m(\eta_c)$ variables
are no longer independent. The size of boxes is proportional to the
number of events in a certain $E_{\gamma 3}$, $m(\eta_c)$ range. One may
note that events are mostly concentrated around the point
$E_{\gamma 3}$=0.503 GeV, $m(\eta_c)$=2.98 GeV. For the case of
$p \overline{p}\rightarrow \pi^{0} \pi^{0}$ background reaction
(see \Reffig{fig:sim:pi0pi0_egamma_metac}) events are more uniformly
distributed along the same line.

\begin{figure}
\begin{center}
\includegraphics[width=0.8\swidth]{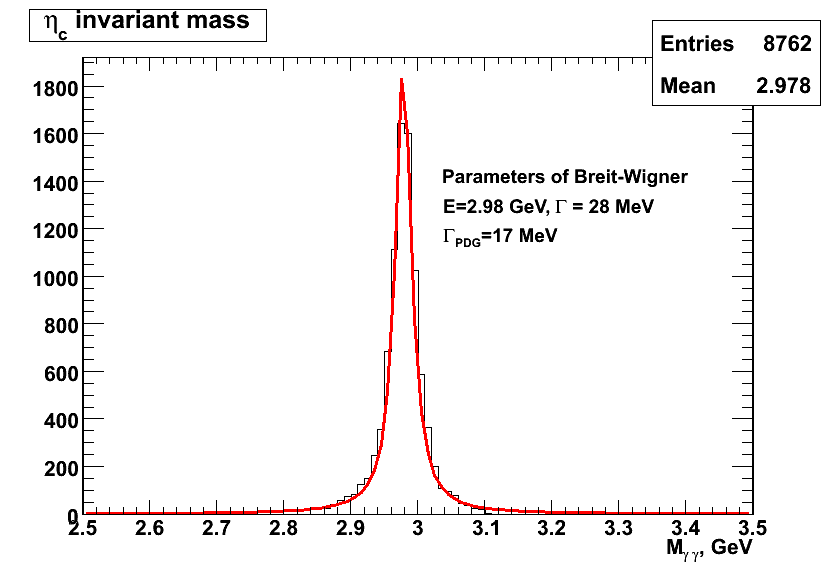}
\caption{Reconstructed invariant mass of $\eta_c$. Solid line is the Breit-Wigner fit with the indicated parameters.}
\label{fig:sim:hc_3gamma_metac}
\end{center}
\end{figure}

\Reffig{fig:sim:hc_3gamma_metac} presents the reconstructed
distribution of the $\eta_{c}$ invariant mass (projection of \Reffig{fig:sim:hc_3gamma_egamma_metac} on the $M(\eta_c)$ axis). A Breit-Wigner formula,
fitted to the simulated data, is superimposed on the figure. The fitted
width parameter $\Gamma$ = 28 MeV is slightly larger than the PDG value
$\Gamma_{PDG}$=17 MeV, which illustrates the influence of instrumental
resolution, $\Delta_{exp} \approx$ 22 MeV, expected with the PANDA-EMC.

It was mentioned in \Refsec{sec:mc:hc_3gamma_bkgr}, that the background cross-sections and the
corresponding intensities of $\gamma$'s are peaked in the forward and
backward directions. The distribution of the reconstructed $\gamma_{1,2}$
in $\cos(\theta)$ in the CM system is shown in
\Reffig{fig:sim:pi0pi0_costheta12} for the
$p \overline{p}\rightarrow \pi^{0} \pi^{0}$ channel. An abrupt drop in
intensity above $|\cos(\theta)|>0.6$ reflects the limits of the
generated events, because of the missing experimental information (see
the remarks related to \Reffig{fig:sim:hc_pi0pi0_crosssection}).
Vertical lines at $|\cos(\theta)|=0.6$ indicate the width of the window.
In case of the signal (\Reffig{fig:sim:hc_3gamma_costheta12}) events
have rather flat distribution in $\cos(\theta)$, therefore applying a
cut eliminating forward and backward angles is a priori expected to
improve the signal to background ratio. The two dips seen in \Reffig{fig:sim:hc_3gamma_costheta12} are related to the
EMC installation: the one at $\approx$ 0.35 reflects irregularity between
the barrel and the forward endcap, the other one at $\approx$ 0.95
is at the angle of transition between the forward endcap and the shashlyk calorimeter in the Forward Spectrometer.

\begin{figure}
\begin{center}
\includegraphics[width=0.8\swidth]{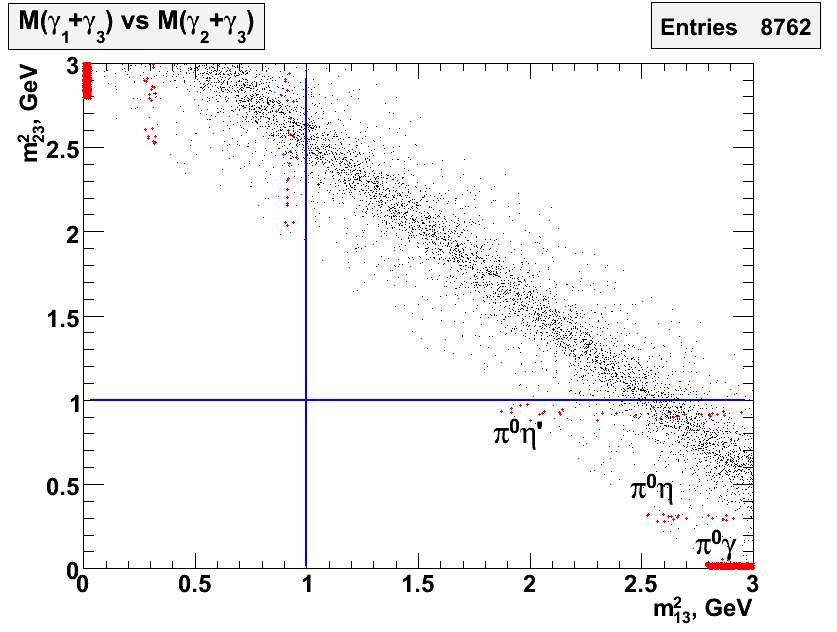}
\caption{Dalitz plot for the $p \overline{p} \rightarrow \gamma \gamma \gamma$ events. Vertical lines denote the positions of cuts to eliminate $\pi^0 \gamma$, $\pi^0 \eta$ and $\pi^0 \eta'$ events originating from the background reactions listed in \Reftbl{tab:sim:hc_3gamma_eff}.}
\label{fig:sim:m12m23}
\end{center}
\end{figure}

\Reffig{fig:sim:m12m23} presents the Dalitz plot for 3$\gamma$ events,
which pass all the previously mentioned cuts. The signal events are
concentrated along the diagonal extending from the top left to the
bottom right. Background events from different reactions are marked as
red dots. The blue lines mark the selection cuts, i.e.
$M(\gamma_1,\gamma_3) > 1.0 GeV$ and $M(\gamma_2,\gamma_3)>1.0 GeV$. The
value of 1 GeV is chosen to eliminate events originating from the
$p\overline{p}\rightarrow \pi^{0}\eta'$ background channel, i.e. by
the mass of $\eta'$. None of the background events passes this
selection cut and only 37$\%$ of the signal events is able to survive,
which results in the total signal efficiency of about 8$\%$.

\subsection {Results}
The selection efficiencies for different cuts are collected in \Reftbl{tab:sim:hc_3gamma_eff}. Efficiencies are cumulative, i.e. applied
one after another. Taking into account the signal cross-section
$\sigma_{p \overline{p} \rightarrow h_c}$=33 nb at resonance, branching
ratio $BR(\eta_c\rightarrow \gamma \gamma)=4.3 \cdot 10^{-4}$ and
background cross-sections from \Reftbl{tab:sim:hc_3gamma_bkgr} one gets
the expected signal to background ratios as listed in
\Reftbl{tab:sim:hc_3gamma_sb}. The expected event rate for running
in high luminosity mode, $L = 2 \cdot 10^{32} cm^{-2}s^{-1}$,
is 20 events/day, and for high resolution mode with
$L =2 \cdot 10^{31} cm^{-2} s^{-1}$, 2.0 events/day, respectively.

\begin{table}
\begin{center}
\begin{tabular}{|l|c|c|c|c|c|c|}
  \hline
  Cut & $h_c$ &$\pi^{0}\gamma$ & $\pi^{0}\pi^{0}$ & $\pi^{0}\eta$ & $\eta \eta$ & $\pi^{0}\eta'$\\
  \hline
  preselection & 0.70 & 0.43 & 0.14 & $8.2\cdot 10^{-2}$ & $4.0 \cdot 10^{-2}$ & $8.5 \cdot 10^{-2}$ \\
  3 $\gamma$ & 0.47 & 0.31 & $1.3 \cdot 10^{-2}$ & $7.5 \cdot 10^{-3}$ & $2.7 \cdot 10^{-3}$ & $8.7 \cdot 10^{-3}$ \\
  $CL>10^{-4}$& 0.44 & 0.30 & $9.9 \cdot 10^{-3}$ & $4.9 \cdot 10^{-3}$ & $7.2 \cdot 10^{-4}$ & $5.7 \cdot 10^{-3}$ \\
  $E_{\gamma}$ [0.4;0.6] GeV& 0.43 & 0.12 & $3.9 \cdot 10^{-3}$ & $2.0 \cdot 10^{-3}$ & $2.8 \cdot 10^{-4}$ & $2.3 \cdot 10^{-3}$ \\
  $|\cos(\theta)|<0.6$ & 0.22 & $9.2 \cdot 10^{-2}$ & $2.7 \cdot 10^{-3}$ & $1.1 \cdot 10^{-3}$ & $7.0 \cdot 10^{-5}$ & $7.5 \cdot 10 ^{-4}$ \\
  $m_{13}^{2}, m_{23}^{2}> 1.0 GeV$  & $8.1 \cdot 10^{-2}$ & 0 & 0 & 0& 0 & 0 \\
  \hline
\end{tabular}
\caption{Selection efficiencies for $h_c \rightarrow 3 \gamma$ and the indicated background reactions.}
\label{tab:sim:hc_3gamma_eff}
\end{center}
\end{table}

\begin{table}
\begin{center}
\begin{tabular}{|l|c|}
  \hline
  Channel & S/B ratio\\
  \hline
  $p\overline{p}\rightarrow\pi^{0}\pi^{0}$ & $>$ 94 \\
  $p\overline{p}\rightarrow\pi^{0}\gamma$ & $>$ 164 \\
  $p\overline{p}\rightarrow\pi^{0}\eta$ & $>$ 88 \\
  $p\overline{p}\rightarrow\eta\eta$ & $>$ 87 \\
  $p\overline{p}\rightarrow \pi^{0}\eta'$ & $>$ 250 \\
  \hline
\end{tabular}
\caption[Signal to background ratio for $h_c \rightarrow 3 \gamma$ and different background channels.]
{Signal to background ratio for $h_c \rightarrow 3 \gamma$ and different background channels.}
\label{tab:sim:hc_3gamma_sb}
\end{center}
\end{table}

\section{$h_c \rightarrow \phi \phi \gamma$ decay mode}\label{sec:MC:hc_phiphi}
As a benchmark channel with a hadronic decay mode of the $\eta_c$ the $\phi \phi$ final state was studied with $BR=2.6 \cdot 10^{-3}$. The $\phi$ was detected through the decay $\phi \rightarrow K^{+} K^{-}$, with $BR=$0.49 \cite{bib:pdg}.

\subsection{Background considerations}
For the exclusive decay mode considered in this study:
\begin{center}
$\overline{p} p \rightarrow h_c \rightarrow \eta_c \gamma \rightarrow \phi \phi \gamma \rightarrow K^{+} K^{-} K^{+} K^{-} \gamma$,
\end{center}
the following three reactions are considered as the main contributors to
the background:\\
\begin{enumerate}
\item
$\overline{p} p \rightarrow K^{+} K^{-} K^{+} K^{-} \pi^{0}$,
\item
$\overline{p} p \rightarrow \phi K^{+} K^{-} \pi^{0}$,
\item
$\overline{p} p \rightarrow \phi \phi \pi^{0}$.
\setcounter{saveenumi}{\theenumi}
\end{enumerate}

With one photon from the $\pi^{0}$ decay left undetected, these reactions
have the same final state particles as the studied $h_c$ decay.\\

Additional possible sources of background are:
\begin{enumerate}
\setcounter{enumi}{\thesaveenumi}
\item
$\overline{p} p \rightarrow \Delta^{++}(1232) \overline{\Delta}^{--}(1232) \pi^{0} \rightarrow p \pi^{+} \overline{p} \pi^{-}
\pi^{0}$,
\item
$\overline{p} p \rightarrow K^{+} K^{-} \pi^{+} \pi^{-} \pi^{0}$,
\item
$\overline{p} p \rightarrow \pi^{+} \pi^{-} \pi^{+} \pi^{-} \pi^{0}$.
\end{enumerate}

The reaction \#4 has been considered as one of the main sources of
background in the Fermilab experiment E835 \cite{bib:sim:hc_E835},
which studied the same decay mode of $h_c$. This reaction has similar
kinematics as the reaction of interest, i.e. 4 charged tracks in the
forward direction. PANDA has significant advantages for discrimination
of this source of background in comparison with E835, being equipped in
magnetic analysis and PID tools. Nevertheless, an attempt was made to
estimate explicitly the suppression factor for the reaction \#4.

The last two reactions could contribute to background because of pion
misidentification as kaons. In addition to PID such events can be
effectively suppressed with the 4C-fit to beam energy-momentum. However,
because of large cross-sections for these reactions, it seemed necessary
to estimate the signal to background ratio for these channels in a
detailed simulation.

\begin{figure}
\begin{center}
\includegraphics[width=0.8\linewidth]{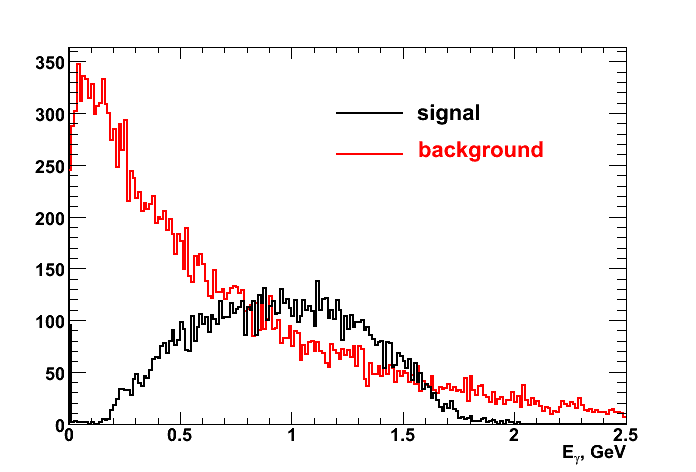}
\caption[Energy distribution of the $\gamma$ for the $\overline{p} p \rightarrow h_c \rightarrow \eta_c \gamma$ and $\overline{p} p \rightarrow \phi \phi \pi^{0}$ reactions.]
{Energy distribution of the $\gamma$ for the $\overline{p} p \rightarrow h_c \rightarrow \eta_c \gamma$ and $\overline{p} p \rightarrow \phi \phi \pi^{0}$ reactions.}
\label{fig:sim:hc_e_gamma}
\end{center}
\end{figure}

\Reffig{fig:sim:hc_e_gamma} presents the distributions of $\gamma$'s in
the laboratory energy for the signal and the background channel \#3.

The energy range of $\gamma$'s from the $h_c$ decay is
$[0.15; 2.0]$ GeV. The distribution for the background has
different features. First, it is wider distributed extending beyond
this energy range. Second, it has an increasing tendency towards zero
energy. If we want to recover $\pi^{0}$'s to separate the signal from
background, we should succeed to lower photon detection threshold as
much as possible. Moreover, when one $\gamma$ from the $\pi^{0}$ decay
has low energy, the other $\gamma$ together with the charged hadrons
has total momentum closer to the total momentum of the initial
$\overline{p} p$ system. Such events pass cuts on the fit probability
of the 4C-fit; this situation increases the relative importance of having low
energy threshold in $\gamma$-ray detection in order to apply a veto on $\pi^{0}$'s in an event
and correspondingly to suppress background due to $\pi^{0}$'s.

There are no experimental measurements, to our best knowledge, of the
cross-sections for the first three background reactions, which are
supposed to be the main contributors to background. The only way to estimate
their cross-sections was found to use the DPM (Dual Parton Model) event
generator \cite{bib:sim:DPM}; $2 \cdot 10^7$ events were generated with DPM at
the $\overline{p}$ beam momentum $p_z=5.609\,  \gevc$, which corresponds to the studied
$h_c$ resonance. The corresponding numbers of events are 115 and 12 for
the first two background channels. No events for the
$\overline{p} p \rightarrow \phi \phi \pi^{0}$ reaction were observed in the generated sample.
With the total $\overline{p} p$ cross-section at this beam momentum of
60 $mb$, the cross-sections for the corresponding background channels
are estimated at 360 $nb$, 36 $nb$ and below 6 $nb$, respectively. For
the $\overline{p} p \rightarrow \Delta^{++}(1232) \overline{\Delta}^{--}(1232) \pi^{0}$
and $\overline{p} p \rightarrow \pi^{+} \pi^{-} \pi^{+} \pi^{-} \pi^{0}$
background channels the cross-section are known from the measurements at
the $\overline{p}$ beam momentum $p=5.7\, \gevc$ \cite{bib:Atherton1974ma}
and \cite{bib:Armenteros:1969}. The values are equal $\sigma=530 \,\mu b$
and $\sigma=750 \,\mu b$, respectively. For the purpose of estimation of
the signal to background ratio it was assumed that the phase space engaged in
the final state is uniformly populated. An estimate of the
$\overline{p} p \rightarrow K^{+} K^{-} \pi^{+} \pi^{-} \pi^{0}$
cross-section was done by extrapolating from a lower energy according to
the total inelastic cross-section. The result is $\sigma=30 \,\mu b$.

\subsection{Analysed events and selection criteria}
The numbers of analysed events are listed in \Reftbl{tab:perf:hc_nevts}. For the $\overline{p} p \rightarrow K^{+} K^{-} \pi^{+} \pi^{-} \pi^{0}$ channel 15 million out of the 20 million of the events were simulated with a filter applied on the invariant mass of the pair of kaons. The events with $m(K^{+} K^{-})$ in the range [0.95; 1.2] \gev were selected. The efficiency of the filter is 29.9 \percent, i.e. only 29.9 \percent of the events produced by the event generator are simulated with GEANT4 transport code with the following event reconstruction. This gives the effective number of simulated events $\sim$ 55 million.\\

\begin{table}
\begin{center}
\begin{tabular}{|l|c|}
  \hline
  Channel & N of events \\
    \hline
  $\overline{p} p \rightarrow h_c \rightarrow \phi \phi \gamma$ & 20 k \\
  $\overline{p} p \rightarrow K^{+} K^{-} K^{+} K^{-} \pi^{0}$ & 6.2\,M \\
  $\overline{p} p \rightarrow \phi K^{+} K^{-} \pi^{0}$ & 200 k\\
  $\overline{p} p \rightarrow \phi \phi \pi^{0}$ & 4.2\,M\\
  $\overline{p} p \rightarrow \Delta^{++}(1232) \overline{\Delta}^{--}(1232) \pi^{0} $ & 100 k \\
  $\overline{p} p \rightarrow K^{+} K^{-} \pi^{+} \pi^{-} \pi^{0}$ & 5 M + 15 M \\
  $\overline{p} p \rightarrow \pi^{+} \pi^{-} \pi^{+} \pi^{-} \pi^{0}$ & 1 M \\

  \hline
\end{tabular}
\caption{The number of analysed events for the $h_c$ decay and the selected background reactions}
\label{tab:perf:hc_nevts}
\end{center}
\end{table}

The following selection criteria were applied:
\begin{enumerate}
\item
$\phi$ candidates were defined as $K^{+}$, $K^{-}$ pairs with the invariant
mass in the window [0.8; 1.2] \gev. Two $\phi$ candidates in one event
with invariant mass in the window [2.6;3.2] \gev defined an $\eta_c$
candidate which, combined with a neutral candidate, formed an $h_c$
candidate.
\item
A 4C-fit to beam energy-momentum was applied to the $h_c$ candidate,
which was stored to a root ntuple together with the updated information
on its decay products.
\item
The following additional cuts are performed at the ntuple level for
background suppression:
\begin{enumerate}
\item
cut on the confidence level of the 4C-fit to beam energy-momentum,
$CL> 0.05$,
\item
$\eta_c$ invariant mass within [2.9; 3.06] GeV,
\item
$E_{\gamma}$ within [0.4; 0.6] GeV,
\item
$\phi$ invariant mass within [0.99; 1.05] GeV,
\item
no $\pi^{0}$ candidate in an event, i.e. no 2$\gamma$ invariant mass
in the range [0.115; 0.15] GeV with two different low energy photon
thresholds: 30 MeV and 10 MeV.
\end{enumerate}
\end{enumerate}

\begin{figure}
\begin{center}
\includegraphics[width=0.8\swidth]{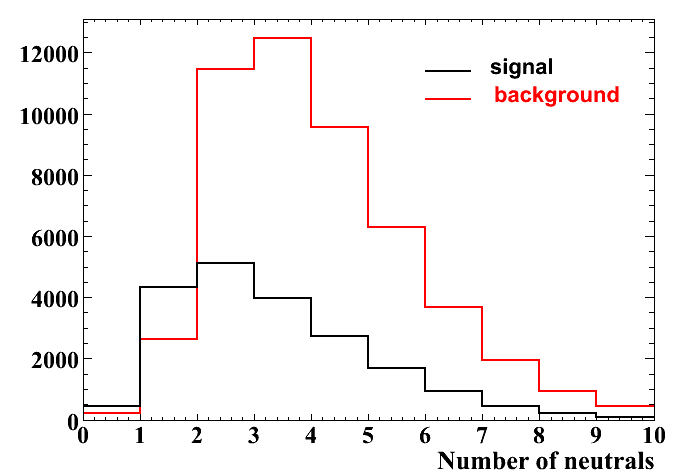}
\caption{The number of reconstructed EMC clusters for the $\overline{p} p \rightarrow h_c \rightarrow \eta_c \gamma$
and $\overline{p} p \rightarrow \phi \phi \pi^{0}$ reactions.}
\label{fig:sim:hc_n_gamma}
\end{center}
\end{figure}

\Reffig{fig:sim:hc_n_gamma} presents the multiplicity distribution of
reconstructed EMC clusters for the signal and one of the background
channels. One may note that the mean number of neutral candidates
exceeds one, the value expected for the signal, or two expected for
the background from $\pi^{0}$ decay. This is the result of hadronic
split-offs, caused by nuclear interactions of charged hadrons with the material of the detector.
Secondary particles produced in such interactions can create additional clusters in EMC, which cannot be associated with primary charged hadrons, and thus are interpreted as neutral candidates.
This effect makes it impossible to select as the signal the events
with only one cluster, because it leads to a significant drop in
efficiency. This observation emphasizes the importance of other
selection criteria, in particular of the veto on $\pi^{0}$ in an event.
The effect of the latter requirement strongly depends on the assumed
low energy photon threshold.

\begin{figure}
\begin{center}
\includegraphics[width=0.8\swidth]{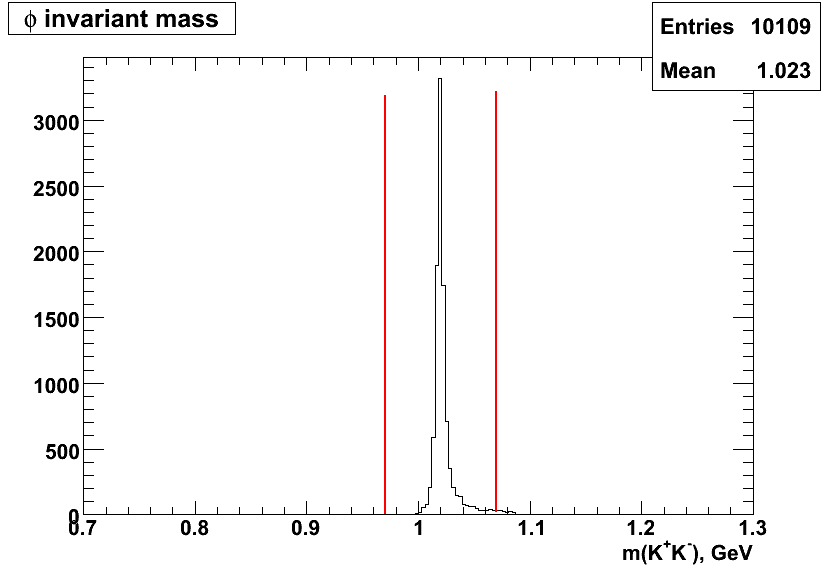}
\caption{Invariant mass of $K^{+}K^{-}$ with mass window on $\phi$ for $h_c \rightarrow \phi \phi \gamma$.}
\label{fig:sim:mphi_hc}
\end{center}
\end{figure}

\begin{figure}
\begin{center}
\includegraphics[width=0.8\swidth]{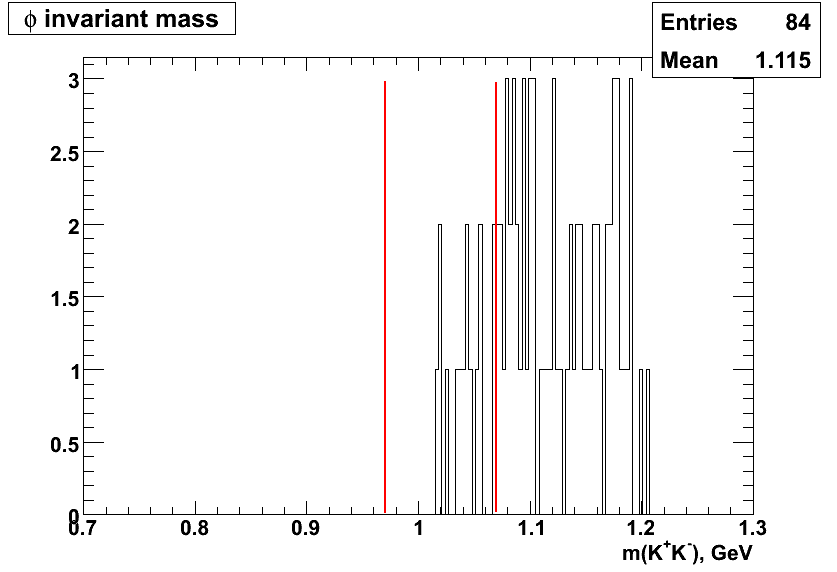}
\caption{Invariant mass of $K^{+}K^{-}$ with mass window set on $\phi$ for the $p \overline{p} \rightarrow K^{+}K^{-}K^{+}K^{-}\pi^{0}$ background.}
\label{fig:sim:mphi_4Kpi0}
\end{center}
\end{figure}

The distribution of $K^{+}$ - $K^{-}$ pair invariant mass is presented
in \Reffig{fig:sim:mphi_hc} for the signal and in
\Reffig{fig:sim:mphi_4Kpi0} for the $\overline{p} p \rightarrow K^{+} K^{-} K^{+} K^{-} \pi^{0}$
background. Windows imposed on the mass of $\phi$ are marked with vertical
lines (in red).

\subsection{Signal to background ratio and low energy $\gamma$-ray threshold}
The efficiencies of various cut selection criteria are listed in
\Reftbl{tab:sim:hc_eff} for the signal and the first three of the
considered background channels.

\begin{table}
\begin{center}
\begin{tabular}{|l|c|c|c|c|c|}
  \hline
  Selection criteria & signal & $4K \pi^{0}$ & $\phi  K^{+} K^{-} \pi^{0} $ & $\phi \phi \pi^{0}$ & $K^+ K^- \pi^+ \pi^- \pi^0$\\
    \hline
  pre-selection & 0.51 & $9.8\cdot 10^{-3}$ & $1.3\cdot 10^{-2}$ & $4.9\cdot 10^{-2}$& $9.0\cdot 10^{-6}$\\
  $CL>0.05$ & 0.36 & $1.5\cdot 10^{-3}$ & $2.0\cdot 10^{-3}$ & $7.0\cdot 10^{-3}$& $4.0\cdot 10^{-8}$\\
  $m(\eta_c), E_{\gamma}$ & 0.34 & $4.1\cdot 10^{-4}$ & $5.2\cdot 10^{-4}$ & $1.8\cdot 10^{-3}$ & 0\\
  $m(\phi)$ & 0.31 & $4.5\cdot 10^{-6}$ & $1.2\cdot 10^{-4}$ & $1.7\cdot 10^{-3}$ & 0\\
  $no \; \pi^0 (30\,MeV)$ & 0.26 & $2.7\cdot 10^{-6}$ & $4.5\cdot 10^{-5}$ & $9.2\cdot 10^{-4}$ & 0\\
  $no \; \pi^0 (10\,MeV)$ & 0.24 & $1.8\cdot 10^{-6}$ & $3.0\cdot 10^{-5}$ & $7.1\cdot 10^{-4}$ & 0\\

  \hline
\end{tabular}
\caption{Efficiency of different event selection criteria.}
\label{tab:sim:hc_eff}
\end{center}
\end{table}

For $\overline{p} p \rightarrow \Delta^{++}(1232) \overline{\Delta}^{--}(1232) \pi^{0} $
and $\overline{p} p \rightarrow \pi^{+} \pi^{-} \pi^{+} \pi^{-} \pi^{0}$ no event
has been found to pass pre-selection.

Assuming the $h_c$ production cross-section of $33 nb$ at resonance, one
obtains the signal to background ratios given in \Reftbl{tab:sim:hc_sb_ratio}.

\begin{table}
\begin{center}
\begin{tabular}{|l|c|}
  \hline
  channel & Signal/background ratio\\
    \hline
  $\overline{p} p \rightarrow K^{+} K^{-} K^{+} K^{-} \pi^{0}$ & $8$ \\
  $\overline{p} p \rightarrow \phi K^{+} K^{-} \pi^{0}$ & 8\\
  $\overline{p} p \rightarrow \phi \phi \pi^{0}$ & $>10$\\
  $\overline{p} p \rightarrow K^{+} K^{-} \pi^{+} \pi^{-} \pi^{0}$ & $>12$ \\
  \hline
\end{tabular}
\caption{Signal to background ratio for different $h_c$ background channels.}
\label{tab:sim:hc_sb_ratio}
\end{center}
\end{table}

For the $\overline{p} p \rightarrow \Delta^{++}(1232) \overline{\Delta}^{--}(1232) \pi^{0} $
and the $\overline{p} p \rightarrow \pi^{+} \pi^{-} \pi^{+} \pi^{-} \pi^{0}$
channels, because of their large cross-sections, one needs around
$10^{8}$ events to make valid conclusions on background suppression at
S/B levels above 1:1, which is not easy to fulfil because of the
required excessive CPU time. The $\overline{p} p \rightarrow K^{+} K^{-} \pi^{+} \pi^{-} \pi^{0}$
channel is more favorable from this point of view, having an order of
magnitude smaller cross-section. Fortunately, it permits to study the
influence of same factors, i.e. PID misidentification and power of the
4C-fit on background suppression. For the latter channel $2 \cdot 10^7$ events have been analysed and the achieved conclusion tells that the expected $S/B$ ratio will exceed 12:1 (see \Reftbl{tab:sim:hc_sb_ratio}).

For the $\overline{p} p \rightarrow \phi \phi \pi^{0}$ background
channel the reduction of low energy $\gamma$-ray threshold from 30 MeV
to 10 MeV gives $20\percent$ improvement in the signal to background
ratio, for the $\overline{p} p \rightarrow \phi K^{+} K^{-} \pi^{0}$
the corresponding improvement is $40\percent$.

It is worthwhile to note that finite energy resolution is taken into account in the digitization of EMC signals which results in around 14 \percent at 10 \mev in Monte Carlo simulations using the adopted software. For comparison the measurements presented in \Refsec{sec:VDG:results} give around 17\percent at this energy. However the readout with two APDs considered for PANDA will increase the collected light and improve resolution especially at these low energies where the statistical term is of higher importance. Taking this remark into account one may conclude that the resolution used in Monte Carlo simulations is consistent with the above presented measurements.

With the final signal selection efficiency of $24\percent$ (see
\Reftbl{tab:sim:hc_eff}) and the assumed luminosity in high luminosity
mode of $L=2 \cdot 10^{32}cm^{-2}s^{-1}$, the expected signal event rate
is 92 events/day. For the high resolution mode with
$L=2 \cdot 10^{31}cm^{-2}s^{-1}$, the expected signal event rate is 9
events/day, respectively.\\

\section{$h_c \rightarrow K_s K^{+} \pi^{-} \gamma$ decay mode}
The decay mode $\eta_c \rightarrow K_s K^{+} \pi^{-}$ has the highest branching ratio among the decay modes of $\eta_c$ (see \Reftbl{tab:sim:etac_decays}). The presence of $K_s$ in the final state allows a clear selection of this channel using the  geometrical constraint on the secondary decay vertex. The lifetime of $K_s$ corresponds to $c\tau$ = 2.68 cm. The dominant decay mode of $K_s$ is $K_s \rightarrow \pi^{+} \pi^{-}$ with branching ratio BR=68.95 \percent.
The complete reaction and decay path is as follows:
\begin{center}
$\overline{p} p \rightarrow h_c \rightarrow \eta_c \gamma \rightarrow K_S K^{+} \pi^{-} \gamma \rightarrow \pi^{+} \pi^{-} K^{+} \pi^{-} \gamma$.
\end{center}
\subsection{Background considerations}
The main source of background considered in this study is $p \overline{p} \rightarrow K_s K^{+} \pi^{-} \pi^{0}$, i.e. it differs from the studied signal by $\gamma$ replaced with $\pi^{0}$. Such a source of background is very typical for the study of electromagnetic transitions, where non-resonant background contains the same hadrons in the final state and one $\gamma$ from $\pi^{0}$ decay mimics the $\gamma$ from the considered electromagnetic transition.
The background cross section is unknown at the beam momentum $p_z=5.609 \, \gevc$, however it was measured at lower beam momentum $p_z=3.66 \, \gevc$ in 1966 in Brookhaven \cite{bib:Armenteros:1969}. It is equal 74 $\mu b$ and a rough estimate of the cross section at $p_z=5.609 \, \gevc$ was obtained by scaling it down with the total inelastic cross section. The obtained value is 60 $\mu b$.

Additional sources of background are $K^{+} K^{-} \pi^{+} \pi^{-} \pi^{0}$ and $\pi^{+} \pi^{-} \pi^{+} \pi^{-} \pi^{0}$ assuming PID misidentification of $\pi$ and $K$. However it will be shown below that the background channel $p \overline{p} \rightarrow K_s K^{+} \pi^{-} \pi^{0}$ cannot be sufficiently suppressed and these two additional channels have not been studied in detail.
\subsection{Event selection and background suppression}
The analysis has been performed with 80 k of the signal events $p \overline{p} \rightarrow h_c \rightarrow K_s K^{+} \pi^{-} \gamma$ and 100 k of the background events $p \overline{p} \rightarrow K_s K^{+} \pi^{-} \pi^{0}$.

The event selection was done in the following steps:
\begin{enumerate}
\item
$K_S$ candidates were selected combining two charged $\pi$ with the requirement of the common vertex and mass window [0.3; 0.8] GeV. $K_S$ were combined with $K^{+}$ and $\pi^{-}$ and provided their invariant mass fallen into the window [2.6; 3.2] GeV they defined an $\eta_c$
candidate. $\eta_c$ combined with a neutral candidate, formed an $h_c$ candidate.
\item
A 4C-fit to beam energy-momentum was applied to the $h_c$ candidate,
with the result stored to a root ntuple together with the updated information
on its decay products.
\item
The following additional cuts are performed at the ntuple level for
background suppression:
\begin{enumerate}
\item
cut on the confidence level of the 4C-fit to beam energy-momentum,
$CL> 0.001$,
\item
$\eta_c$ invariant mass within [2.9; 3.1] GeV,
\item
$E_{\gamma}$ within [0.4; 0.6] GeV,
\item
no $\pi^{0}$ candidates in the event, i.e. no 2$\gamma$ invariant mass
in the range [0.115; 0.15] GeV with two different low energy photon
thresholds: 30 MeV and 10 MeV.
\end{enumerate}
\end{enumerate}

\subsection{Results}
The efficiency of different selection criteria for the signal and background are presented in the \Reftbl{tab:sim:hc_KsKPi_eff}.
\begin{table}
\begin{center}
\begin{tabular}{|l|c|c|}
  \hline
  Selection criteria & signal & $K_s K^{+} \pi^{-} \pi^{0}$\\
    \hline
  pre-selection & 0.41 & $2.82 \cdot 10^{-2}$\\
  $CL>0.001$ & 0.27 & $4.29 \cdot 10^{-3}$\\
  $m(\eta_c), E_{\gamma}$ & 0.26 & $1.17 \cdot 10^{-3}$\\
  $no \; \pi^0 (30\,MeV)$ & 0.22 & $7.90 \cdot 10^{-4}$\\
  $no \; \pi^0 (10\,MeV)$ & 0.20 & $5.00 \cdot 10^{-4}$\\
  \hline
\end{tabular}
\caption{Efficiency of different event selection criteria for $K_S K^{+} \pi^{-} \gamma$ decay mode of $h_c$.}
\label{tab:sim:hc_KsKPi_eff}
\end{center}
\end{table}

The signal to background ratio 1:240 has been obtained with the signal efficiency of 20 \percent, which is prohibitively small, in our point of view, to perform this experiment. The signal to background ratio is improved by 20 \percent by reducing the $\gamma$-ray registration threshold from 30 MeV to 10 MeV, however this does not reverse the conclusion on difficulties with $h_c$ registration in this final state. This estimate is based on the assumption that the background cross section decreases with energy as the inelastic $p \overline{p}$ cross section. If it drops faster the signal to background ratio can be more favorable for performing this experiment.

\begin{figure}
\vspace{-10pt}
\begin{center}
\includegraphics[width=1.0\swidth]{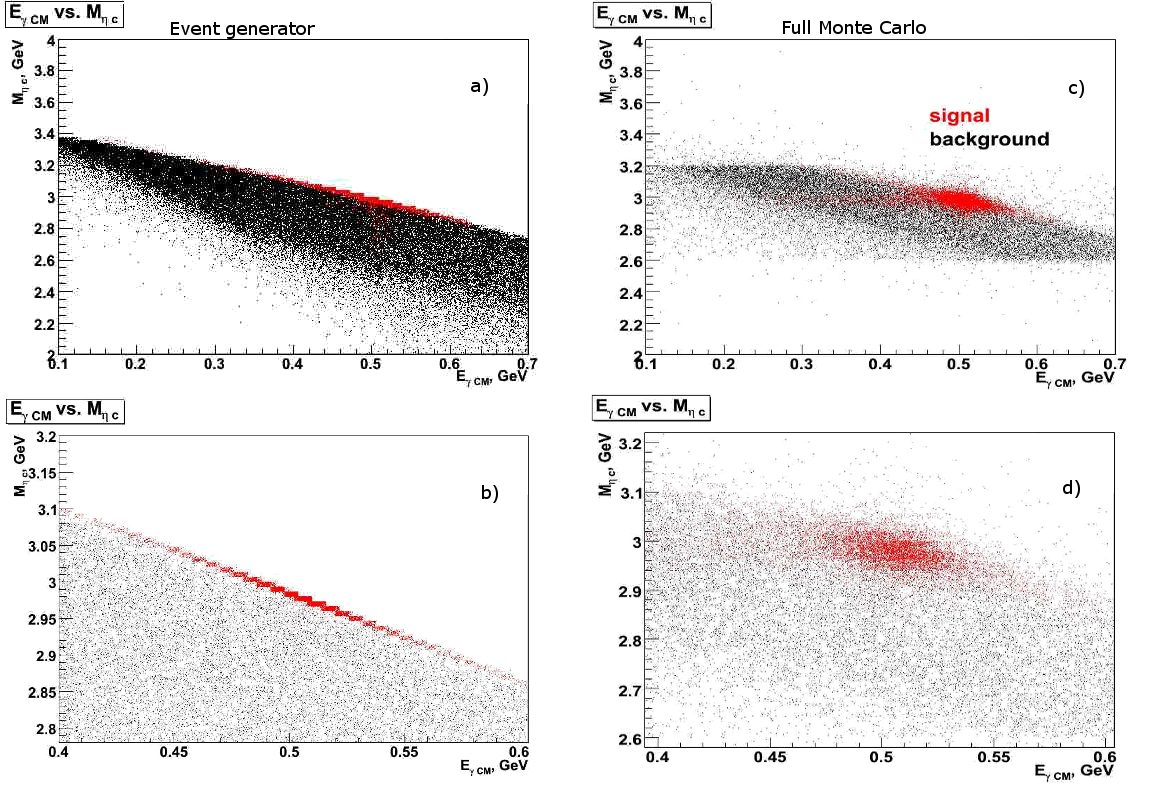}
\caption{Signal and background event distributions in the $m(\eta_c)$ vs $E_{\gamma}$ plane for the direct output of the event generator [a) and b)] and after full Monte Carlo analysis [c) and d)].}
\vspace{-10pt}
\label{fig:sim:metac_egamma_KsKPi}
\end{center}
\end{figure}

Additional explanation of the difficulty with signal and background separation follows from the inspection of \Reffig{fig:sim:metac_egamma_KsKPi}. This Fig. demonstrates the 2-dimensional distributions of the signal and background events in the $m(\eta_c)$ vs $E_{\gamma}$ plane. The panels a) and b) contain the events coming directly from the event generator, i.e. with no detector effects included. The panel b) is an expanded part of a) containing the region of signal events. One is prone to conclude from \Reffig{fig:sim:metac_egamma_KsKPi} b) that pure kinematics might suffice to separate the signal and background events. However, the finite detector resolution smears the region that corresponds to the signal so that it overlaps strongly with the background [see \Reffig{fig:sim:metac_egamma_KsKPi} d)].

\section{$h_c \rightarrow K^{0*} \overline{K^{0*}} \gamma$ decay mode}
The analysis of this decay mode is very similar to the two hadronic final states presented above. The branching ratio for the $\eta_c \rightarrow K^{0*} \overline{K^{0*}}$ decay mode is $BR=8.5\cdot 10^{-3}$. The main decay mode of the $K^{0*}$ resonance is $K^{0*} \rightarrow K^{+} \pi^{-}$ with the branching ratio BR=0.67.

The complete reaction and decay path is:
\begin{center}
$\overline{p} p \rightarrow h_c \rightarrow \eta_c \gamma \rightarrow K^{0*} \overline{K^{0*}} \gamma \rightarrow K^{+} \pi^{-} K^{-} \pi^{+} \gamma$.
\end{center}

\subsection{Background considerations}
The two background channels were considered for this decay mode of $h_c$:
\begin{center}
$\overline{p} p \rightarrow K^{+} K^{-} \pi^{+} \pi^{-} \pi^{0}$,
\end{center}
and
\begin{center}
$\overline{p} p \rightarrow K^{0*} K^{-} \pi^{+} \pi^{0} \rightarrow K^{+} K^{-} \pi^{+} \pi^{-} \pi^{0}$,
\end{center}
i.e. one intermediate $K^{0*}$ resonance is engaged in the latter background channel. The background cross sections were estimated in the same way as for the $K_S K^{+} \pi^{-} \gamma$ decay mode. The measurements at the beam momentum $p=3.66 \, \gevc$ give $\sigma_{p\overline{p}\rightarrow K^{+} K^{-} \pi^{+} \pi^{-} \pi^{0}}=34 \mu b$ \cite{bib:Armenteros:1969}. Scaling down according to the inelastic cross-section gives $\sigma = 30 \,\mu b$ at $p=5.6 \, \gevc$. At the 3.66 \gevc antiproton beam momentum 30 \percent of reactions goes via one $K^{0*}$ resonance. Assuming the same fraction at 5.6 \gevc momentum one gets an estimate $\sigma_{p\overline{p}}\rightarrow K^{0*}K^{-}K^{+}\pi^{0}=8.4 \, \mu b$. No evidence for $K^{0*} \overline{K^{0*}}$ emission was observed in \cite{bib:Armenteros:1969}.

\subsection{Event selection and background suppression}
The number of generated events for the signal and two background channels is collected in \Reftbl{tab:perf:hc_KstarKstar_nevts}.

\begin{table}
\begin{center}
\begin{tabular}{|l|c|}
  \hline
  Channel & N of events \\
    \hline
  $\overline{p} p \rightarrow h_c \rightarrow K^{0*} \overline{K^{0*}} \gamma$ & 80 k \\
  $\overline{p} p \rightarrow K^{0*} K^{-} \pi^{+} \pi^{0}$ & 1.1 M \\
  $\overline{p} p \rightarrow K^{+} K^{-} \pi^{+} \pi^{-} \pi^{0}$ & 200 k \\
  \hline
\end{tabular}
\caption {The numbers of analysed events for $h_c \rightarrow K^{0*} \overline{K^{0*}} \gamma$ decay and background reactions}
\label{tab:perf:hc_KstarKstar_nevts}
\end{center}
\end{table}

The event selection was done in the following steps:
\begin{enumerate}
\item
$K^{*0}$ candidates were selected combining $K^{+}$ and $\pi^{-}$ pairs with the requirement for a common vertex and invariant mass in the window [0.7; 1.1] GeV. $\overline{K^{*0}}$ is produced in the charge conjugated combination within the same mass window. The pair of $K^{*0}$ and $\overline{K^{*0}}$ with invariant mass in the window [2.6; 3.2] GeV defines an $\eta_c$ candidate. $\eta_c$ combined with a neutral candidate, formed an $h_c$ candidate.
\item
A 4C-fit to beam energy-momentum was applied to the $h_c$ candidate,
which was stored to a root ntuple together with the updated information
on its decay products.
\item
The following additional cuts are performed at the ntuple level for
background suppression:
\begin{enumerate}
\item
cut on the confidence level of the 4C-fit to beam energy-momentum,
$CL> 0.1$,
\item
$\eta_c$ invariant mass within [2.93; 3.03] GeV,
\item
$E_{\gamma}$ within [0.4; 0.6] GeV,
\item
invariant mass of $K^{*0}$ and $\overline{K^{*0}}$ within [0.86; 0.94] GeV mass window.
\item
no $\pi^{0}$ candidates in the event, i.e. no 2$\gamma$ invariant mass
in the range [0.115; 0.15] GeV with two different low energy photon
thresholds: 30 MeV and 10 MeV.
\end{enumerate}
\end{enumerate}

\subsection{Results}
The efficiency of different selection criteria for the signal and background are presented in the \Reftbl{tab:sim:hc_KstarKstar_eff}.
\begin{table}
\begin{center}
\begin{tabular}{|l|c|c|c|}
  \hline
  Selection criteria & signal & $K^{0*} K^{-} \pi^{+} \pi^{0}$ & $K^{+} K^{-} \pi^{+} \pi^{-} \pi^{0}$ \\
    \hline
  pre-selection & 0.43 & $1.4 \cdot 10^{-2}$ & $6.6 \cdot 10^{-3}$\\
  $CL>0.1$ & 0.31 & $2.3 \cdot 10^{-3}$ & $1.0 \cdot 10^{-3}$\\
  $m(\eta_c), E_{\gamma}$ & 0.27 & $3.5 \cdot 10^{-4}$ & $1.9 \cdot 10^{-4}$\\
  $m(K^{0*})$, m($\overline{K^{0*}})$ & 0.13 & $4.5 \cdot 10^{-5}$ & 0\\
  $no \; \pi^0 (30\,MeV)$ & 0.11 & $2.4 \cdot 10^{-5}$ & 0\\
  $no \; \pi^0 (30\,MeV)$ & 0.10 & $1.3 \cdot 10^{-5}$ & 0\\
  \hline
\end{tabular}
\caption{Efficiency of different event selection criteria for the $K^{0*} \overline{K^{0*}} \gamma$ decay mode of $h_c$.}
\label{tab:sim:hc_KstarKstar_eff}
\end{center}
\end{table}

The obtained signal to background ratio is 1:10 for the $K^{0*} K^{-} \pi^{+} \pi^{0}$ and for the $K^{+} K^{-} \pi^{+} \pi^{-} \pi^{0}$ channel the limit $>$ 1:15 was established. So the conclusion is that this decay mode presents difficulties for an experimental observation.

\section{Determination of the $h_c$ width}\label{sec:MC:hc_width}

In this section we investigate the performance of PANDA in
determination of the $h_c$ width. For this purpose we performed Monte
Carlo simulations of energy scans around the resonance. Events were
generated at 10 different energies around the $h_c$ mass, each point
corresponding to 5 days of measurements in high resolution mode.

The expected shape of measured resonance in
$\overline{p} p \rightarrow h_c \rightarrow \eta_c \gamma$ is the
convolution of the Breit-Wigner resonance curve with the
normalised beam energy distribution and an added background term. The
expected number of events at the $\textit{i}$-th data point is:
\begin{equation}
\label{eq:MC:width_nu}
\nu_{i}=[\varepsilon \times \int Ldt]_{i} \times [\sigma_{bkgd}(E)+\frac{\sigma_{p}\Gamma^{2}_{R}/4}{(2\pi)^{1/2}\sigma_{i}} \times
\int\frac{e^{-(E-E')^{2}/2\sigma^{2}_{i}}}{(E'-M_{R})^{2}+\Gamma_{R}^{2}/4}dE'],
\end{equation}
where $\sigma_{i}$ is the beam energy resolution at the $\textit{i}$-th
data point, $\Gamma_{R}$ and $M_{R}$ the resonance width and mass,
$\sigma_{p}$ incorporates branching ratios for the formation and decay
(see \Refeq{eq:MC:sigma}), the factor in square brackets in front of the r.h.s. of
\Refeq{eq:MC:width_nu} is the product of $\varepsilon$, an overall efficiency and
acceptance factor and the integrated luminosity at the $\textit{i}$-th
point of measurements. To extract the resonance parameters the
likelihood function, -ln$\mathcal{L}$, is minimized assuming Poisson
statistics, where:
\begin{equation}
\mathcal{L}=\prod_{j=1}^{N}\frac{\nu_{j}^{n_{j}}e^{-\nu_{j}}}{n_{j}!}.
\end{equation}

\begin{figure}
\begin{center}
\includegraphics[width=0.8\swidth]{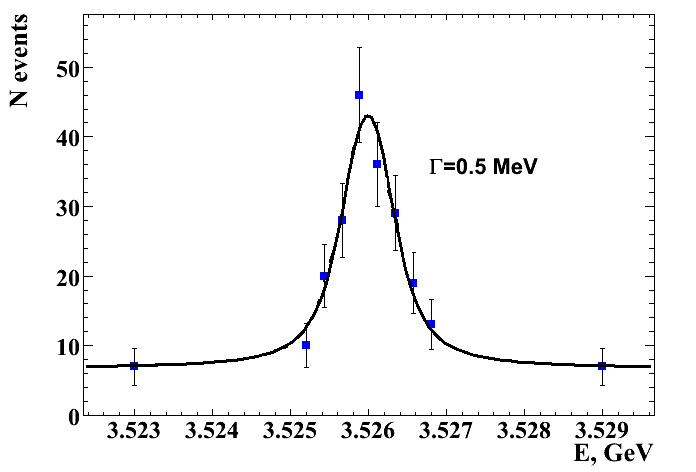}
\caption{Fit of the Breit-Wigner distribution convoluted with gaussian resolution of the beam momentum (solid line) to the "experimental points" generated assuming that the $h_c$ resonance width is $\Gamma$ = 0.5 \mev.}
\label{fig:sim:hc_width05MeV}
\end{center}
\end{figure}

\begin{figure}
\begin{center}
\includegraphics[width=0.8\swidth]{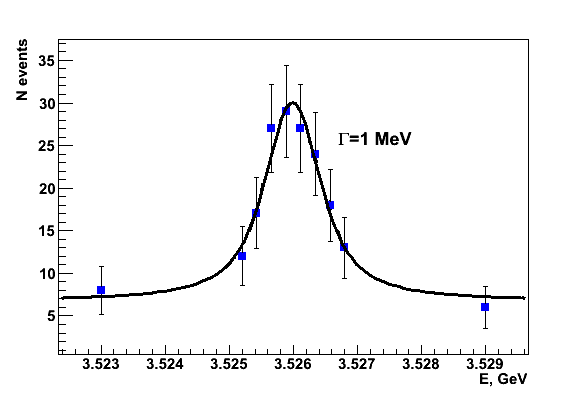}
\caption{Fit of the Breit-Wigner distribution convoluted with gaussian resolution of the beam momentum (solid line) to the "experimental points" generated assuming that the $h_c$ resonance width is $\Gamma$ = 1.0 \mev.}
\label{fig:sim:hc_width1MeV}
\end{center}
\end{figure}

For our simulation we assumed a signal to background ratio of 8:1 and we
used the signal reconstruction efficiency of the
$h_c \rightarrow \eta_c \gamma \rightarrow \phi \phi \gamma$ channel (see \Refsec{sec:MC:hc_phiphi}).
The simulated data were fitted to the expected signal shape with four
free parameters: $E_R$, $\Gamma_{R}$, $\sigma_{bkgd}$, $\sigma_p$. The
background was assumed energy independent. The study has been
repeated for three different $\Gamma_{R}$ = 0.5, 0.75 and 1.0 MeV. The
results of the fit for 0.5 MeV and 1.0 MeV are presented in
\Reffig{fig:sim:hc_width05MeV} and \Reffig{fig:sim:hc_width1MeV},
respectively. The extracted $\Gamma_{R}$'s ($\Gamma_{R,\, reco}$) with errors ($\Delta \Gamma_R$) are summarized in
\Reftbl{tab:sim:hc_width}. As a conclusion one may state that with the designed parameters of the beam energy resolution the precise determination of the width of the $h_c$ resonance is feasible with PANDA. However the scan should be performed in High Resolution mode of HESR which provides lower luminosity and therefore is rather time consuming.

\begin{table}
\begin{center}
\begin{tabular}{|c|c|c|}
  \hline
  $\Gamma_{R, MC}$, MeV & $\Gamma_{R, reco}$, MeV & $\Delta \Gamma_{R}$, MeV\\
  \hline
  1 & 0.90 & 0.22 \\
  0.75 & 0.71 & 0.17\\
  0.5 & 0.52 & 0.14\\
  \hline
\end{tabular}
\caption{Reconstructed $h_c$ width ($\Gamma_{R,\, reco}$) by fitting to the "experimental points" generated in Monte-Carlo simulations described in \Refsec{sec:MC:hc_width}, assuming that the width is $\Gamma_{R,\, MC}$.}
\label{tab:sim:hc_width}
\end{center}
\end{table}

\section{Reconstruction of the angular distribution of photons in the $h_c$ decay}\label{sec:MC:angdist}

From the measurements of $h_c$ in the CLEO experiment \cite{bib:sim:hc_CLEO} the decay of $h_c$ is consistent with the hypothesis of spin-parity $J^{PC}=1^{+-}$, however with rather high uncertainty. The present Monte Carlo study has been performed to determine the minimum duration of an experiment with the PANDA detector which would determine angular distribution of photons with sufficient precision that could serve a proof of $J^{PC}=1^{+-}$.

\begin{figure}
\begin{center}
\includegraphics[width=0.8\swidth]{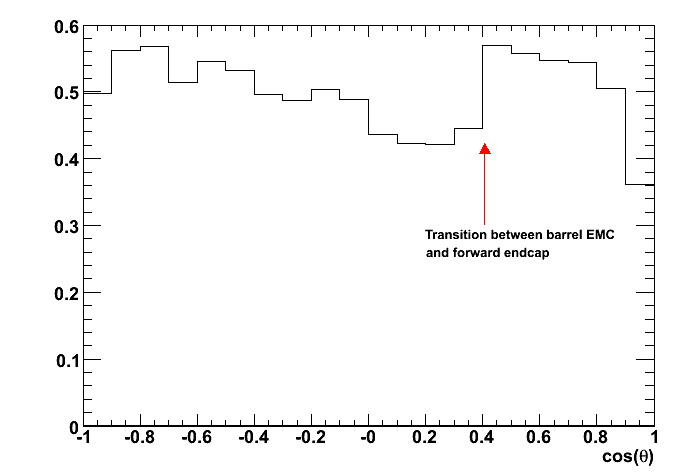}
\caption{Efficiency of $h_c$ reconstruction in the 20 intervals of $\cos{\theta}$.}
\label{fig:sim:accepatnce}
\end{center}
\end{figure}

To study the feasibility of reconstruction of the angular distribution in $h_c\rightarrow\eta_c + \gamma$ decay, the proper angular distribution was implemented into Monte Carlo simulations. EvtGen event generator provides the possibility to define angular distribution of the decay. The $\hat{W}(\theta)\sim \sin ^2 \theta$ dependence was implemented for $p \overline{p}\rightarrow h_c\rightarrow\eta_c + \gamma$ reaction which follows from \Refsec{sec:theory:angular_dist}, where $\theta$ is the polar angle of $\eta_c$ emission with respect to the proton momentum in the $\overline{p}p$ center-of-mass system. The $\eta_c \rightarrow \phi \phi$ channel was selected for reconstruction of the angular distribution as the most promising for $h_c$ detection (see \Refsec{sec:MC:hc_phiphi}) from the point of view of signal/background ratio.
It should be taken into account that efficiency of the EMC is not uniform in $\cos{\theta}$ to reconstruct properly the angular distribution. Events with phase space uniform distribution for $h_c\rightarrow\eta_c + \gamma$ decay have been generated and reconstructed to determine correction factors taking into account deviations from uniform detection as a function of the polar angle. Efficiency of $h_c$ reconstruction as a function of $\cos{\theta}$ is presented in \Reffig{fig:sim:accepatnce}. The characteristic drop in efficiency can be seen around $\cos{\theta}=0.4$. This region corresponds to the transition between the barrel EMC and the forward endcap.

\begin{figure}
\begin{center}
\includegraphics[width=0.8\swidth]{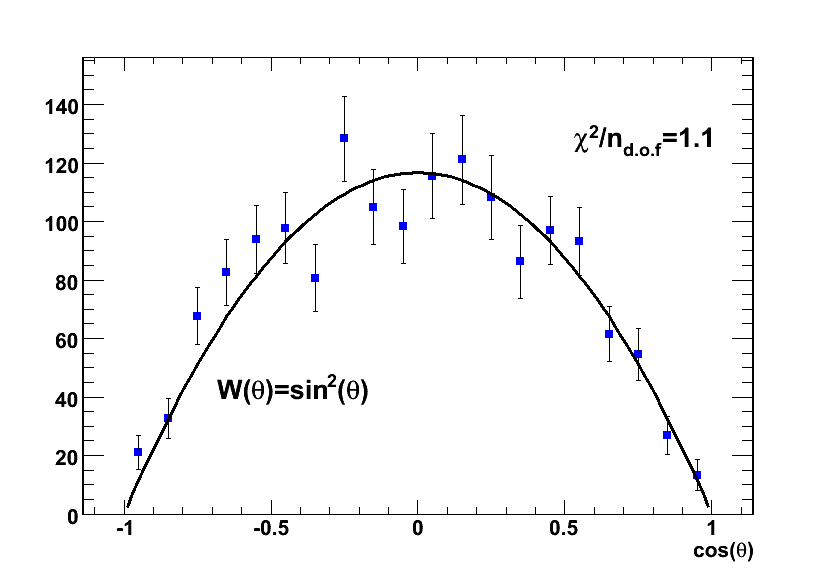}
\caption{The distribution of photons emitted in $\overline{p}p \rightarrow h_c \rightarrow \eta_c + \gamma$ in the 20 intervals of $\cos{\theta}$, assuming $J^{PC}=1^{+-}$ for $h_c$, which leads to E1 emission in the transition $h_c \rightarrow \eta_c$.}
\label{fig:sim:angular_distr}
\end{center}
\end{figure}

The reconstructed angular distribution of $\gamma$-rays from the $p \overline{p}\rightarrow h_c\rightarrow\eta_c + \gamma$ reaction, corrected for acceptance of the detector, is presented in \Reffig{fig:sim:angular_distr} as a distribution in $\cos{\theta}$. The presented data points correspond to 10 days of running with the estimated production rate of 92 events/day in the High Luminosity mode. The entire angular range in $\cos{\theta}$ was divided into 20 intervals. The fit with $\hat{W}(\theta)= \sin ^2 \theta$ hypothesis gives $\chi^2/n_{d.o.f}=1.1$. We conclude that the accumulated statistics is sufficient to identify $J^{PC}=1^{+-}$ for the $h_c$ resonance, uniquely.

%% file: conclusions.tex
\chapter{Conclusions}\label{chap:Conclusions}

The PANDA experiment at the FAIR facility is in the stage of finalizing its design with the beginning of construction expected in the nearest one or two years. With the start of operation expected in 2015 it will contribute to many topics in hadron physics and the high precision charmonium spectroscopy is a one of the major tasks of PANDA. Electromagnetic transitions to lower lying states are among the main decay modes with the aid of which the states of charmonium can be identified. This emphasizes the importance of an electromagnetic calorimeter (EMC) as one of the main components of the PANDA detector. The electromagnetic calorimeter of PANDA will be based on lead tungstate (PWO) scintillators readout with avalanche photodiodes (APDs).

In this thesis the results of measurements of energy resolution of lead tungstate scintillators readout with APDs are presented. Energy resolution was measured in the energy range from 6 to 17 \mev and around 25 \mev, which is close to the low energy threshold assumed for the PANDA EMC. Two series of measurement have been performed: using protons and low energy $\gamma$-rays produced in radiative capture reactions.

Protons with energy around 25 \mev accelerated with the C-30 cyclotron, were used to measure energy and time resolution of small size PWO-II scintillators $20\times 20 \times 40 \, mm^3$ readout with several types of avalanche photodiodes. One of the purposes of these measurements was to gain a practical experience of work with this scintillator having a rather low light yield and to draw conclusions on how the readout performance is improved when the APD is cooled to -20\degrees{C}. The energy resolution measured with two types of API APDs 17\percent and 22\percent appeared better then 35\percent measured with the Hamamatsu APD, which is to large extent an effect of twice smaller active area of the latter. However the Hamamatsu S8664-1010 APD was selected for the PANDA experiment because of its mechanical design and lower operating voltage are better fitting large scale applications. Simulation of light propagation in a scintillator and its APD detection performed with the Litrani program showed that the setup was not optimum from the point of view of light collection, which resulted in high contribution of the statistical term to energy resolution. At the same time the time resolution below 1 ns was demonstrated for PWO-II scintillator with APD readout. The results of these measurements were taken into account during construction of the setup for measurements of energy resolution with low energy photons produced in radiative capture reactions.

The setup was constructed to measure energy resolution of cooled PWO scintillator with Hamamatsu APD readout at the three discrete $\gamma$-ray energies of 6.13 \mev, 12.8 \mev and 17.2 \mev, the latter two produced in $^{11}B(p,\gamma)^{12}C$ reaction and the former one in $^{19}F(p,\alpha)^{16}O^{*}$ on the target contaminant. The crystals of the size $20\times 20 \times 200 \, mm^{3}$ were used, which is close to the final design of crystals for the PANDA EMC. The goal temperature $-25\degrees C$ has not been achieved due to limitation of the cooling capacity of the used JULABO F32-ME Refrigerated/Heating Circulator and a measurement at $-20\degrees C$ has been performed. The energy resolution $19\pm 3 \,\%$, $15\pm 2 \,\%$ and $9.6\pm 1.4 \,\%$ has been obtained for the three energies, respectively. These confirm nicely that the smooth energy dependence of resolution, established in the energy range 40.9-674.5 \mev by the experiment performed at the MAMI facility in Mainz (Germany) [\Refeq{eq:sigmaE_mami}] gives a valid extrapolation of resolution into this low energy range.

The Monte Carlo simulations aimed to study the performance of the PANDA detector in detection of the charmonium state $h_c$ demonstrated feasibility of the study of this state with high precision with PANDA. With the main decay channel $h_c\rightarrow \eta_c + \gamma$ it can be registered exclusively in either neutral or hadronic final states of $\eta_c$. Several possible decay modes of $\eta_c$ have been considered ($K_s K^{+} \pi^{-}$, $\overline{K^{*}}K^{*}$, $\phi\phi$ and $\gamma \gamma$). It was demonstrated that the most promising decay mode for $h_c$ observation is $\phi\phi\gamma$, where signal to background ratio around 8:1 is expected. With the signal selection efficiency of 24\percent around 92 events per day are expected in the high-luminosity mode. An alternative channel for $h_c$ registration is a neutral final state $h_c\rightarrow \eta_c +\gamma \rightarrow \gamma \gamma \gamma$, however with lower predicted event rate due to small efficiency and low branching ratio of the $\eta_c\rightarrow \gamma \gamma$ decay. A study of the PANDA sensitivity in determination of the $h_c$ width has been performed and estimates of precision and needed running time for the three hypotheses concerning the $h_c$ width: $\Gamma$=0.5 \mev, 0.75 \mev and 1 \mev were obtained. The feasibility of reconstruction of the angular distribution of the $h_c$ decay photons with the required precision for confirmation of the spin-parity assignment $J^{PC}=1^{+-}$ was also demonstrated. 

%% file: acknowledgments.tex
\chapter*{Acknowledgements}
First of all, I would like to thank my supervisor, Dr. hab. Boguslaw Zwieglinski for his help and efforts in preparation of this Thesis and providing me a possibility to work for the PANDA experiment, which was a valuable and exciting experience.

Many thanks to my colleagues from the Andrzej Soltan Institute for Nuclear Studies who helped me with my work and especially to Dr. Andrzej Korman for his help with measurements on the Van-de-Graaff accelerator and to Dr. Jolanta Wojtkowska for her help with measurements at the C-30 cyclotron. I also appreciate the technical support rendered by Mr. Wladyslaw Mielczarek.

I am grateful to Professor Rainer Novotny from the Giessen University who allowed me to participate in several measurements with PWO scintillators organised by him in Mainz University. I learned a lot about good organization in experimental physics and gained from him a lot of practical experience.

Many thanks are due to the group from the Bochum University, namely: Bertram Kopf, Marc Pelizaeus, Matthias Steinke and Jan Zhong from whom I learned a lot about Monte Carlo simulations and data analysis.

Thanks to the rest of my colleagues from the PANDA collaboration, whom I do not mention here by names, for numerous interesting discussions.